\documentclass{aa}

\usepackage[normalem]{ulem}

\usepackage{siunitx}
\usepackage{amsmath}
\usepackage{enumitem}
\usepackage{times}
\usepackage{placeins}

\usepackage{soul}
\usepackage{graphicx}
\usepackage{xcolor}
\usepackage{xspace}
\usepackage{natbib,twoopt}
\usepackage[breaklinks=true]{hyperref} 
\bibpunct{(}{)}{;}{a}{}{,}             

\usepackage{csquotes}

\usepackage[figuresright]{rotating}
\usepackage{tabularx}

\usepackage{float} 
\usepackage{subfig}
\usepackage{amssymb} 

\usepackage{wrapfig} 
\usepackage{longtable} 

\usepackage{wallpaper} 
\usepackage{geometry} 
\usepackage{tikz} 
\usetikzlibrary{positioning}
\usepackage{datetime} 

\usepackage{multirow} 
\usepackage{pdflscape} 
\newcommand{\SAAB}[0]{\textit{S\underline{A}B}\xspace}
\newcommand{\SABB}[0]{\textit{SA\underline{B}}\xspace}
\newcommand{\SA}[0]{\textit{SA}\xspace}
\newcommand{\SAB}[0]{\textit{SAB}\xspace}
\newcommand{\SB}[0]{\textit{SB}\xspace}
\newcommand{\nAG}[0]{\textbf{\textit{N/A}}\xspace}

\usepackage{txfonts}

\begin{document}

\title{The Gas Morphology of Nearby Star-Forming Galaxies}


\author{Sophia K. Stuber\inst{\ref{i1}}
          \and Eva Schinnerer\inst{\ref{i1}}
          \and Thomas G. Williams\inst{\ref{i2}}
          \and Miguel Querejeta\inst{\ref{i3}}
          \and Sharon Meidt\inst{\ref{i4}}
          \and Eric Emsellem\inst{\ref{i5}, \ref{i6}}
          \and Ashley Barnes\inst{\ref{i5}, \ref{i7}}
          \and Ralf S.\ Klessen\inst{\ref{i8}, \ref{i9}}
          \and Adam K. Leroy \inst{\ref{i10}}
          \and Justus Neumann \inst{\ref{i1}}
          \and Mattia C.~Sormani \inst{\ref{i8}}
          \and Frank Bigiel \inst{\ref{i7}}
          \and M\'elanie Chevance\inst{\ref{i8}, \ref{i11}}
          \and Danny Dale\inst{\ref{i12}}
          \and Christopher Faesi\inst{\ref{i18}}
          \and Simon C.~O.~Glover\inst{\ref{i8}}
          \and Kathryn Grasha\inst{\ref{i13}, \ref{i14}}
          \and J.~M.~Diederik Kruijssen\inst{\ref{i11}}
          \and Daizhong Liu\inst{\ref{i15}}
          \and Hsi-an Pan\inst{\ref{i16}}
          \and Jerome Pety\inst{\ref{i19},\ref{i20}}
          \and Francesca Pinna\inst{\ref{i1}}
          \and Toshiki Saito\inst{\ref{i17}}
          \and Antonio Usero\inst{\ref{i3}}
          \and Elizabeth J. Watkins\inst{\ref{i7}}
          }

\institute{
    Max-Planck-Institut für Astronomie, Königstuhl 17, 69117 Heidelberg Germany\label{i1}
    \and 
    Sub-department of Astrophysics, Department of Physics, University of Oxford, Keble Road, Oxford OX1 3RH, UK\label{i2}
    \and 
    Observatorio Astron{\'o}mico Nacional (IGN), C/Alfonso XII 3, Madrid E-28014, Spain\label{i3}
    \and 
    Sterrenkundig Observatorium, Universiteit Gent, Krijgslaan 281 S9, B-9000 Gent, Belgium\label{i4}
    \and 
    European Southern Observatory, Karl-Schwarzschild-Stra{\ss}e 2, 85748 Garching, Germany\label{i5}
    \and 
    Univ Lyon, Univ Lyon1, ENS de Lyon, CNRS, Centre de Recherche Astrophysique de Lyon UMR5574, F-69230 Saint-Genis-Laval France\label{i6}
    \and 
    Argelander-Institut für Astronomie, Universität Bonn, Auf dem Hügel 71, 53121 Bonn, Germany\label{i7}
    \and 
    Universit\"{a}t Heidelberg, Zentrum f\"{u}r Astronomie, Albert-Ueberle-Str. 2, 69120 Heidelberg, Germany\label{i8}
    \and 
    Universit\"{a}t Heidelberg, Interdisziplin\"{a}res Zentrum f\"{u}r Wissenschaftliches Rechnen, Im Neuenheimer Feld 205, 69120 Heidelberg, Germany \label{i9}
    \and 
    Department of Astronomy, The Ohio State University, 140 West 18th Ave, Columbus, OH 43210, USA\label{i10}
    \and 
    Cosmic Origins Of Life (COOL) Research DAO, coolresearch.io\label{i11}
    \and 
    Department of Physics \& Astronomy, University of Wyoming, Laramie, WY 82071, USA\label{i12}
    \and 
    Research School of Astronomy and Astrophysics, Australian National University, Canberra, ACT 2611, Australia\label{i13}  
    \and 
    ARC Centre of Excellence for All Sky Astrophysics in 3 Dimensions (ASTRO 3D), Australia\label{i14}
    \and 
    Max-Planck-Institut f\"{u}r extraterrestrische Physik, Giessenbachstra{\ss}e 1, D-85748 Garching, Germany\label{i15}
    \and 
    Department of Physics, Tamkang University, No.151, Yingzhuan Road, Tamsui District, New Taipei City 251301, Taiwan\label{i16}
    \and 
    National Astronomical Observatory of Japan, 2-21-1 Osawa, Mitaka, Tokyo, 181-8588, Japan\label{i17}
    \and 
    Department of Physics, University of Connecticut, Storrs, CT, 06269, USA\label{i18} 
    \and 
    IRAM, 300 rue de la Piscine, 38400 Saint Martin d'H\`eres, France\label{i19}
    \and 
    LERMA, Observatoire de Paris, PSL Research University, CNRS, Sorbonne Universit\'es, 75014 Paris\label{i20}
    }

\date{Accepted for publication in A\&A on May 22, 2023}

\abstract{

The morphology of a galaxy stems from secular and environmental processes during its evolutionary history. Thus galaxy morphologies have been a long used tool to gain insights on galaxy evolution. 
We visually classify morphologies on cloud-scales based on the molecular gas distribution of a large sample of 79 nearby main-sequence galaxies, using 1\arcsec\ resolution \mbox{CO(2--1)} ALMA observations taken as part of the PHANGS survey.
To do so, we devise a morphology classification scheme for different types of bars, spiral arms (grand-design, flocculent, multi-arm and smooth), rings (central and non-central rings) similar to the well-established optical ones, and further introduce bar lane classes. 

In general, our cold gas based morphologies agree well with the ones based on stellar light. Both our bars as well as grand-design spiral arms are preferentially found at the higher mass end of our sample.
Our gas-based classification indicates a potential for misidentification of unbarred galaxies in the optical when massive star formation is present. Central or nuclear rings are present in a third of the sample with a strong preferences for barred galaxies (59\%).

As stellar bars are present in $45\pm5\%$ of our sample galaxies, we explore the utility of molecular gas as tracer of bar lane properties.
We find that more curved bar lanes have a shorter radial extent in molecular gas and reside in galaxies with lower molecular to stellar mass ratios than those with straighter geometries. 
Galaxies display a wide range of CO morphology, and this work provides a catalogue of morphological features in a representative sample of nearby galaxies. 
}

   \maketitle
%

\section{Introduction}
\label{sec:Motivation}
The classification of galaxies based on their morphological appearance has been a fundamental topic in astronomy since the early 20th century \citep[e.g., the Hubble tuning fork;][]{hubble_no_1926, hubble_realm_1936}.
It turns out that there is a strong (anti-) correlation between the occurrence of features such as spiral arms and stellar bars and the star formation rate (SFR) or the star formation history \citep[][]{oswalt_galaxy_2013}, making such features important parameters to study. 

Historically, most observations of galaxy morphology are performed on optical or infrared images, identifying stellar bars \citep[e.g.,][]{hubble_realm_1936, de_vaucouleurs_revised_1963,buta_classical_2015}, spiral arms \citep[e.g.,][]{elmegreen_flocculent_1982} and rings \citep[][]{de_vaucouleurs_third_1991, schwarz_response_1981} in the stellar distribution of different types of disk galaxies \citep[including e.g., dwarf and anemic disk galaxies; ][]{de_vaucouleurs_classification_1959, van_den_bergh_new_1976}.  
Over time, the morphological information has been expanded to higher redshifts \citep[e.g.,][]{sheth_barred_2003, moran_wide-field_2007, tsukui_spiral_2021}, to specific sub-samples  \citep[e.g., isolated galaxies;][]{buta_comprehensive_2019} or to different identification methods, such as crowd-sourcing \citep[e.g., ][]{lintott_galaxy_2008} or machine-learning algorithms \citep[e.g., distinguishing spiral from elliptical or merging galaxies;][]{ciprijanovic_deepadversaries_2022, dominguez_sanchez_sdss-iv_2022}. 

In addition to this wide variety of work based on the stellar distribution, observations of other phases of the interstellar medium (ISM) have been proven fruitful \citep[e.g., studying star formation in bars in H$\alpha$:][]{fraser-mckelvie_sdss-iv_2020}. 
Comparisons between galaxy morphologies in stellar light and other ISM phases can provide insight into how morphological features decouple and where new generations of stars may form. 

Especially cold molecular gas and its dissipational non-linear response to the gravitational potential may provide a sharpened picture of the underlying morphology and allow for important insights into what drives the secular evolution of galaxies. 
The variation in azimuthal contrasts in CO and $3.6\,\mu$m emission (tracing the stellar contribution) suggests that molecular gas in nearby galaxies is organized by large-scale stellar dynamical features \citep{meidt_organization_2021}. 
However, molecular clouds may not populate all orbits or orbit phases equally. 
Instead, the dissipative gas can reveal higher order features like orbit intersections due to non-axisymmetric potentials that give rise to shocks and radial gas flows. 
This can highlight dust lanes similar to dust observations \citep[e.g.,][]{comeron_curvature_2009, Thilker_2023}.
Further, galaxies out of equilibrium can result in a delayed or enhanced gas response that introduces departures from the underlying stellar morphology in the gas organization. 
Establishing when and where the two morphologies agree or disagree (as a function of galactic environment, within and among galaxies) indirectly probes the dominant factors responsible for establishing the morphology of the gas.

While the most abundant molecule, H$_2$, is difficult to observe due to its large rotational constant and lack of a permanent dipole moment, the second most abundant molecule,
CO, has proven to be a good tracer of spiral structure \citep[e.g., BIMA SONG; HERACLES;][]{helfer_bima_2003, leroy_heracles_2009}, and its emission can be linked to the overall H$_2$ mass \citep[][]{bolatto_co--h2_2013}. 
Still, morphological surveys based on cold molecular gas in the local universe have suffered until now from either small sample sizes and/or low $\sim$kpc resolutions and sensitivity \citep[e.g.,][]{sheth_molecular_2000, helfer_bima_2003, bolatto_edge-califa_2017}.
With  1" (${\sim}100\,$pc) resolution \mbox{CO(2--1)} observations with ALMA, the Physics at High Angular resolution in Nearby GalaxieS survey \citep[PHANGS,][]{leroy_phangsalma_2021, leroy_phangsalma_2021-1} allows us to perform classifications based on the molecular gas morphology for 90 nearby main sequence (MS) galaxies. 
We aim to learn 
what the relative roles of the background potential and gas properties are in driving global and local gas morphologies.
This systematic study of CO morphologies which we put in relation to stellar morphologies provides a much-needed test of CO as a tracer of global morphology.

\subsection{Previous classifications}

Although the Hubble tuning fork \citep{hubble_no_1926, hubble_realm_1936} still remains one commonly used basis for current classifications, several additional classifications were made over time. 
This includes subdivisions of bar-classes into 
strong (\SB) and weak bars \citep[e.g., bars with intermediate apparent bar strengths, \SAB][]{de_vaucouleurs_revised_1963}, 
or even mixed classes like \SAAB (a bar showing less clear traces of a bar, such as bars at low contrast or a bar with broad oval shapes) or \SABB \citep[clear and well defined bar but visually weaker-looking bar than class \SB;][]{buta_classical_2015}. 
Stellar bars are known to funnel gas to the center of a galaxy, as revealed by correlations between the presence of a stellar bar and the central molecular gas concentration \citep[][]{sakamoto_bar-driven_1999, sheth_secular_2005}.
Still, the exact lifetime of these bars, their formation mechanism or even the fraction of galaxies that host stellar bars is still highly debated \citep[e.g.,][]{erwin_dependence_2018, combes_starbursts_2006}, making this one of the most important features to study.

Other works focused on classifying the number of spiral arms present in a galaxy, introducing the terms \textit{grand-design} for a symmetric two-armed structure, \textit{flocculent} for unclear patterns with fluffy segments \citep[e.g.,][]{elmegreen_flocculent_1982}, or in later works \textit{multi-arm} for, e.g., three-armed or four-armed systems. The number of arms is also an important parameter in recent studies such as \citet{buta_classical_2015}, or the citizen-science project of Galaxy Zoo \citep[][]{lintott_galaxy_2008}.
As the formation process of these different types of arm could reflect different formation scenarios, it is an important aspect to investigate.  Symmetric two-armed structures could represent global oscillation modes driven by self-gravity \citep{lin_spiral_1964}, while flocculent structure might be rather the result of stochastic processes \citep[see review by ][]{sellwood_spirals_2022} or feedback \citep{dobbs_simulations_2018}.

Classifications of different types of rings are introduced by \citet{de_vaucouleurs_third_1991} into their scheme, which are also considered throughout different wavelengths \citep[e.g., ][]{schwarz_response_1981, treuthardt_kinematically_2007, comeron_ainur_2010}. 
Some rings (e.g., central rings) can be formed in the centers of galaxies, e.g., when gas occupies the central bar orbits, while so-called \textit{inner} rings appear outside of the bar, at the inner 4:1 resonance. 
Even further out in the disk, \textit{outer} rings are usually found at ${\sim}$twice the bar radius at the outer Lindblad resonance (OLR), or the outer 4:1 resonance \citep[e.g.,][]{comeron_arrakis_2014, oswalt_galaxy_2013}. 
Although the exact formation mechanisms of rings are still unclear, recent findings that rings are more common in green valley galaxies compared to star-forming galaxies \citep{kelvin_galaxy_2018, smith_effect_2022} suggest that they might be associated with quenching and are potential predictors of a galaxy's evolutionary state.

We organize this paper as follows. A detailed description of the data used can be found in Section~\ref{sec:PHANGS}.
We present the adopted scheme for the morphological classification based solely on the molecular gas distribution in Section~\ref{sec:MorphMethods}, including the definition, selection and classification process of morphological features.
In Section~\ref{sec:MorphologyResults} we will analyse the results of the morphological classification for their usability and derive feature percentages, followed by a comparison to literature findings in Section~\ref{sec:Discussion}, as well as a search for trends with galaxy properties. We summarize our results in Section~\ref{sec:Summary}.

\section{Data}
\label{sec:PHANGS}

To study the statistics of morphologies based on the gas distribution, we require not only a representative sample of main sequence galaxies, but also high-resolution and high-sensitivity maps of the cold gas distribution that allow for a detailed classification of features present in the gas disk. 
The Physics at High Angular resolution in Nearby GalaxieS (PHANGS\footnote{\url{https://sites.google.com/view/phangs/home}, \label{ft:1}}) survey with its PHANGS-ALMA dataset fulfills all these requirements with a sample of CO(J=2--1) observations at ${\sim}1^{\prime\prime}$ resolution for 90 nearby main sequence galaxies taken by the Atacama Large Millimeter/Submillimeter Array (ALMA) \citep[][]{leroy_phangsalma_2021}.

\subsection{PHANGS-ALMA sample}
\label{sec:PHANGS:sample}

The PHANGS-ALMA sample contains close to all nearby massive,  ALMA-visible and actively star-forming galaxies \citep{leroy_phangsalma_2021-1}. 
The 90 galaxies of the sample are selected to be moderately face-on ($i < 75\deg$) with distances $d$ less than ${\sim}20\,$Mpc, to achieve Giant Molecular Cloud (GMC) scale resolutions of ${\sim}100\,$pc (${\sim}1$") with ALMA. 
For M$_\star \gtrsim 10^{9.3}\,M_\odot$, the sample completeness is almost excellent at $5 < d < 12\,$Mpc and very good up to $17\,$Mpc \citep[see][]{leroy_phangsalma_2021-1}.
The selection function does not include any information about morphological features such as spiral arms or bars, making it the ideal sample for our study.
The galaxies were originally selected to be close to the 
main sequence of star forming galaxies. 
For more details see \citet{leroy_phangsalma_2021-1}.

\subsection{PHANGS-ALMA observations and datasets}
\label{sec:PHANGS:data}

The PHANGS-ALMA dataset contains high-resolution \mbox{CO(2--1)} observations
 made using ALMA 12\,m and 7\,m arrays combined with the total power antennas which ensures recovery of emission arising from all spatial scales. 
Each target is covered by several pointings that add to a large, multifield mosaic with standard calibrations from the observatory. 
Imaging follows a standard procedure as described in more detail in \citet{leroy_phangsalma_2021}. 
The science-ready data products of all 90 PHANGS-ALMA galaxies are publicly available on the project website (footnote~\ref{ft:1}). 

We drop galaxies with a flux recovery less than 30 percent from the analysis to ensure good quality of the used images
(compare equation 12 of \citealt{leroy_phangsalma_2021-1}). 
This threshold removes 11 galaxies from our sample. 
We list the 79 remaining galaxies and their properties used in Table~\ref{tab:AppendixPHANGSProperties}.

\begin{figure}[t]
    \centering
    \includegraphics[width=0.5\textwidth]{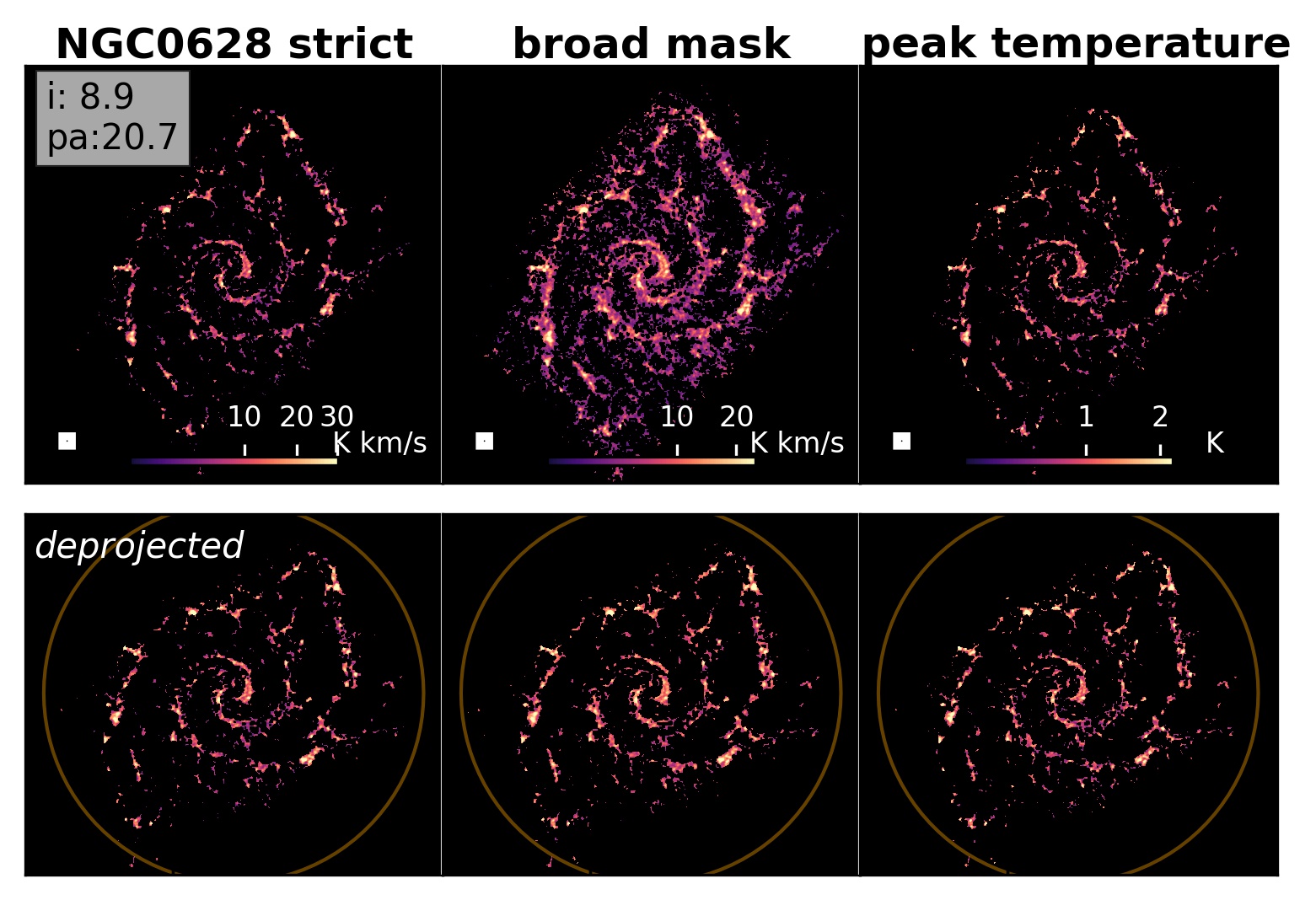}
    \caption{Example CO maps for NGC~0628. From left to right: Strict mom-0 map, broad mom-0 map and peak temperature map (top panels), as well as their deprojected and derotated versions (bottom panels). A circle with $0.5\cdot R_{25}$ (orange) is added in the lower panels, as well as the inclination (incl) and position angle (pa) used during deprojection and derotation (top left corner on upper left panel). This galaxy is classified as \textbf{G}-\textbf{A}-\textbf{nR}}
    \label{fig:exampleCOmom0}
\end{figure}

For this study, we use intensity-integrated maps (mom-0 maps) which are integrated along the spectral dimension, as well as their corresponding noise maps.
Specifically, we utilize \textit{strict} and \textit{broad} maps which are differently masked mom-0 maps, and peak temperature maps as explained below \citep[see ][]{leroy_phangsalma_2021}. 
The so-called \textit{strict} mask has a higher confidence to include real emission, but also lower completeness.
The \textit{broad} mask provides a high completeness map at the expense of slightly lower signal-to-noise ratio. 
\textit{T$_\mathrm{peak}$} maps show the peak brightness temperature along each line of sight.
Such maps show the gas morphology in a different way, as in these maps the broad line widths commonly present in centers do not over emphasize the central emission compared to fainter features in the disk.
For a detailed explanation on the masking process, we refer the reader to section~7 of \citet{leroy_phangsalma_2021}.
Figure~\ref{fig:exampleCOmom0} displays an example of these maps, as well as their images deprojected by inclination and de-rotated by position angle (bottom row) in which the typical scale length of half the 25th magnitude isophotal B-band radius  ($R_\mathrm{25}$) is shown as an orange circle as a scale reference.

\subsection{Additional data products}
\label{sec:Data:Additional}

For the analysis of the survey, we make use of published global galaxy properties from \citet[][]{leroy_phangsalma_2021-1}, including galaxy orientations (right ascension, declination, inclination, position angle, $R_{25}$), distances, stellar masses, SFRs, and several more gas parameters in Table~\ref{tab:AppendixPHANGSProperties} in Appendix~\ref{Appendix:PHANGSProperties}. 
To analyze deviations between our CO morphologies and literature morphologies based on other tracers, we utilize available maps from different tracers, such as near infrared $3.6\, \mu$m \textit{Spitzer} IRAC images, dust contamination maps, H$\alpha$ images and so-called environmental masks as described below.

\paragraph{\textit{Spitzer} images}
We use \textit{Spitzer} IRAC 3.6$\mu m$ images from the Spitzer Survey of Stellar Structure in Galaxies \citep[S$^4$G; ][]{sheth_spitzer_2010}, which are often used for morphological classifications \citep[e.g.,][]{buta_classical_2015}.
These images depict the infrared emission of the stellar distribution, and can be contaminated with emission from hot dust associated with star-forming sites. 
\citet{querejeta_spitzer_2015} applied an independent component analysis method to separate the IRAC 3.6$\mu m$ maps  into stellar and dust emission (using the neighbouring 4.5$\mu m$ band). 
Here, we use those dust contamination maps (in short \textit{ICA} maps) at a resolution of ${\sim}1.7''$.

To remove contamination by spiral arms when studying bar lanes in our galaxies, we make use of the environmental masks created by  \citet{querejeta_stellar_2021}. 
These masks are based on the photometric \textit{Spitzer} 3.6\,$\mu m$ images, and indicate different morphological regions such as bars \citep[aided by bar measurements from ][]{herrera-endoqui_catalogue_2015}, spiral arms, and centers. 
For our purpose the \textit{simple} environmental masks are well suited, which capture the dominant environment per pixel and do not account for overlaps of different environments.

\paragraph{H$\alpha$ images}
To compare our cold molecular gas based morphologies to a star formation tracer, we obtain continuum subtracted H$\alpha$ images from the \textit{Spitzer} Infrared Nearby  Galaxies Survey \citep[SINGS;][]{kennicutt_sings_2003} as well as the Survey for Ionization in Neutral Gas Galaxies \citep[SINGG;][]{meurer_survey_2006} plus newly obtained H$\alpha$ images by PHANGS (Razza et al. in prep.). 
These maps are available for most PHANGS galaxies. 
We make use of them especially for the seven cases with complex and ambiguous morphologies from Section~\ref{sec:MorphologyResults:Missingbars}: 
For NGC~1385 and NGC~5068 images are obtained from the Wide-Field Imager (WFI) narrowband and R-band imaging from SINGS, with astrometry and photometry calibrated using the Gaia DR1 catalogues. 
For IC~1954, NGC~1511, NGC~1559, NGC~4781 and NGC~4941 observations are made with the directCCD instrument of the duPont telescope. 
All those images are sky-subtracted following standard procedure as described in \citet[][]{schinnerer_gasstar_2019}.

\section{Methodology}
\label{sec:MorphMethods}

We devise a visual classification scheme for CO similar to those used at other wavelengths, with the goal of identifying whether the appearance of features is sensitive to the tracers. 
As an example, the \textit{stronger bar effect} describes that bars observed in infrared tend to be classified stronger than in optical images (e.g. a shift from class \SAB to class \SABB) \citep[][]{buta_mid-infrared_2010}.

As we focus on the identification of morphological features, we rely on human classifiers. 
We note, however, that there are more quantitative characterization methods for features (e.g., isophote shapes, Fourier amplitudes \citealt[e.g.,][]{aguerri_population_2009}, transverse-to-radial force ratio maps \citealt[][]{lee_bar_2020}), but those are probed on smooth features and not clumpy molecular gas.

\subsection{Feature selection}
\label{sec:Morphology:FeatureMotivation}

As a first step we devise a classification scheme. 
We base the following set of classes and features on the above mentioned findings of optical/near-IR studies and the expected characteristics of the cold molecular gas.
\begin{itemize}[noitemsep,topsep=0pt,leftmargin=2\parindent]
\item We classify the presence of \textbf{bars}, as these features are assumed to be one of the major drivers of secular evolution \citep[e.g.,][]{kormendy_secular_2004}. 
\item As dense gas is funneled along the bar, we are able to observe these features - so-called dust lanes or \textbf{bar lanes} - in our gas distribution well. 
Bar lanes have been studied in simulations in great detail \citep[e.g.,][]{athanassoula_existence_1992, athanassoula_morphology_1992, englmaier_two_1997}, but observational studies to confirm or disprove their results are still rare. 
We introduce classes that both describe the intensity profile of gas on the bar, as well as the shape and curvature of the bar lanes (see Figure~\ref{fig:ClassificationAtlas}). 
\item \textbf{Spiral arms} are another prominent feature in disk galaxies, linking morphology to disk dynamics. 
The arms consist of an overdensity of different disk components such as young stars, old stars, dust and gas \citep[e.g., see summary by][]{sellwood_spirals_2022, pour-imani_strong_2016}, making them ideal targets for multi-wavelength observations. 
We make use of literature arm classes slightly adapted to our molecular medium (see Section~\ref{sec:Morphology:Methodology:Outer}), which allows for a direct comparison to literature results. 
\item  Another commonly classified feature that allows for comparison with studies throughout different wavelengths are \textbf{rings} \citep[e.g., ][]{schwarz_response_1981, treuthardt_kinematically_2007, comeron_ainur_2010}. 
We do not differentiate between rings and pseudorings, which are broken or partial rings made up of spiral arms \citep{buta_galactic_1996}.
Instead, we differentiate between rings inside the bar (\textbf{central rings}), and rings further out in the disk (\textbf{non-central rings}). 
\end{itemize}
With the high-resolution CO observations at hand, we classify the presence of these features to gain insights into their formation mechanisms and how the host galaxy might influence their presence.

\subsection{Classification scheme}
\label{sec:Morphology:Methodology}

\begin{figure}[]
    \centering
    \includegraphics[width = 0.46\textwidth]{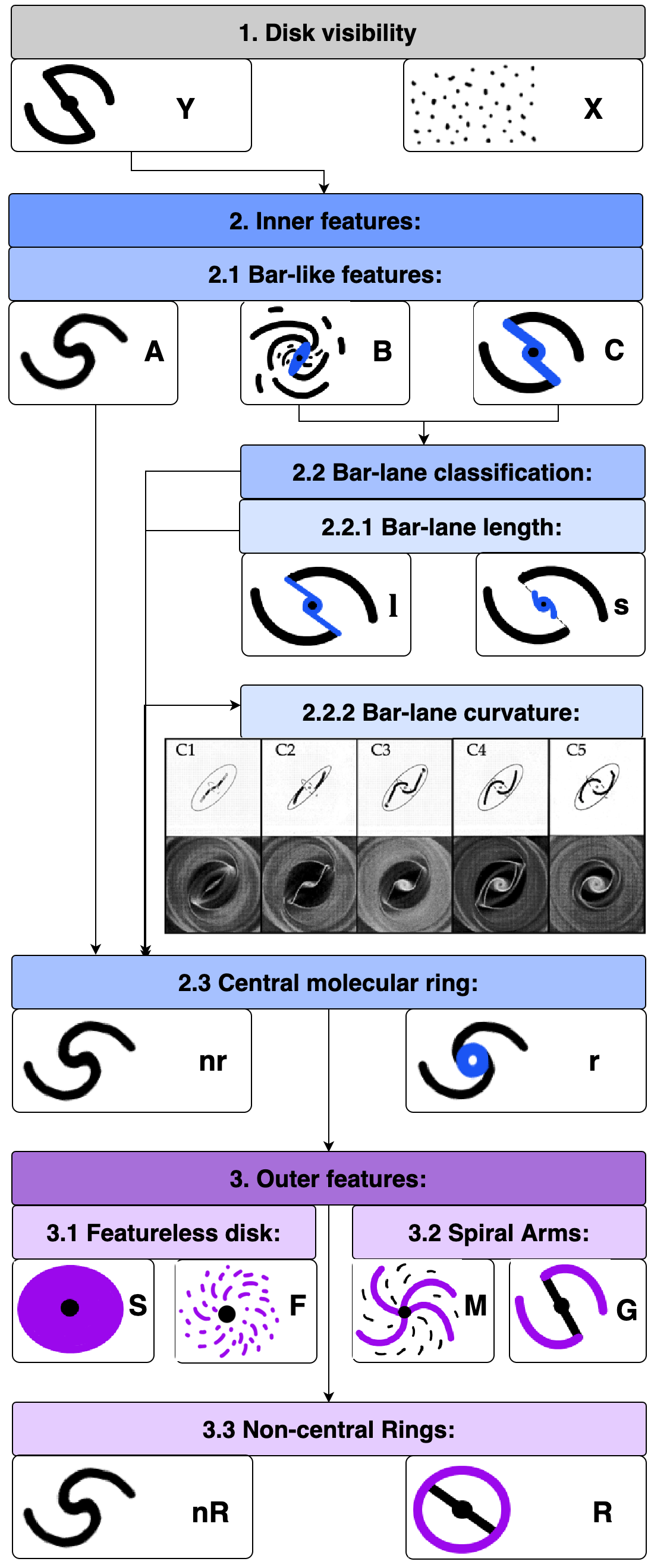}
    \caption{Flow chart of the morphological classification developed for PHANGS-ALMA \mbox{CO(2--1)} images. Colors represent the three major groups of classes. Arrows indicate dependencies of sub-classes on parent classes and the order of classification. 
    Bar lane curvatures show a selection of peak density maps (top row) and gas distributions (bottom row) of simulated face-on bars (at 45$^\circ$ from the horizontal) from \citet{athanassoula_existence_1992, athanassoula_morphology_1992}. These image are used as orientation to classify the shape of molecular bar lanes found from class \textbf{c1} to \textbf{c5}. They were selected to represent physically plausible cases that can be found in our sample, and have a visually increasing curvature from \textbf{c1} to \textbf{c5}.}
    \label{fig:ClassificationAtlas}
\end{figure}

Figure~\ref{fig:ClassificationAtlas} shows a flow chart of our adopted classification. 
We group the classes into three major classes that are disk visibility, inner bar-related features and other features including spiral arms and non-central rings to which we refer to as \textit{outer} features. 
Arrows indicate the order in which the classification was applied for each galaxy, as well as potential dependencies between classes. 
A more detailed definition of the individual symbols can be found below. 
Examples of PHANGS galaxy and their classes can be found in Appendix~\ref{Appendix:Morphexamples}.

\subsubsection{Disk visibility}
\label{sec:Morphology:Methodology:Visibility}

The molecular disk of the galaxy is visually identifiable and thus, morphological classification is possible.
Visibility (class \textbf{Y}) is ensured via two criteria: 
\begin{itemize}[noitemsep,topsep=0pt,leftmargin=2\parindent]
    \item[a)] The S/N ratio is high enough to identify a disk that does not appear to originate from random noise;
    \item[b)] The galaxy must not be edge-on and must enable the viewer to visually separate features that would otherwise be projected onto one elongated straight feature\footnote{Although most PHANGS galaxies are selected to be close to face-on, some highly inclined additions exist.}.
\end{itemize}
If these criteria are not fulfilled (class \textbf{X}), the galaxy is not used for further analysis.

\subsubsection{Inner features: Bars and central rings}
\label{sec:Morphology:Methodology:Bars}
We focus on bar-like features found in the inner regions of the disk. 
If a bar-like feature is identified, we furthermore attempt to classify bar lane properties. 

\paragraph{Bar-like features}
A bar is defined as either 
a perfectly straight elongated feature that extends more or less symmetrically outward to both sides of the center or 
two elongated curved lanes that extend to first order symmetrically from the center with increasing offset from a straight line towards the center (see schematics in Figure~\ref{fig:ClassificationAtlas}).  
We classify the object as: 
\begin{itemize}[noitemsep,topsep=0pt,leftmargin=2\parindent]
    \item \textbf{C}: A clear bar-like feature is visible, the region surrounding the elongated object is cleared of most of its gas or there is a strong contrast between emission on the bar-like feature and emission around it. 
    \item \textbf{B}: The elongated feature is less clear and the surrounding region is not cleared of most of its gas
    \item \textbf{A}: No such features are visible in the disk.
\end{itemize}

\paragraph{Bar lanes}
For class \textbf{C} and some cases of \textbf{B}, the bar is made up of bar lanes, which are two parallel or curved lines with an offset in the central region from a straight line connecting the bar ends and the center. We classify these bar lanes as following: 
\begin{itemize}[noitemsep,topsep=0pt,leftmargin=2\parindent]
    \item \textbf{Length of bar lanes:} If the emission on the bar lanes connects with the outer bar ends (e.g., the radius where the spiral arms begin, an outer ring exists, or, if none of the previous are present, the cleared region around the bar lane ends), we classify the bar lanes as \textbf{l} (long). 
    Class \textbf{s} (short) is applied if the emission on each bar lane ceases before reaching more than half of the bar semi-length, or experiences a sudden drop in the intensity along the bar lanes.
\item \textbf{Curvature and shape of bar lanes:} We visually compare the shape of the bar lanes to a set of five shapes (\textbf{c1} to \textbf{c5}) based on simulations from \citet{athanassoula_existence_1992, athanassoula_morphology_1992} as shown in Figure~\ref{fig:ClassificationAtlas} (see examples in Appendix~\ref{Appendix:Morphexamples}). 

\end{itemize}
For all galaxies with visible disks we further classify the existence of a \textbf{central ring or pseudoring:} A closed or nearly closed circular ring or ellipse at the very center of the disk is classified as \textbf{r} if present, \textbf{nr} if not.

\subsubsection{Outer features: Spiral arms and rings}
\label{sec:Morphology:Methodology:Outer}
We identify non-central rings located outside the bar radius, as well as spiral arms.

\paragraph{Spiral arm features}
\begin{itemize}[noitemsep,topsep=0pt,leftmargin=2\parindent]
    \item \textbf{Grand design (G):} Two spiral arms with roughly equally high intensity, thickness and extent in the outer disk. 
    \item \textbf{Multi-arm (M):} One or more than two spiral arms with roughly equally high flux, thickness and extent. 
    \item \textbf{Flocculent (F):} No clear spiral arms are visible, the disk consists of randomly distributed short unconnected arm segments.
    \item \textbf{Smooth (S):} No spiral arms are visible, the disk is rather smooth than clumpy. 
\end{itemize}

\paragraph{Non-central ring}
Independently of the above-mentioned features, we classify the presence of a ring  or pseudoring in the outer disk similarly to the central ring with \textbf{R} if present or \textbf{nR} if not present for all galaxies.

\subsection{Classification procedure}
\label{sec:Morphology:Classification}

In total, ten co-authors (from here-on \textit{inspectors}) followed the classification scheme to assign classifications to all galaxies. 
An example of the images used for classification is shown in Figure~\ref{fig:exampleCOmom0}.
A feature is classified if it is visible in at least one map. 

This galaxy in Figure~\ref{fig:exampleCOmom0} (NGC~628) is classified by all 10 inspectors as having a clearly visible CO disk (\textbf{Y}) with grand design \textbf{G} spiral arm.
It is classified by 9 out of 10 as a disk without bar-like feature (\textbf{A}) and as having no non-central ring (\textbf{nR}).
We provide additional examples of the various morphologies among our galaxies (e.g., \textbf{F},  \textbf{M} \textbf{C},  \textbf{B}, \textbf{R} and several more) in Appendix~\ref{Appendix:Morphexamples}.

\subsection{Classification agreement}
\label{sec:Morphology:Agreement}

We assign a class to a galaxy only if at least a given fraction (specified below) of inspectors agrees on the classification. 
We refer to classifications that reached the required threshold as \textit{verified classification}.
Classifications that do not reach the required threshold are referred to as \textit{\nAG} (non-agreement). 
The final verified classifications for each galaxy can be found in Table~\ref{tab:Allclasses} in Section~\ref{sec:MorphologyResults}.

\begin{figure}[t]
    \centering
    \includegraphics[width = 0.5\textwidth]{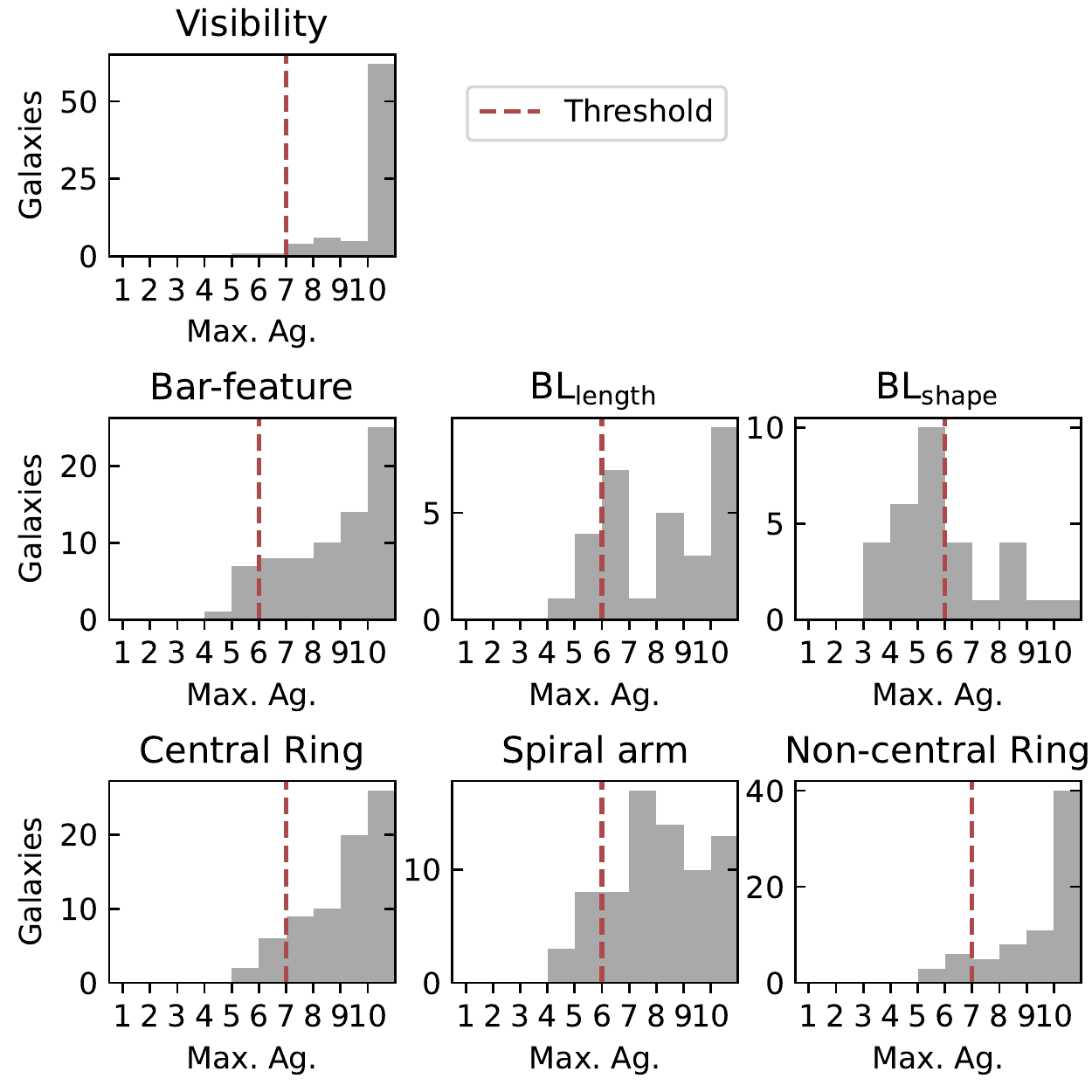}
    \caption{Histogram of maximum agreement per galaxy per class from ten inspectors. From left to right and top to bottom we show the maximum agreement distribution in the upper panels for: the disk visibility, bar-like features, bar lane lengths (BL$_\mathrm{length}$), bar lane shapes (BL$_\mathrm{shape}$), central ring features, spiral arm features, and non-central rings. 
    For each feature we require a minimum agreement threshold (red dashed line) to consider a classification as reliable. For features with only two options to chose from this is set to 70\%, for other features 60\% }
    \label{fig:Agreement}
\end{figure}

Figure~\ref{fig:Agreement} shows the distribution of the maximum agreement\footnote{Maximum agreement refers to the number of people that chose the most selected option.} achieved of all 79 galaxies for all features considered. 
We excluded non-classifications which can occur for sub-classes that depend on a parent class\footnote{
As an example, when most inspectors agree that a galaxy is classified as unbarred (\textbf{A}), also most people will collectively not classify bar lanes. This could be misinterpreted as good agreement, however, this is simply what we call a \textit{non-classification} and is directly dependent on the agreement achieved for the parent class.}.
To derive uncertainties on the fraction of features found in a given (sub-)sample of galaxies, we perturb the threshold of inspectors that need to agree on a single feature by $\pm 10\%$. 
This is equal to one volunteer changing their classification and can in- or decrease the amount of galaxies with such a feature.

For the visibility, central- and non-central ring class, one could only choose between two possible classes (either disk-visible, \textbf{Y}, or not, \textbf{X}, and either \textbf{r} (\textbf{R}), if (non-) central ring present, or \textbf{nr} (\textbf{nR}), if not), and we require agreement threshold of 70\% among the inspectors. 
The probability to receive at least 70\% agreement from a random binominal distribution of two equal options is 34\%, which is reasonable considering that those features (visibility and rings) are the most easiest to identify in our gas distribution. 
We further confirm that increasing (or decreasing) this threshold does not significantly change our results, and capture this with our uncertainties. 

For all other quantities, which have more than two possible classes to chose from and thus, an expected larger scatter, we require an agreement of 60\% on the same class to adopt it for the galaxy. 
The probability to obtain at least 60\% agreement from a random distribution of three (or more) equal options is 23\% (or less), which we consider reasonable. 
The bar lane shape classes (see Section~\ref{sec:MorphMethods:barlanes}) are an exception and treated differently.
These threshold allow us to include only galaxies with clearly recognizable features, which lowers potential biases that arise when searching for trends with galaxy properties in Section~\ref{sec:MorphologyResults} later on.

For the visibility, central and non-central rings and bar-features, the maximum agreement is the highest among all classes with a peak at 10/10. 
For the visibility of a disk, we find an average maximum agreement of 95\% with a standard deviation of 11\%, i.e., on average less than one person disagrees with the classification of the other nine. 
For most galaxies, between ${\sim}$8 and 10 people agreed on a single class.
For the main features, bars and spiral arms, the average maximum agreement is 82\% and 75\%, respectively, in galaxies with visible disks.
Since there were more classes to choose from to classify these features, the standard deviations are expectedly higher with 18\% and 17\%.
We consider this very good agreement among the inspectors.

\subsection{Note on bar lane shape classes}
\label{sec:MorphMethods:barlanes}

The maximum agreement for the shape of the bar lanes peaks at only 5 out of 10 people, likely due to the high number of options to choose from (five compared to two or three for other features:\textbf{c1, c2, c3, c4, c5}).
Often, inspectors agreed on neighbouring classes (e.g. most people chose either \textbf{c1} or \textbf{c2}), but the required agreement of 6/10 was not reached for a single class. 
As the bar lane shape classes have an internal order, and are selected to represent an actual physical range with, e.g., increasing curvature from class \textbf{c1} to \textbf{c5}, we introduce intermediate classes. 
If no single shape class achieves the required minimum agreement of 6/10, but two neighbouring classes achieve the required agreement, the final class is the mean of their numerical values. 
As an example, three people selecting class \textbf{c1} (corresponding to a numerical value of 1.0) and another three people \textbf{c2} (numerical 2.0) will result in a final class \textbf{c1.5}. 
For each galaxy we confirm that there is only one absolute maximum of agreement found when combining two neighbouring classes. 
In addition to the visual bar lane classifications, we will introduce a method to measure bar lane properties in Section~\ref{sec:BarlaneAnalysis}.

\section{Results}
\label{sec:MorphologyResults}

We provide verified morphologies of all 79 galaxies on which the CO-based classification was performed in Table~\ref{tab:Allclasses}. Classifications without the required agreement as described in Section~\ref{sec:Morphology:Agreement} are indicated by \textit{\nAG}.
Some sub-classes are dependent on their parent class and can result in empty entries (`-') which we refer to as non-classifications.
We further provide the statistical distribution of each class and sub-class among our sample.

\begin{figure*}[t]
    \centering\includegraphics[width = 
 1.15\textwidth, angle = 90]{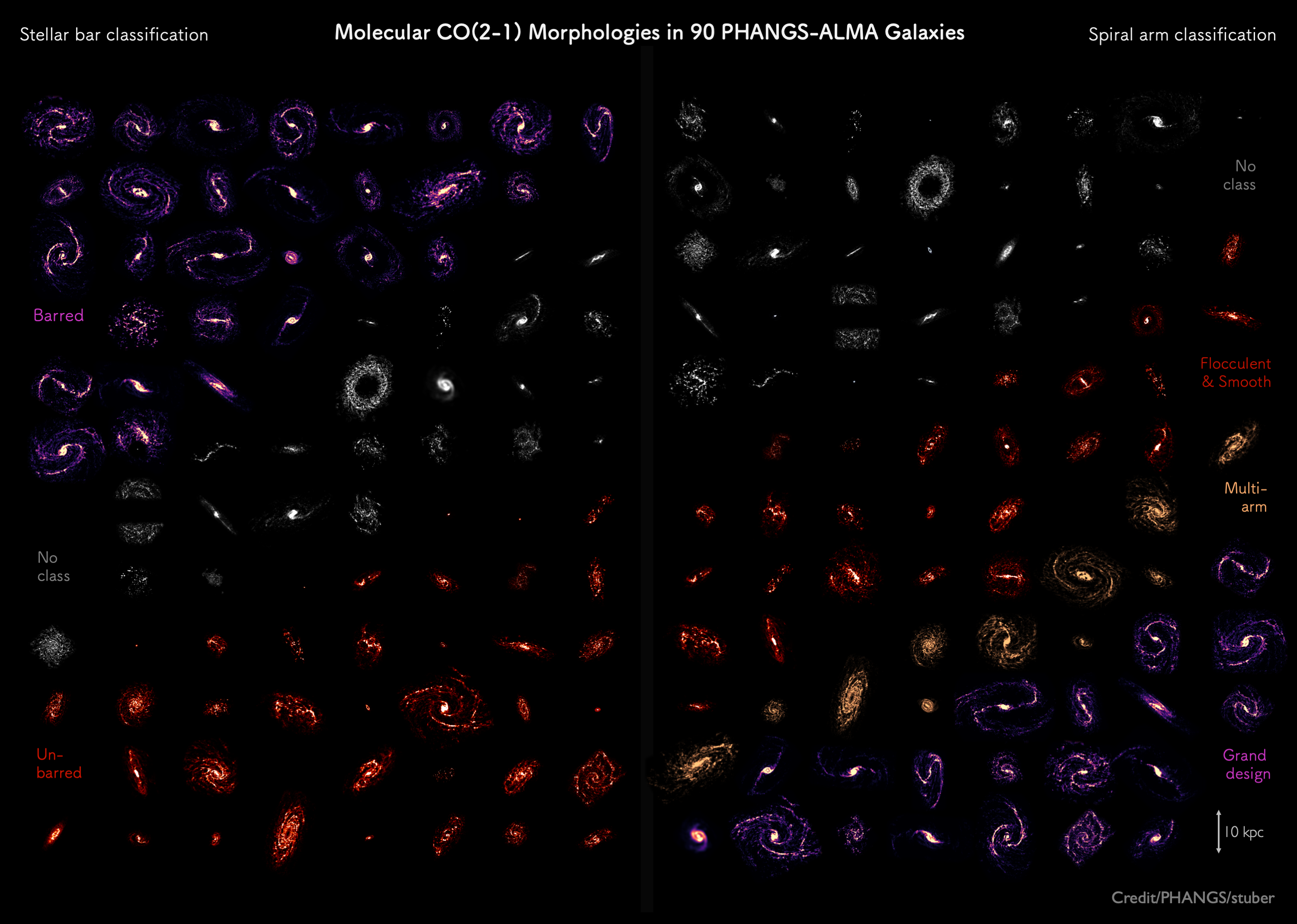}
    \caption{Bar (left) and spiral arm (right) classifications of PHANGS galaxies. This contains the full sample of 90 galaxies including the galaxies deemed insufficient for classification.}
    \label{fig:Gallery}
\end{figure*}

\subsection{Visibility}
\label{sec:MorphologyResults:Visibility}

Out of the 79 galaxies used for this classification, 72 have sufficient visibility of their disk (i.e., detected CO emission)  - we will refer to these as \textit{valid disks}. 
Out of the remaining 7 galaxies, 5 lack sufficient CO emission and are at the higher end of the samples inclination ($60 \leq i \leq 90$) and are therefore not suitable for this survey. 
For 2 galaxies (NGC~3599, NGC~5128) the required agreement among inspectors could not be reached, thus no verified classification is available. 
We do not find any significant trend with galaxy distance, opposing a resolution dependency.

\subsection{Statistics of morphological features}
\label{sec:MorphologyResults:BarsArms}

\begin{figure}[t]
    \centering
    \includegraphics[width = 0.49\textwidth]{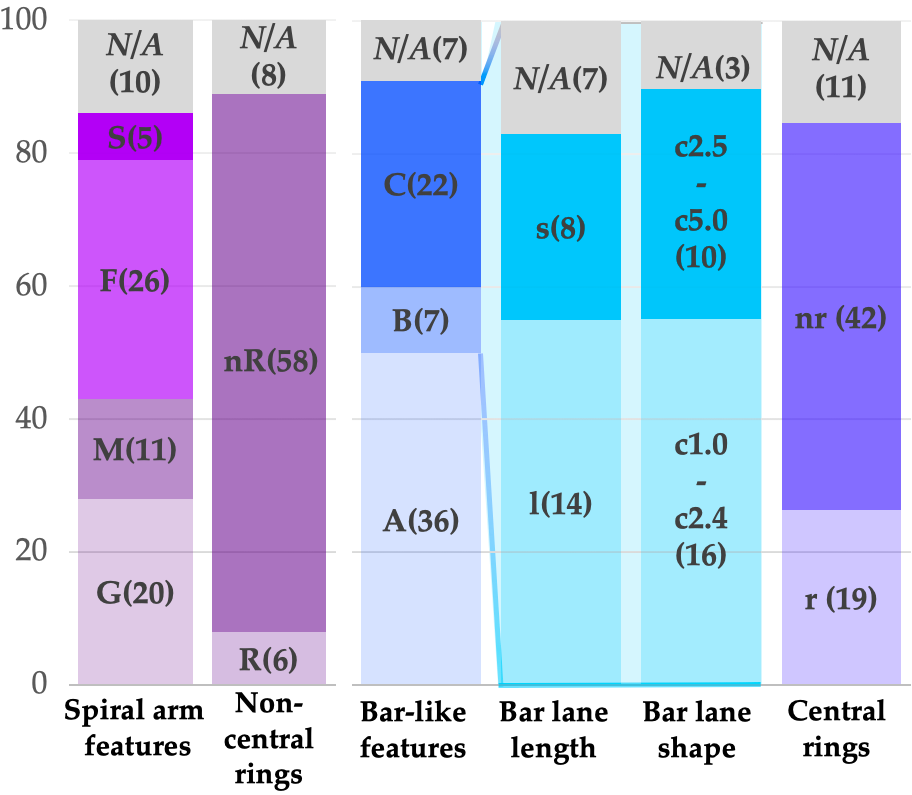}
    \caption{Percentages of morphological classes of all PHANGS galaxies classified as having a valid disk (i.e., sufficient CO emission) as well as galaxies with verified classification. 
    Galaxy numbers are listed in the chart in the respective class and the percentages have an uncertainty of ${\sim}5$\%. 
    Galaxies without verified classification are referred to as \nAG. 
    For barred galaxies (\textbf{B} and \textbf{C}) we show (blue) bar lane length classes. Bar lane shapes are divided into two regimes, \textbf{c1.0}-\textbf{c2.4} (light blue) and \textbf{c2.5}-\textbf{5.0} (dark blue). 
    }
    \label{fig:MorphResults:StatisticsI}
\end{figure}

We analyze the distribution of all features among our sample according to our CO-based classifications.
Figure~\ref{fig:MorphResults:StatisticsI} shows the overall distribution of features in PHANGS galaxies. 
The uncertainty of the individual sub-samples is derived by perturbing the required agreement threshold for the respective class by $\pm1$ (see Section~\ref{sec:Morphology:Agreement}). 
Overall we find a relative uncertainty of ${\sim}10$\% for the total amount of galaxies with bar or spiral arm classification and an uncertainty of ${\sim}5$\% for the individual sub-classes. 

\paragraph{Statistics of spiral arms}
Among the 72 valid galaxy disks, $28\pm 2$\% of galaxies (20 galaxies) host grand-design spiral arms, $15\pm3$\% (11) multi-arms, $36\pm6$\% (26) have flocculent disks and $7\pm2$\% (5) smooth disks. For $14\pm 9\%$ (10) of galaxies no verified (sufficient agreement) class could be achieved.
Out of the 62 galaxies with valid disks \textit{and} verified classification, these numbers correspond to $32\pm2$\%, $18\pm3$\%, $42\pm6$\%, $8\pm2$\% for grand design, multi-arm, flocculent, and smooth disk.

\paragraph{Statistics of bar-like features}

Among the 72 valid galaxy disks, $50\pm6$\% (36 galaxies) have class \textbf{A} (unbarred), $10\pm2$\% (7) \textbf{B} (weakly barred),  $31\pm3$\% (22) \textbf{C} (strongly barred) and $10\pm9$\% (7) do not reach sufficient agreement.
These numbers correspond to $55\pm7$\% class \textbf{A}, $11\pm2$\% class \textbf{B} and  $34\pm3$\% class \textbf{C} for galaxies with both valid disks \textit{and} verified classifications.

We define the total bar fraction $f_\mathrm{bar}$ as the ratio of class \textbf{B} and \textbf{C} galaxies divided by the total number of galaxies with both sufficient visibility and agreement.
We find $f_\mathrm{bar} = 45 \pm 5\%$, which we will compare to literature results in Section~\ref{sec:MorphologyResults:Barfraction}.

\paragraph{Statistics of non-central rings:}

Of our 72 valid galaxy disks, $8 \pm 1$\% ($6$ galaxies) have a non-central ring (\textbf{R}), $81 \pm 8$\% ($58$) do not have a non-central ring (\textbf{nR}), and $11 \pm 8$\% ($8$) have no verified classification. 
When considering only galaxies with verified classifications those numbers correspond to \textbf{R:} $9 \pm 2$\% and \textbf{nR:} $91 \pm 9$\%. 
Non-central rings are rings and pseudorings within our FoV in the outer regions of the disk at radii larger than the bar radii. 
In our sample of galaxies, such rings tend to be a rare feature. 

\paragraph{Statistics of bar lane lengths:}

Out of the 29 galaxies classified as barred (either \textbf{B} or \textbf{C}), $55 \pm 12$\% (14) received a bar lane classification of \textbf{l} (long) and $28 \pm 7$\% (8) received \textbf{s} (short). 
The majority ($67\pm15$\%) of CO bar lanes with verified classification are larger than half the full bar size (\textbf{l}) and only $33\pm 8$\% have CO bar lanes that are shorter than half the full bar length (\textbf{s}).

\paragraph{Statistics of bar lane shapes:}
Since we transformed the range of bar lane shapes into numerical values in Section~\ref{sec:MorphMethods:barlanes}, we have a large range of values between 1.0 and 5.0. 
Of all 29 barred galaxies, only one was classified as not having resolved bar lanes (galaxy of class \textbf{B}), for another two the sufficient agreement was not achieved (one galaxy of class \textbf{B}, one of class \textbf{C}) and for 26 galaxies (21 \textbf{C} and 5 \textbf{B} galaxies) shapes between 1.0 and 5.0 were assigned. 
Out of these 26 galaxies, more than half (16 galaxies) are classified with shapes less than 2.5 (straighter shape), and the other 10 galaxies have shapes of 2.5 or larger (more curved shapes). The average scatter is 0.7.

\paragraph{Statistics of central ring features:}

We find that $31\pm4\%$ ($22$) of valid galaxies exhibit a central ring \textbf{r}, $60\pm6\%$($43$) do not (\textbf{nr}), and for $10\pm5\%$ ($7$) the required agreement was not achieved.

\begin{table*}
    \centering
    \caption{Morphological classifications for 79 galaxies.}
\label{tab:Allclasses}
\begin{scriptsize}
\begin{tabular}{l|cc|cc|cc|cc|cc|cc|cc} 

\hline 

Galaxy & \multicolumn{2}{c}{Visibility} & \multicolumn{2}{c}{Bar} & \multicolumn{2}{c}{BL$_\mathrm{len}$} & \multicolumn{2}{c}{BL$_\mathrm{shape}$} & \multicolumn{2}{c}{Central ring} & \multicolumn{2}{c}{Spiral arm} & \multicolumn{2}{c}{Non-central ring}\\

& \multicolumn{2}{c}{(\textbf{X},\textbf{Y})}  & \multicolumn{2}{c}{(\textbf{A},\textbf{B},\textbf{C})} & \multicolumn{2}{c}{(\textbf{s},\textbf{l})} & \multicolumn{2}{c}{(\textbf{c:}1-5)} & \multicolumn{2}{c}{(\textbf{r},\textbf{nr})} & \multicolumn{2}{c}{(\textbf{G},\textbf{M},\textbf{F},\textbf{S})} & \multicolumn{2}{c}{(\textbf{R},\textbf{nR})} \\

(1) & \multicolumn{2}{c}{(2)} & \multicolumn{2}{c}{(3)} & \multicolumn{2}{c}{(4)} & \multicolumn{2}{c}{(5)} & \multicolumn{2}{c}{(6)} & \multicolumn{2}{c}{(7)} & \multicolumn{2}{c}{(8)} \\

\hline

Circinus & Y & 100\% & \nAG & 50\% & - & - & - & - & r & 80\% & F & 60\% & nR & 90\% \\
IC~1954 & Y & 100\% & A & 70\% & - & - & - & - & nr & 100\% & M & 80\% & nR & 100\% \\
IC~5273 & Y & 100\% & \nAG & 50\% & - & - & - & - & nr & 100\% & F & 100\% & nR & 100\% \\
NGC~0253 & Y & 100\% & B & 60\% & \nAG & 50\% & 2.5 & 40\% & nr & 90\% & G & 100\% & \nAG & 60\% \\
NGC~0300 & Y & 70\% & A & 70\% & - & - & - & - & nr & 70\% & F & 70\% & \nAG & 60\% \\
NGC~0628 & Y & 100\% & A & 90\% & - & - & -& - & nr & 90\% & G & 100\% & nR & 90\% \\
NGC~0685 & Y & 100\% & B & 60\% & l & 80\% & 1 & 80\% & nr & 100\% & \nAG & 50\% & nR & 100\% \\
NGC~1068 & Y & 100\% & \nAG & 50\% & - & - & - & - & r & 100\% & G & 90\% & nR & 70\% \\
NGC~1087 & Y & 100\% & B & 100\% & l & 60\% & 1.5 & 30\% & nr & 90\% & \nAG & 50\% & nR & 100\% \\
NGC~1097 & Y & 100\% & C & 90\% & l & 100\% & 2.5 & 50\% & r & 100\% & G & 80\% & nR & 80\% \\
NGC~1300 & Y & 100\% & C & 100\% & l & 90\% & 1.5 & 50\% & r & 100\% & G & 100\% & nR & 80\% \\
NGC~1317 & Y & 100\% & B & 70\% & - & - & - & - & \nAG & 60\% & M & 70\% & \nAG & 60\% \\
NGC~1365 & Y & 100\% & C & 90\% & s & 80\% & 2.5 & 40\% & r & 90\% & G & 70\% & nR & 100\% \\
NGC~1385 & Y & 100\% & A & 100\% & - & - & -& - & nr & 100\% & F & 100\% & nR & 100\% \\
NGC~1433 & Y & 100\% & C & 90\% & s & 60\% & 4.0 & 60\% & r & 90\% & \nAG & 40\% & R & 80\% \\
NGC~1511 & Y & 90\% & A & 90\% & - & - & - & - & nr & 90\% & F & 80\% & nR & 90\% \\
NGC~1512 & Y & 100\% & C & 90\% & s & 80\% & 4.0 & 60\% & r & 100\% & \nAG & 50\% & R & 100\% \\
NGC~1546 & Y & 100\% & A & 100\% & - & - & - & - & nr & 70\% & \nAG & 50\% & nR & 90\% \\
NGC~1559 & Y & 100\% & A & 100\% & - & - & - & - & nr & 100\% & F & 100\% & nR & 100\% \\
NGC~1566 & Y & 100\% & C & 70\% & s & 60\% & 4.5 & 40\% & r & 80\% & G & 100\% & nR & 100\% \\
NGC~1637 & Y & 100\% & B & 60\% & l & 60\% & 1.5 & 50\% & nr & 90\% & G & 70\% & nR & 100\% \\
NGC~1672 & Y & 100\% & C & 100\% & l & 80\% & 1.5 & 30\% & r & 100\% & G & 60\% & nR & 90\% \\
NGC~1792 & Y & 100\% & A & 70\% & -& - & - & - & \nAG & 60\% & M & 70\% & nR & 100\% \\
NGC~1809 & Y & 70\% & A & 70\% & - & - & - & - & nr & 70\% & F & 70\% & nR & 70\% \\
NGC~2090 & Y & 100\% & A & 100\% & -& - & - & - & nr & 90\% & \nAG & 40\% & \nAG & 60\% \\
NGC~2283 & Y & 100\% & \nAG & 50\% & - & - & - & - & nr & 100\% & G & 70\% & nR & 100\% \\
NGC~2566 & Y & 100\% & C & 100\% & \nAG & 50\% & 1.5 & 50\% & r & 100\% & G & 90\% & \nAG & 60\% \\
NGC~2903 & Y & 100\% & C & 100\% & l & 80\% & 1.5 & 50\% & r & 80\% & G & 70\% & nR & 90\% \\
NGC~2997 & Y & 100\% & A & 80\% & - & -& - & - & r & 100\% & G & 90\% & nR & 100\% \\
NGC~3059 & Y & 100\% & B & 80\% & \nAG & 50\% & 1.0 & 60\% & nr & 100\% & F & 80\% & nR & 100\% \\
NGC~3137 & Y & 100\% & A & 100\% & - & - & - & - & nr & 100\% & \nAG & 50\% & nR & 100\% \\
NGC~3351 & Y & 100\% & C & 90\% & s & 100\% & 4.5 & 50\% & r & 90\% & F & 60\% & R & 100\% \\
NGC~3489 & Y & 100\% & A & 60\% & - & - & - & - & r & 70\% & \nAG & 50\% & nR & 100\% \\
NGC~3507 & Y & 100\% & C & 60\% & s & 60\% & 1.5 & 30\% & nr & 80\% & G & 90\% & nR & 100\% \\
NGC~3511 & Y & 100\% & A & 100\% & - & - & - & - & nr & 80\% & F & 100\% & nR & 100\% \\
NGC~3521 & Y & 100\% & A & 100\% & - & - & - & - & nr & 80\% & M & 80\% & nR & 90\% \\
NGC~3596 & Y & 100\% & A & 90\% & - & - & - & - & \nAG & 60\% & M & 70\% & nR & 100\% \\
NGC~3599 & \nAG & 50\% & - & - & - & - & - & -& - & - & - & - & - & - \\
NGC~3621 & Y & 100\% & A & 100\% & - & - & - &  & nr & 100\% & F & 100\% & nR & 100\% \\
NGC~3626 & Y & 100\% & A & 100\% & - & - & - &  & nr & 90\% & F & 100\% & nR & 100\% \\
NGC~3627 & Y & 100\% & C & 100\% & l & 100\% & 2.0 & 60\% & nr & 90\% & G & 80\% & nR & 100\% \\
NGC~4207 & X & 90\% & - & - & - & - & - & - & - & - & - & - & - & - \\
NGC~4254 & Y & 100\% & A & 80\% & - & -& - & - & nr & 100\% & M & 90\% & nR & 100\% \\
NGC~4293 & X & 90\% & - & - & - & - & - & - & - & - & - & - & - & - \\
NGC~4298 & Y & 100\% & A & 60\% & - & - & - & - & nr & 100\% & F & 80\% & nR & 100\% \\
NGC~4303 & Y & 100\% & C & 100\% & l & 100\% & 3.0 & 80\% & r & 100\% & M & 70\% & nR & 100\% \\
NGC~4321 & Y & 100\% & C & 100\% & l & 100\% & 3.5 & 40\% & r & 100\% & G & 100\% & nR & 100\% \\
NGC~4424 & X & 80\% & - & - & - & - & - & - & - & -& - & - & - & - \\
NGC~4457 & Y & 100\% & A & 80\% & -& - & - & - & nr & 70\% & M & 60\% & nR & 80\% \\
NGC~4459 & Y & 100\% & A & 100\% & -& - & - & - & r & 90\% & S & 100\% & \nAG & 50\% \\
NGC~4476 & Y & 100\% & A & 100\% & - & - & - & - & r & 70\% & S & 70\% & R & 90\% \\
NGC~4477 & Y & 80\% & A & 60\% & -& - & - & - & r & 80\% & S & 60\% & nR & 70\% \\
NGC~4535 & Y & 100\% & C & 100\% & l & 100\% & 2.0 & 80\% & \nAG & 50\% & G & 90\% & nR & 100\% \\
NGC~4536 & Y & 100\% & \nAG & 40\% & -& - & - & - & \nAG & 60\% & \nAG & 40\% & nR & 100\% \\
NGC~4540 & Y & 100\% & A & 90\% & - & -& - & - & nr & 100\% & F & 80\% & nR & 100\% \\
NGC~4548 & Y & 100\% & C & 100\% & l & 100\% & 1.0 & 70\% & nr & 90\% & G & 90\% & nR & 100\% \\
NGC~4569 & Y & 100\% & A & 80\% & - & - & - & - & nr & 80\% & F & 80\% & nR & 100\% \\
NGC~4579 & Y & 100\% & C & 60\% & s & 60\% & \nAG & 40\% & \nAG & 60\% & M & 80\% & \nAG & 60\% \\
NGC~4596 & Y & 80\% & A & 90\% & - & - & - & - & nr & 80\% & S & 90\% & nR & 90\% \\
NGC~4654 & Y & 100\% & B & 90\% & \nAG & 40\% & \nAG & 50\% & nr & 100\% & M & 70\% & nR & 100\% \\
NGC~4689 & Y & 100\% & A & 100\% & - & - & -& - & nr & 100\% & M & 60\% & nR & 100\% \\
NGC~4731 & X & 70\% & - & - & - & - & - & - & - & - & - & - & - & - \\
NGC~4781 & Y & 100\% & A & 80\% & - & - & - & - & nr & 90\% & F & 80\% & nR & 100\% \\
NGC~4826 & Y & 100\% & A & 100\% & - & - & - & - & nr & 90\% & \nAG & 50\% & nR & 80\% \\
NGC~4941 & Y & 100\% & C & 70\% & s & 90\% & 4.5 & 50\% & \nAG & 60\% & F & 70\% & R & 80\% \\
NGC~4945 & X & 80\% & - & - & - & - & - & - & - & - & - & - & - & - \\
NGC~4951 & Y & 100\% & \nAG & 50\% & - & - & - & - & nr & 100\% & F & 100\% & nR & 100\% \\
NGC~5068 & Y & 70\% & A & 70\% & - & - & - & - & nr & 70\% & F & 70\% & nR & 70\% \\
NGC~5128 & \nAG & 60\% & - & - & - & - & - & - & - & - & - & - & - & - \\
NGC~5134 & Y & 90\% & A & 80\% & - & - & - & - & nr & 90\% & F & 70\% & nR & 90\% \\
NGC~5236 & Y & 100\% & C & 100\% & l & 100\% & 1.5 & 50\% & r & 70\% & G & 80\% & nR & 100\% \\
NGC~5248 & Y & 100\% & \nAG & 50\% & -& - & - & - & r & 90\% & G & 80\% & nR & 100\% \\
NGC~5530 & Y & 100\% & A & 100\% & - & -& - & - & nr & 100\% & F & 90\% & nR & 100\% \\
NGC~5643 & Y & 100\% & C & 90\% & l & 100\% & 1.0 & 100\% & nr & 90\% & F & 70\% & \nAG & 50\% \\
NGC~6300 & Y & 100\% & C & 90\% & l & 90\% & 1.0 & 90\% & nr & 100\% & F & 60\% & R & 100\% \\
NGC~7456 & Y & 80\% & A & 80\% & - & - & - & - & nr & 90\% & F & 80\% & nR & 80\% \\
NGC~7496 & Y & 100\% & C & 80\% & l & 70\% & 1.0 & 80\% & nr & 90\% & F & 70\% & nR & 80\% \\
NGC~7743 & Y & 80\% & A & 80\% & -& -& - & - & r & 70\% & S & 60\% & nR & 70\% \\
NGC~7793 & Y & 90\% & A & 90\% & -& - & - & - & nr & 80\% & F & 90\% & nR & 90\% \\
\end{tabular}
\tablefoot{For each galaxy (1) we list verified (sufficient agreement) CO-based classifications  if available with their maximum agreement, otherwise \textit{\nAG}. We list classifications for the visibility (2), bar-features (3), bar lane length (4) and shapes (5), central rings (6), non-central rings (8) and spiral arms (7) (see Figure~\ref{fig:ClassificationAtlas}). Some classifications are dependent on parent classes and result in class '-' if the required parent class is not selected. }
\end{scriptsize}
\end{table*}

\section{Discussion}
\label{sec:Discussion}

We classified PHANGS-ALMA galaxies based on their visual appearance as observed in the \mbox{CO(2--1)} emission line and guided by a simple scheme similar to literature classifications of optical/near-IR imaging but tailored to the molecular medium. 
In addition, we introduced a set of classes for bar lanes.

There is very good agreement between classifications done by 10 inspectors for most galaxies, proving that CO is a suitable morphology tracer.
We will compare our classifications to those from the literature based on infrared or optical images to test consistency. 
We further search for trends with host galaxy properties that could point to the formation or evolution of morphological features.

\subsection{Comparison of main features to optical/IR classifications}
\label{sec:MorphologyResults:BarsArmsLiterature}
The most prominent features are spiral arms and bars. 
We compare our results to the classifications from IRAC $3.6\,\mu$m  for S$^4$G \citep{sheth_spitzer_2010} by \citet{buta_classical_2015} when available and by \citet[][]{de_vaucouleurs_third_1991} otherwise (=RC3). 
These optical/infrared classifications probe the underlying older stellar population, but they are affected by young star-forming sites \citep[e.g., contamination by host dust;][]{querejeta_spitzer_2015}. 
All relevant properties for our sample are listed in Table~\ref{tab:AppendixPHANGSProperties}.

As our classification system is not identical, we equate our class \textbf{A} (no bar-like object present) directly to the traditional class \SA, our class \textbf{C} (strong evidence for bar-like object) to \SB. 
For spiral arms \textbf{G} is compared to literature \textit{G} (grand-design), \textbf{M} (multi-arm) to literature \textit{M} and \textbf{F} (flocculent) to literature \textit{F}.

We note that the distance distribution of our sample is small, and includes only galaxies up to $\sim 24$\,Mpc (with a physical resolution of up to $\sim 180\,$pc and angular resolution of $\sim 1\arcsec$.). This is well matched to the $\sim 1.7\arcsec\ $ resolution from \citet{buta_classical_2015}. This allows us to identify resolution-independent influences on our classifications that stem from the molecular gas medium itself or from host galaxy properties.

\subsubsection{Spiral arms in CO vs optical/IR}
\label{sec:MorphologyResults:ArmsLiterature}

\begin{figure}[t]
    \centering
    \includegraphics[width = 0.45\textwidth]{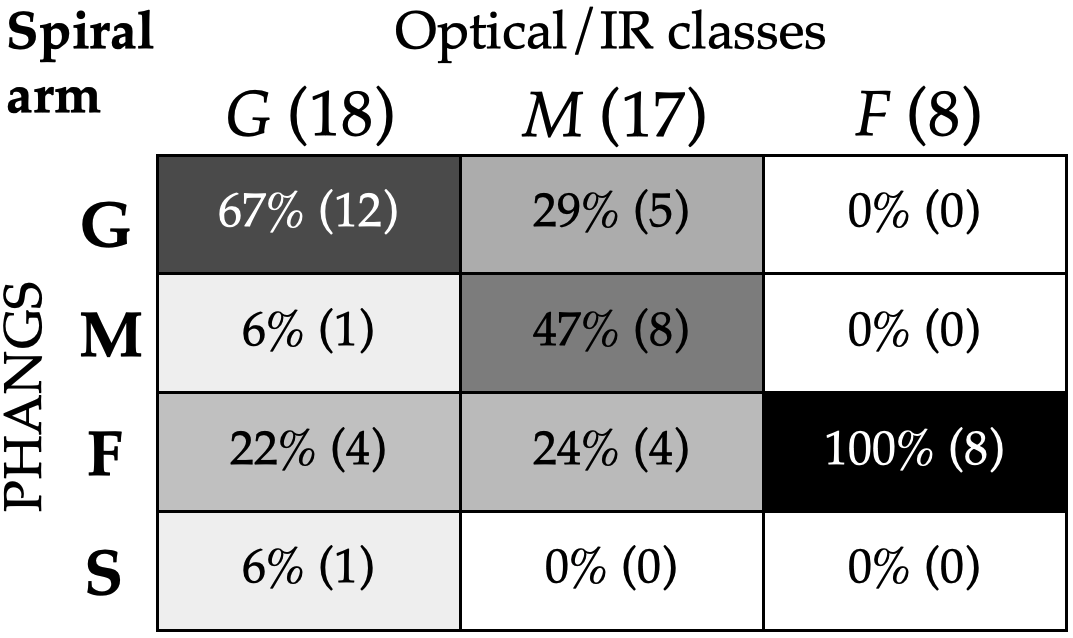}
    \caption{Confusion matrix of morphological spiral arm classes (G: Grand-design, M: Multi-arm, F: Flocculent) from this work compared to literature classes for 43 galaxies. Literature classes are from Buta et al. (2015) or \citet[][]{de_vaucouleurs_third_1991}, based on optical/infrared images. 
    The values of each column sum up to 100\%.
    We note that class \textbf{S} (smooth) is not available in the literature.}
    \label{fig:ConfusionMatrixArm}
\end{figure}

We compare the distribution of our arm classifications to optical/infrared literature classes for 43 galaxies that have both PHANGS and literature classes available (Figure~\ref{fig:ConfusionMatrixArm}). 
Grand-design ($12/18 = 67\%$) spirals as well as flocculent galaxies ($8/8 = 100\%$) are well identified in both stellar and gas distribution with an excellent agreement ($77\pm6$\% of \textbf{G}-\textit{G} and \textbf{F}-\textit{F}).
Relative uncertainties are between $3$ and $6$\% on the number of galaxies for each combination of our and literature classes.

Multi-armed spirals (\textbf{M}) show the least consistency between CO and near-IR morphologies, implying that they are the most wavelength dependent. 
Not only the agreement with the literature is lower ($47\%$), but also the agreement among the inspectors (on average $74\pm 8$\% for \textbf{M}, compared to $85\pm 12$\% and $85\pm 13$\% for \textbf{G} and \textbf{F}).
This could be due to both the generally reduced arm-interarm contrast 
\citep[e.g., ][]{elmegreen_grand_2011} and tighter pitch angles \citep[e.g., S$^4$G study by ][]{diaz-garcia_shapes_2019} of multi-arm galaxies compared to grand-design spirals. 
\citet[][]{savchenko_multiwavelength_2020} find no pitch angle difference between those arm types, but increased arm widths for grand-design arms that could lead to higher gas densities that ease the CO-based arm classification. 
The dissipative gas is closely connected to the underlying gravitational potential and might respond only to the strongest underlying features. As an example, a stellar multi-arm pattern that consists of two stronger stellar features might be visible as grand-design in the gas distribution. A transient stellar arm feature, on the other hand, might be visible as flocculent disk in the gas distribution.

We find four galaxies that are classified as 
flocculent (\textbf{F}) in our data, but as grand-design (\textit{G}) according to the literature data. 
These might correspond to \textit{quiescent} gas-stripped spiral \citep[e.g.][]{sellwood_spirals_2022}. 
These galaxies have average stellar masses and SFRs similar to those of the full sample (compare Section~\ref{sec:MorphologyResults:GalaxyProps}). 
We investigate the \textit{G}+\textbf{F} cases in more details as well as show their CO and 3.6$\mu$m distribution in Appendix~\ref{sec:Appendix:MorphologySpecialcasesG-F}.
Most of these galaxies have the bulk of the stellar disk's morphological features either outside the CO disk, as the extent of the CO disk is small compared to the optical one, or outside of the FoV, which is selected based on the smaller CO disk.
This makes the restricted FoV one of the major limitations of the CO data. 
One galaxy (NGC~4569) however, known to be a passive spiral, shows how these deviating CO classifications can help unravel the underlying physical evolutionary state of the galaxy.

\subsubsection{Bars in CO vs optical/IR}

\begin{figure}[t]
    \centering
    \includegraphics[width = 0.45\textwidth]{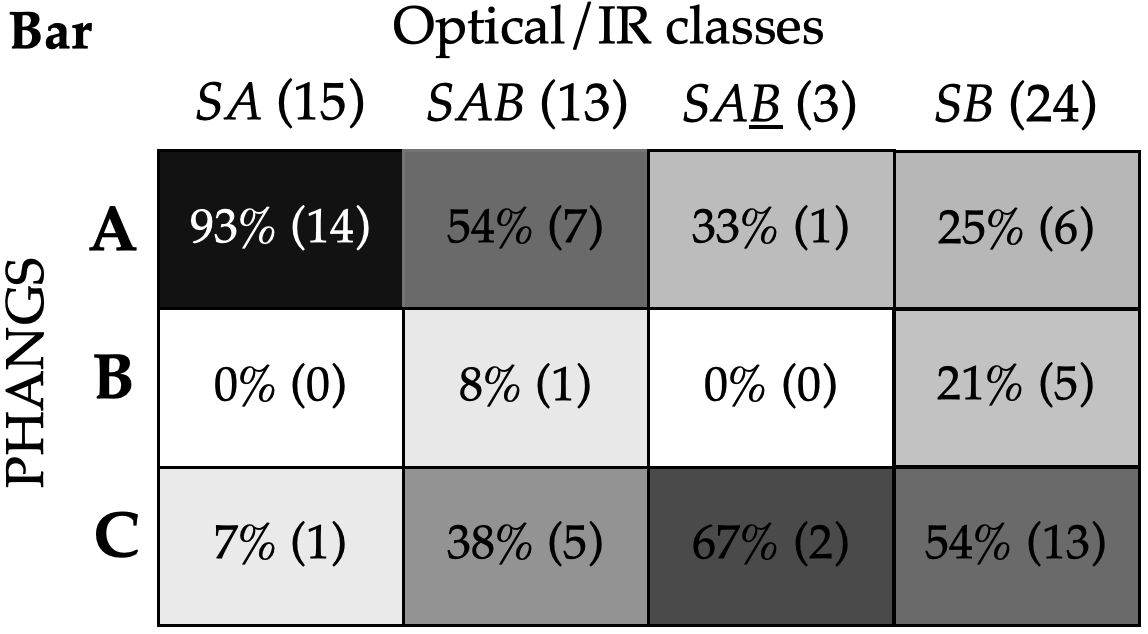}
    \caption{Confusion matrix of morphological bar classes from this work compared to  literature classes from Buta et al. (2015) for S$^4$G using IRAC $3.6\,\mu$m images and \citet[][]{de_vaucouleurs_third_1991} otherwise (=RC3), based on optical and infrared images. 
    The values of each column sum up to 100\%.
    We note that none of our galaxies has a literature bar class of \SAAB, thus no column is added.}
    \label{fig:ConfusionMatrixBar}
\end{figure}

We list the distribution of our bar-like classes compared to the literature classes in Figure~\ref{fig:ConfusionMatrixBar} for those 55 galaxies that have both literature and our verified classes. 
Our relative uncertainties are between $2$ and $17$\% for the bar-like classes. 

We find good agreement with the literature classifications for unbarred galaxies ($14/15 = 93 \pm 10\%$) when comparing class \SA to PHANGS class \textbf{A}, as well as a moderate agreement for strongly barred galaxies (\SB to PHANGS class \textbf{C}: $13/24= 54\pm 2\%$). 
Largest discrepancies are found for intermediate classes, such as \SAB or \SABB.  
Given that the classification schemes of \SAB and \SABB are not the same as PHANGS class \textbf{B}, we consider this an overall good agreement.

For the 55 PHANGS galaxies with both literature and our verified CO bar classes, we infer bar fractions of $f_\mathrm{lit} {\sim}73\%$ (40/55 class \SAB, \SABB or \SB) and $f_\mathrm{CO} {\sim}49\%$ (27/55, class \textbf{B} and \textbf{C}) respectively.

\subsubsection{Are we missing bars?}
\label{sec:MorphologyResults:Missingbars}

We investigate possible reasons for the low bar fraction from the CO morphology, by analyzing galaxies with the largest discrepancies between their bar classes based on the gas and stellar distribution below. 

\paragraph{\SB+\textbf{A}}
\label{sec:MorphologyResults:SB_A}
Six galaxies are classified as strongly barred (\SB) in the literature, but as unbarred based on CO morphology (compare Table~\ref{tab:Allclasses} and \ref{tab:AppendixPHANGSProperties}).
We list average properties of those galaxies, and galaxies with available PHANGS verified bar classifications, and averages of barred (\textbf{B},\textbf{C}) and unbarred (\textbf{A}) PHANGS galaxies in Table~\ref{tab:SB+Agalaxies}. 
A more detailed list of properties for each \SB+\textbf{A} galaxy, literature studies and multi-wavelength images are provided in Appendix~\ref{sec:Appendix:MorphologySpecialcasesSB-A}.
\SB+\textbf{A} galaxies have significantly lower stellar masses than the average PHANGS galaxy, as well as the average barred and unbarred PHANGS galaxy. 
\SB+\textbf{A} galaxies further have SFRs lower than barred galaxies, but higher than unbarred galaxies. 
We find no difference in bar classification agreement of those galaxies compared to average (barred/unbarred) PHANGS galaxies. 
The average molecular to stellar mass ratio for the \SB+\textbf{A} galaxies is slightly lower compared to the full sample.
A lower molecular mass could imply a lower surface density.
We rule out that CO sensitivity is not high enough in our data. We confirm this by taking images of barred galaxies at higher stellar masses (thus with higher luminosity) and subtract increasing levels of noise from the data.  But even at high noise clipping, as long as the disk is still visible, the structure of the bar can also be clearly identified. 

We conclude that, these galaxies could reveal gasless bars, in which further gas inflow is ceased.  
In some cases, an assembly of star-forming clumps might also mimic short elongated structures, that are easily misinterpreted as bars. 
However, in all of these cases the disks are chaotic and structures do not follow clear orbits, making these cases challenging for visual classification.

\begin{table*}
\caption{Average properties of all \SB+\textbf{A} galaxies}\label{tab:SB+Agalaxies}
\centering
\begin{tabular}{llllllll}
\hline\hline
\noalign{\smallskip}
 & N & $\langle \mathrm{log}_{10}  \mathrm{M}_\ast / \mathrm{M}_\odot \rangle$ & $\langle \mathrm{log}_{10}  \mathrm{SFR}/ \mathrm{M_\odot \, \mathrm{yr}^{-1} \rangle}$ & $\langle \mathrm{AG}_\mathrm{Bar}\rangle$ & $\langle \mathrm{AG}_\mathrm{Arm} \rangle$ & $\langle \mathrm{T}_\mathrm{Hubble} \rangle$ & $\langle \mathrm{log}_{10}  \mathrm{M}_\mathrm{mol} / \mathrm{M}_\ast \rangle$\\
\noalign{\smallskip}
\hline
\noalign{\smallskip}
\SB+\textbf{A} & 6 &  9.83 & -0.011  & 85\%& 85\%& $5.2 \pm 0.4$ & $-0.84$\\
PHANGS & 65 & 10.32& -0.013  & 86\%  & 76\%& $3.5 \pm 0.1$ & $-1.22$\\
\textbf{A} & 36 & 10.14 & -0.292 & 86\% & 77\% & $3.4 \pm 0.1$ & $-1.33$\\
\textbf{B,C} & 29 & 10.54 & 0.333 & 86\% & 75\% & $3.6\pm 0.1$ & $-1.10$\\
\hline 
\end{tabular}
\tablefoot{We list average properties of all \SB+\textbf{A} galaxies, all PHANGS galaxies with verified bar classes and barred (\textbf{B},\textbf{C}) and unbarred (\textbf{A}) galaxies. Stellar masses and SFRs have a statistical uncertainty of ${\sim}0.1\,$dex, and for the bar and arm agreement of ${\sim}5-10$\%.}
\end{table*}

\paragraph{\SA+\textbf{C}}
\label{sec:MorphologyResults:SA_C}

\begin{figure}
    \centering
    \includegraphics[width = 0.45 \textwidth]{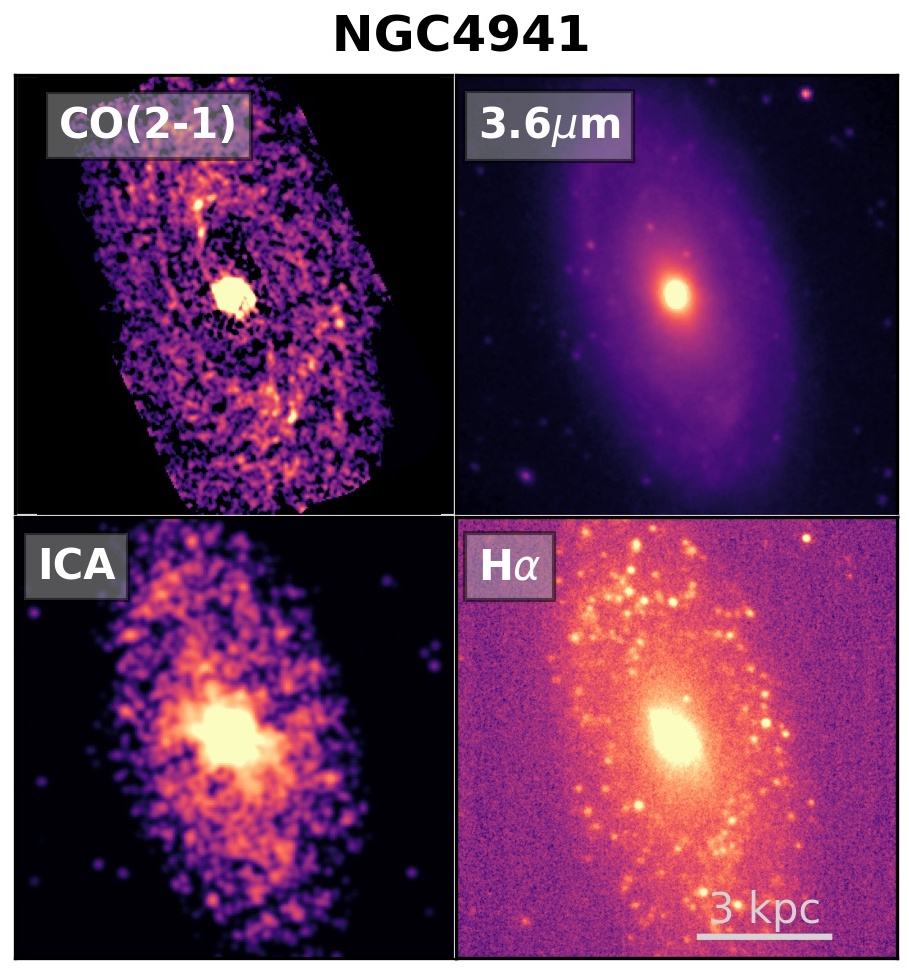}
    \caption{Image of NGC~4941 at different wavelengths. From left to right and top to bottom: broad CO mom-0 map, \textit{Spitzer} $3.6 \mu$m image, ICA s2 image, H$\alpha$ image. Some images are saturated in the center to emphasize faint emission that could indicate the presence of a bar or bar lanes.}
    \label{fig:NGC4941multiICA}
\end{figure}

NGC~4941 shows strong evidence for a bar-like structure in CO (class \textbf{C}), which is not apparent in the stellar light distribution (\SA). 
Its SFR is below the PHANGS average and it has low Hubble type (NGC~4941: log$_{10} \, M_\ast / M_\odot = 10.17$, log$_{10}$ SFR$/ M_\odot\,\mathrm{yr}^{-1} = -0.35$, $T = 2.1 \pm 0.6$, PHANGS average: log$_{10} M_\ast / M_\odot {\sim}10.5$, log$_{10}$ SFR $/ M_\odot\, \mathrm{yr}^{-1} {\sim}+0.42$, $T {\sim}3.6$). 
Figure~\ref{fig:NGC4941multiICA} shows images of the galaxy in different wavelengths.
Two thin bar lanes are visible in CO, which are telling signs of a bar, which are less clear at other wavelengths.
Literature stellar light classifications have been inconclusive as other studies classify this galaxy as \SAB \citep[RC3 catalogue; ][]{de_vaucouleurs_third_1991, corwin_corrections_1994}, and find visual \citep[][]{erwin_double-barred_2004, menendezdelmestre_nearinfrared_2007} and kinematic \citep{lang_phangs_2020} evidence for a stellar bar.
Since this is the only case for which we found an additional bar, this does not account for significant ($>3\sigma$) differences in any of our percentages.
This galaxy emphasizes what benefit the usage of CO data can add.

\subsection{Bars and spiral arms along the main sequence}
\label{sec:MorphologyResults:GalaxyProps}

The morphologies of galaxies are tied to their global properties, such as stellar mass, color, or gas reservoir \citep[e.g.,][]{biviano_arm_1991, sparke_galaxies_2007, lokas_bar-like_2021-1}. 
Since we find a generally good correspondence of CO to optical morphologies, we search here for further trends, that might reveal aspects of the formation and evolution of morphological features in the underlying host galaxy disk. 
Since the only restriction for our sample galaxies is that they must be main sequence galaxies, we will analyze for trends with stellar mass and/or SFR first. 

\begin{figure*}
    \centering
     \makebox[\textwidth][c]{\includegraphics[width=0.99\textwidth]{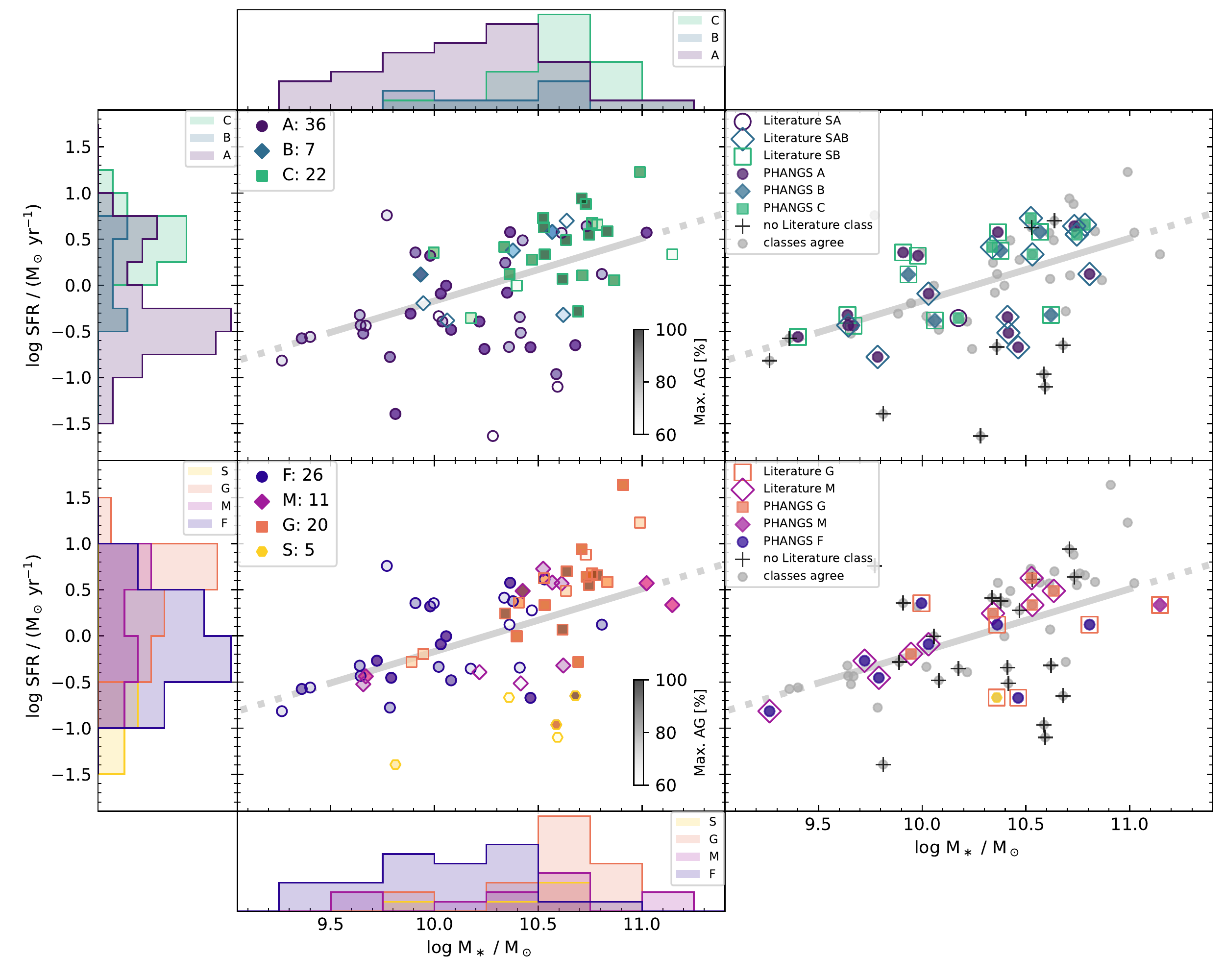}}
    \caption{Distribution of PHANGS bar-like features (upper panels) and outer (arm-) features (lower panels) along the star-forming main sequence according to \citet{leroy_z_2019} (gray solid line, extrapolated to larger and smaller stellar masses with a dotted gray line). 
    \textbf{Left panels}: 
    The galaxies are colored by their classes (upper: A,B,C, lower: F,M,G,S) and opacity indicates the confidence of each classification. An example of the confidence range is indicated as grey scale bar in the left panels. 
    Histograms indicate the distribution of galaxy stellar masses and SFRs for each class. 
    \textbf{Right panels:}
    Galaxies with deviations of one or more than one classes from the literature classification, as described in the text, are highlighted by their PHANGS class  (colored filled symbol) and their corresponding literature class (colored open symbol). 
    Colors and shapes are the same for the left panels. 
    For comparison, galaxies with similar classification (grey points) as well as galaxies without a literature class (black plus) are shown. 
    We note that among our galaxies none have literature classes \SAAB and all literature flocculent classifications agree with PHANGS classifications. PHANGS armclass S does not have a corresponding literature class.}
    \label{fig:MonsterMS}
\end{figure*}

Figure~\ref{fig:MonsterMS} shows the distribution of our verified bar and arm classes along the star-forming main sequence highlighted by their classification confidence (left panels, opacity). 
We highlight deviations to literature classes (right panels) that are not part of the following classes, that we consider as good agreement: \textbf{G}-\textit{G}, \textbf{M}-\textit{M} and \textbf{F}-\textit{F}, as well as \textbf{A}-\SA/-\SAAB, \textbf{B}-\SAAB/-\SAB/-\SABB, \textbf{C}-\SABB/-\SB.
Histograms further indicate the distribution of individual features based on their host galaxy's stellar mass and SFRs.

Bars are preferentially identified at higher stellar masses and higher global SFRs in the sample.
We find significant differences between the distribution of stellar masses of galaxies classified as \textbf{A} compared to those classified as \textbf{C} (KS-test $p_{AC}(\mathrm{M}_\ast) < 0.01\%$), whereas we do not find significant differences between \textbf{A} and \textbf{B} (KS-test $p_{AB}(\mathrm{M}_\ast) = 42\pm5\%$) or \textbf{B} and \textbf{C} (KS-test $p_{BC}(\mathrm{M}_\ast) = 9\pm5\%$), likely due to the small sample size of 7 \textbf{B} galaxies. 
This is consistent with previous studies \citep[e.g.,][]{masters_galaxy_2012, Tawfeek_2022} where bar fraction increases strongly with stellar mass up to log$_{10} M_\ast/M_\odot \sim~ 11.4$. 
Galaxies at higher stellar masses tend to become kinematically cool earlier than lower mass galaxies \citep[e.g.,][]{sheth_hot_2012}, and can therefore become bar-unstable earlier. 

Further, significant trends are also found between \textbf{A} and \textbf{C} for their distribution of $\Delta$MS, SFR, numerical Hubble type, which are not seen for other class combinations.  
Simulations find that higher molecular gas fractions can suppress the formation of bars \citep[][]{lokas_effect_2020}, can result in weaker bars \citep{athanassoula_bar_2013}, or in a decreased bar fraction with both higher gas fraction and lower stellar masses \citep{zhou_barred_2020, rosas-guevara_buildup_2019}.
We do find significant trends between \textbf{A} and \textbf{C} and their distribution of the total molecular gas mass, but not the fraction of molecular to stellar mass. 
In short, our barred galaxies tend to live above the main sequence, at higher SFRs and have more molecular gas in the disk.

Similarly, grand design spirals (\textbf{G}) are hosted by galaxies with higher stellar mass and higher SFRs than flocculent (\textbf{F}) galaxies which tend to exist in lower stellar mass and lower SFR galaxies, in agreement with findings at $3.6\,\mu$m in S$^{4}$G by \citet{bittner_how_2017}. 
We find significantly low $p$-values (< 5\%) for KS-tests for galaxies of class \textbf{G} and \textbf{F} for $\Delta$MS, stellar mass, SFR, atomic gas mass as well as molecular gas mass. Similarly to bars, no trends with gas fractions (e.g., molecular to stellar mass) are found. 

Grand-design spirals tend to lie slightly above the main sequence, have higher SFRs and live in disks with high amounts of absolute molecular gas masses. 
Similarly, \citet{Sarkar_Identification_2022} find a slightly increased total HI mass and higher surface brightness in grand-design galaxies compared to flocculent ones, in a set of ${\sim}$1220 galaxies recently classified by neural networks. 
We find no significant trends when comparing class \textbf{G} and \textbf{M}, in agreement with \citet{bittner_how_2017} as well as no significant trends when comparing \textbf{M} and \textbf{F} except for stellar mass.

The link to stellar mass could reflect the importance of bulges for the propagation of density waves \citep[e.g., review by ][]{sellwood_spirals_2022}, or the dependence of arm number on stellar surface density \citep[e.g. as predicted in the theory of global modes;][]{bertin_modal_1989}. 
The fact that this structure is echoed in the CO morphology may be a testament to the high molecular gas masses that are also characteristic of more massive galaxies. 

We find that both galaxies hosting bars, as well as galaxies hosting grand-design spirals populate the high stellar mass end of our sample. 
Consequently, ${\sim}52\%$ (15/29) of barred galaxies in our sample host a grand-design spiral and ${\sim}75\%$ (15/20) of galaxies with grand-design spiral arms are barred.
The $p$-value of the KS test comparing stellar masses of a sub-sample of galaxies classified as bar-class \textbf{A} and spiral arm class \textbf{F} (17 galaxies) to a sub-sample of bar-class \textbf{C} and spiral arm class \textbf{G} galaxies (13 galaxies) is well below 5\% ($p_{AF-CG} = < 0.01 \pm <0.01$\%). 
This is in agreement with the findings of \citet{elmegreen_flocculent_1982}, who showed that most grand design spirals were in barred galaxies. 
This finding would suggest that the presence of a bar, e.g., via its bar quadrupole moment, could trigger the formation of grand-design structures, as suggested by several works \citep[e.g., see summary by ][]{sellwood_spirals_2022}. 
On the other hand \citet{diaz-garcia_shapes_2019} suggest that both features might coincidentally coexist, as they do not find trends between bar strengths and pitch angles in S$^4$G.

\subsection{Bar fraction}
\label{sec:MorphologyResults:Barfraction}
Among our sample of main sequence galaxies, we obtain a total bar fraction of $f_\mathrm{bar} = 45 \pm 5\%$. 
Historically, the bar fraction has been found to have a value of ${{\sim}}60{-}70\%$ \citep[][]{sheth_barred_2003, sheth_evolution_2008, oswalt_galaxy_2013} in optical and IR surveys, although more recent studies using multi-wavelength data report lower bar fractions \citep[e.g.,][]{erwin_dependence_2018}. 
Our fraction is similar to the results of optical surveys (${\sim}45\%$ from low-inclined nearby ($0.01 < z < 0.04$) spiral and S0 galaxies from SDSS $r$-band observations; \citet{aguerri_population_2009}, 44\% in B-band; \citet[][]{marinova_characterizing_2007} or 47\% in I-band; \citet[][]{reese_photometric_2007}). 
Near-infrared observations tend to find significantly higher fractions \citep[e.g., 60-70\% from S$^4$G $3.6\,\mu$m;][]{buta_classical_2015}.
The overall fraction of our strongly barred galaxies (class \textbf{C}) is ${\sim}34\%$, similar to  ${\sim}36\%$ from the Galaxy Zoo project \citep[][]{lintott_galaxy_2008}, but twice as high compared to ${\sim}16\%$ in isolated galaxies observed in mid-IR \citep[][]{buta_comprehensive_2019}.
In fact, discussions on the bar fraction as well as its dependency on galaxy properties such as stellar mass or SFR have not yet come to an agreement. 
Bar fraction determinations are sensitive to the selected bar-identification technique, e.g., ellipse-fitting vs. Fourier decompositions especially in late type galaxies \citep[e.g., ][]{aguerri_population_2009}. 
Our results suggest that the choice of tracer is another important parameter to consider.

\begin{figure}[t]
    \includegraphics[width = 0.43\textwidth]{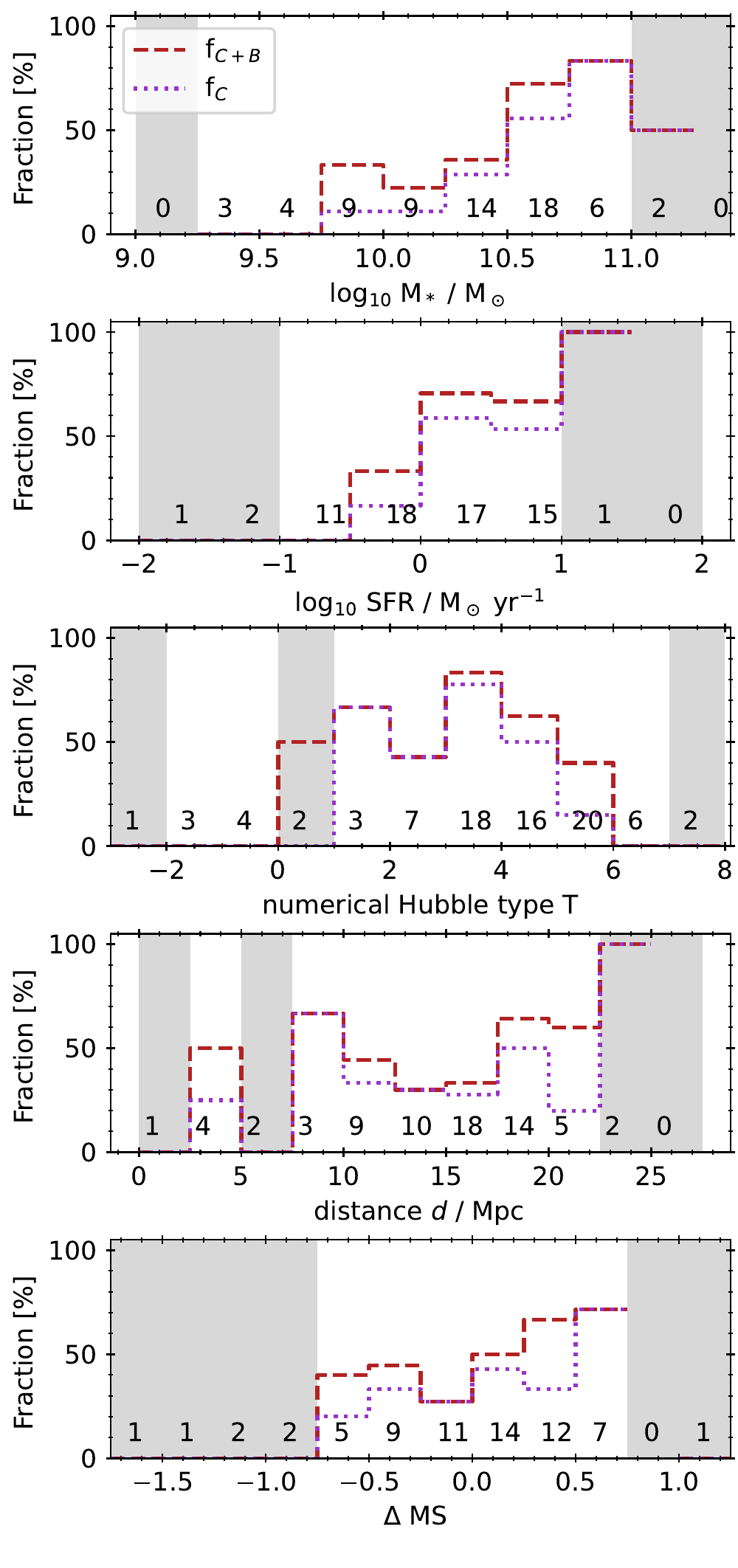}
    \caption{Fraction of all barred galaxies f$_{C+B}$ (red, dashed) and strong bars (\textbf{C}) f$_{C}$ (purple, dotted) as function of stellar mass (top left), SFR (top center), numerical Hubble type (top right) as well as galaxy distance (bottom left) and offset from the main sequence $\Delta\,$MS (bottom center). Numbers indicate the total number of galaxies per bin. Bins with less than 3 galaxies are shaded in grey.}
    \label{fig:Barfractions}
\end{figure}

As the PHANGS sample is considered representative for nearby galaxies within its selection range \citep[e.g., galaxies at $12 < d < 17\,$Mpc;][]{leroy_phangsalma_2021-1}, we search for trends with host galaxy properties that might impact our bar fraction, namely stellar mass, SFR, Hubble type and distance. 
Figure~\ref{fig:Barfractions} displays a clear increase in bar fraction (\textbf{C} and \textbf{B} or just \textbf{C}) with stellar mass, in agreement with \citet[e.g., Galaxy Zoo; OmegaWINGS;]{masters_galaxy_2012, Tawfeek_2022} but in disagreement with \citet{erwin_dependence_2018}, who find a decrease in bar fraction at stellar masses larger than log$_{10} M_\ast/M_\odot \sim 9.7$ in S$^4$G.
Low mass galaxies tend to have smaller bar sizes \citep[${\sim}1.5\,$kpc at $\log M_\ast /M_\odot {\sim}9.5$, which is the lower stellar mass end of our sample;][]{erwin_dependence_2018}. 
We reject resolution biases given our ${\sim}100$\,pc resolution and the absence of a trend between bar fraction and distance in our sample. 
We find an increased bar fraction at higher total (molecular) gas mass, which we consider a secondary correlation due to the relation between stellar mass and gas mass, as we do not find any trend with relative (molecular) gas fractions. 
Although our sample does not probe far beyond the main sequence, we find a weak increase in bar fraction with positive offset from the main sequence $\Delta\,$MS, potentially linking SFR to increased bar presence, or vice-a-versa, probing a future direction of studies for a sample with a larger range of $\Delta\,$MS.

There is no trend with numerical Hubble type, as the bar fraction is roughly constant between type 0 to 6, similar to  \citet{erwin_dependence_2018}, but in disagreement with the strong correlation between bar fraction and Hubble type reported by \citet{Tawfeek_2022}.
Since our sample does not contain galaxies with Hubble types ${\gtrsim}7$ (Sd), we might underestimate the bar fraction, as \citet{buta_classical_2015} find an increased bar fraction for Scd-Sm galaxies of ${\sim}81\%$ compared to a fraction of only ${\sim}55\%$ in S0-Sc galaxies.

\subsection{Bar lane classification}
\label{sec:MorphResults:Bltrends}
The shape and length of bar lanes provide insight into the underlying bar structure \citep[e.g., ][]{athanassoula_morphology_1992, athanassoula_existence_1992} and  
the bar lane geometry is valuable information for estimating bar inflows \citep[e.g.,][Sormani \& Barnes in prep.]{sormani_mass_2019}.
Literature studies further suggest a link between the curvature and shape of bar lanes and bar strengths \citep[e.g., ][]{comeron_curvature_2009}.
Here, we examine the morphology of bar lanes, namely the shapes (between straight \textbf{c1} and most curved \textbf{c5}) and relative length of the CO bar lane with respect to the full bar size (short \textbf{s}, long \textbf{l}) as traced by CO emission and test how well bar lane properties agree with the predictions. 
The classes for the shape are selected to represent a visual increase in curvature, therefore we refer to these classes as shapes or curvatures equally.

\subsubsection{Visual shape and length of bar lanes}
\label{sec:MorphResults:Bltrends:shapelength}

\begin{figure}[t]
    \centering
    \includegraphics[width = 0.5\textwidth]{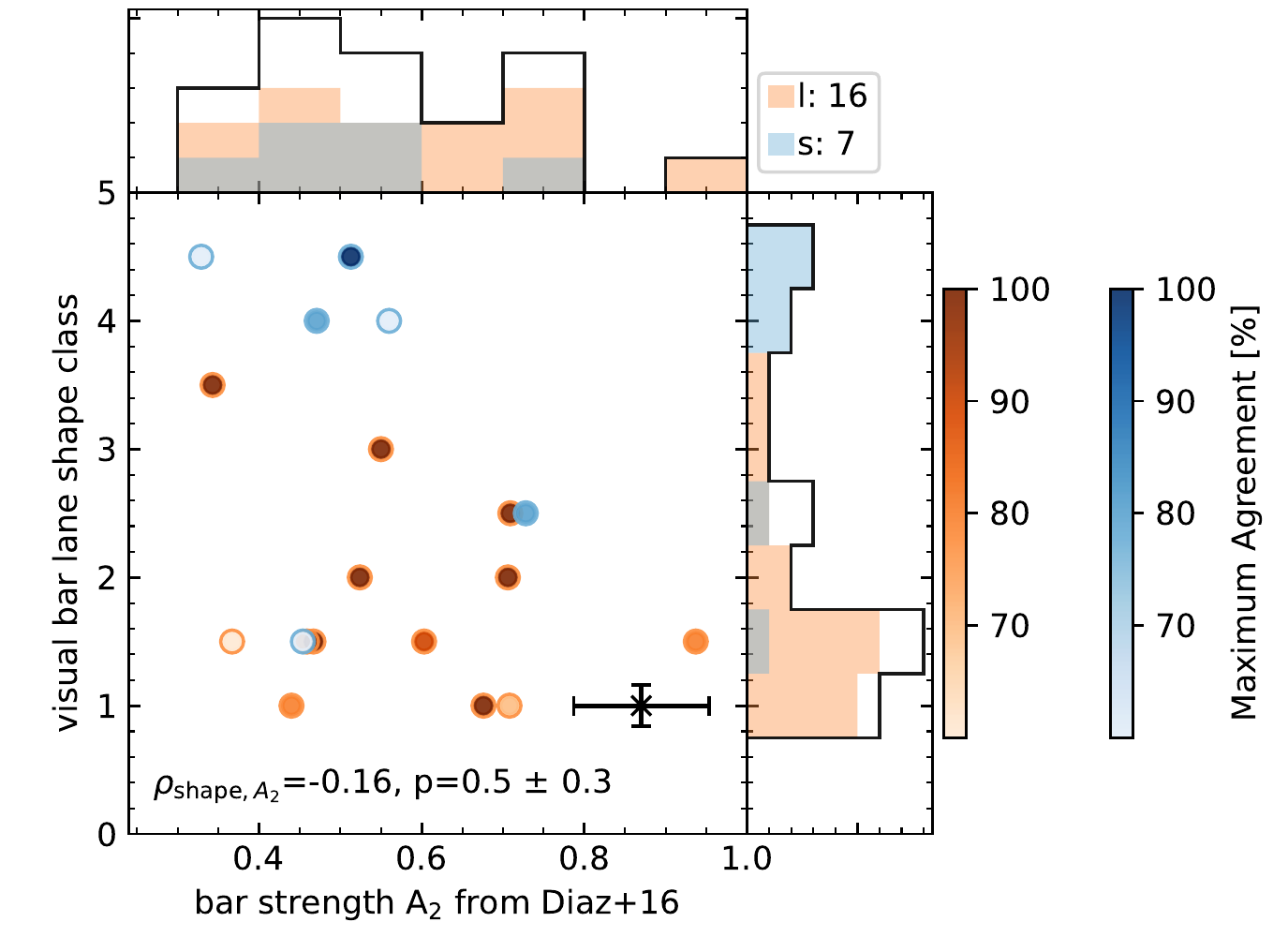}
    \caption{Bar lane curvature (shape) of barred PHANGS galaxies (from \textbf{c1}=1.0 to \textbf{c5}=5.0) as function of bar strength $A_2$ from \cite{diaz-garcia_characterization_2016}, colored by bar lane length (short: \textbf{s}), long: \textbf{l}). 
    Out of 28 with bar-lane classifications no bar strengths were available for 3 galaxies, and for 2 galaxies no high-quality curvature classes are available. 
    The bar lane shapes of \textbf{s} and \textbf{l} galaxies are significantly different (KS-test p-value $<5\%$).
    No significant trends are found between literature bar strengths of \textbf{s} and \textbf{l} galaxies, nor between bar strengths and bar lane shapes.  
    }
    \label{fig:BLDiaz16}
\end{figure}

We find that short bar lanes (class \textbf{s}) are in generally more curved than longer bar lanes (class \textbf{l}) (compare to right histogram of Figure~\ref{fig:BLDiaz16}).
KS tests reveal that \textbf{s} and \textbf{l} bar lanes have distinct curvatures/shapes; KS-test $p$-value of $0.6 \pm 4\%$\footnote{Uncertainties of KS-test $p$-values (and Spearman test $p$-values) are derived by perturbing the values with their intrinsic uncertainties (e.g., $\Delta$\textbf{c}$= 0.16$) and applying bootstrapping. Then we calculate the coefficients on the perturbed and bootstrapped values and repeat the process 1000 times. The standard deviation of the resulting $p$-values reveals the uncertainty.},

The curvature of bar lanes is often found to be inversely proportional to the bar strength \citep[e.g., ][]{athanassoula_morphology_1992, comeron_curvature_2009}, i.e., bar lanes in strong bars are nearly straight, and weaker bars have lanes with higher curvature. 
Figure~\ref{fig:BLDiaz16} shows our bar lane shapes as function of literature bar strength A$_2$ from \citet[][]{diaz-garcia_characterization_2016} (available for 19 of the 26 barred galaxies with verified bar lane classifications) with conservative uncertainties of $15$\% on the values, color-coded by bar-lane length classification.
Trends between A$_2$ values of \textbf{l} and \textbf{s} galaxies, as well as between bar lane shapes and  A$_2$ are not statistically robust, but consistent with literature trends \citep[e.g., lack of high curvature - high bar strength values][]{comeron_curvature_2009}.
We do not find significant trends with other bar strength indicators such as Q$_b$ or the bar ellipticity from \citet[][]{diaz-garcia_characterization_2016}. 

Several studies showed that bars generally grow stronger over time \citep{athanassoula_what_2003, sellwood_secular_2014, fragkoudi_revisiting_2021}. 
Our data potentially indicates that there are no strong and highly curved bars with short gas bar lanes, and therefore potentially no older bars with short and highly curved bar lanes.

We find no significant trends with neither visual shape nor visual length of bar lanes and Hubble type, molecular gas mass, molecular gas mass to stellar mass ratio or atomic gas mass, or atomic to stellar mass ratios on the bar lane geometries. 
Since all our bars are well resolved with at least $\sim 13$ resolution elements at our $1\arcsec\ $ resolution, we do not find any trend with distance.

\subsubsection{Measuring bar lane lengths}
\label{sec:BarlaneAnalysis}

We developed a quantitative approach (see Appendix \ref{sec:Appendix:Barlanemasks}) to assess 
the reliability of the visual classification. 
We focus on class \textbf{C} (22 galaxies), as the bars are better defined i.e, the area surrounding the bar is more cleared and the bar lanes are less thick than for \textbf{B} galaxies.
Out of those we have 21 with bar lane length classifications and bar lane shape classifications.

\paragraph{Generation of bar lane masks}
\label{sec:BarlaneMethod:masks}

\begin{figure}[t]
    \centering
    \includegraphics[width = 0.49\textwidth]{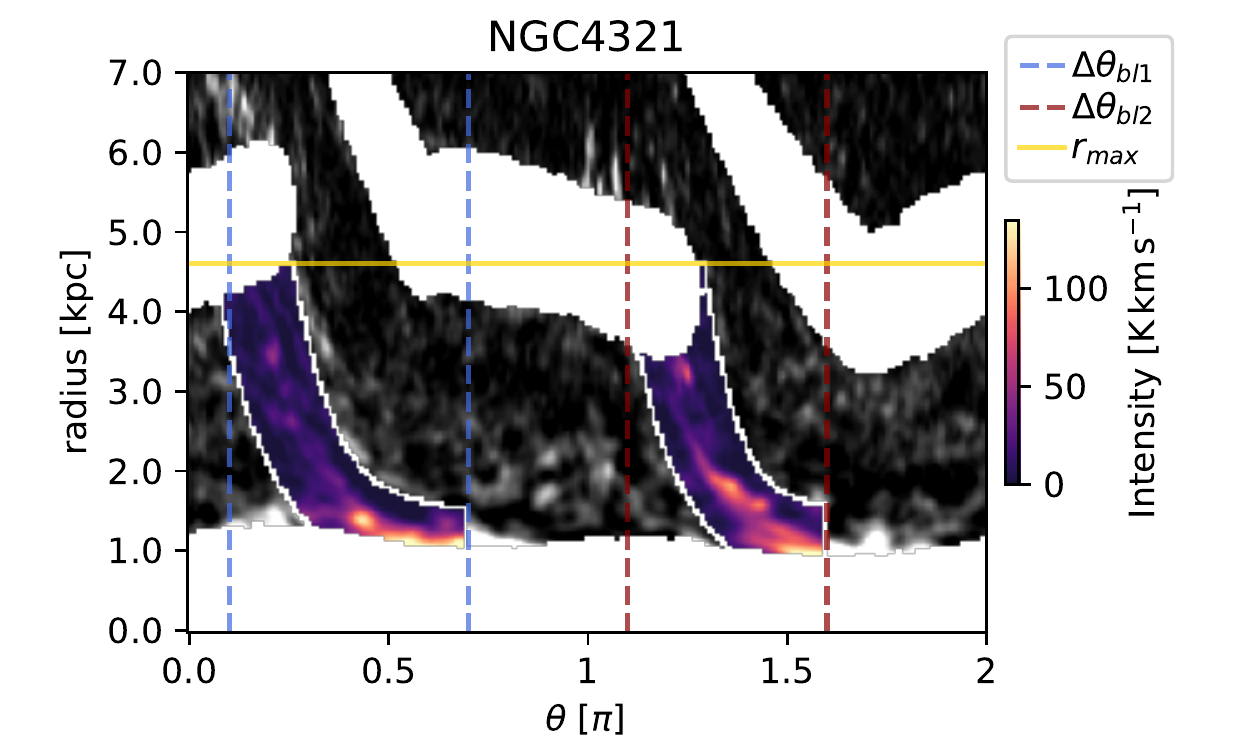}
    \caption{CO polar image of NGC~4321 with the bar lane masks on top. 
    The center is masked out with an environmental mask, as well as the spiral arms. Emission at radii larger than the full bar length (yellow horizontal line) is also masked out (see Table~\ref{tab:Barlanelist}).  
    The selected azimuthal range for each bar lane (first bar lane ID~1: $\Delta \theta_{bl1}$ blue dashed line, second bar lane ID~2: $\Delta \theta_{bl2}$ red dashed line) used for the fitting is indicated. The bar lane mask area is defined by two shifted exponential profiles is shown in colorscale, the area outside in greyscale.
    The starting angle corresponds to a position angle of 90 degrees (measured clockwise from north) in the deprojected map, further environmental region masks are applied.}
    \label{fig:PolarimageFullmask}
\end{figure}

In deprojected mom-0 maps we mask out emission of spiral arms, central regions and outside deprojected bar radii from \citet{diaz-garcia_characterization_2016}. 
In polar projection, we then isolate both bar lanes by selecting an angular range by eye (see Figure~\ref{fig:PolarimageFullmask}) 
and refer to each bar lane according to its order in azimuthal direction with ID~1 (blue) or ID~2 (red) respectively. 
Next, we fit the bar lanes with a simple functional (exponential profile).
Our mask is described by two functions of this shape, which are shifted along the normal to the best-fit exponential function, as seen in Figure~\ref{fig:PolarimageFullmask}.

\paragraph{Radial intensity profiles along the masks}
\label{sec:BarlaneMethods:Radialprofiles}

We measure the length of the bar lanes by finding the radii where the mean bar lane intensity drops below noise level.  
All radial intensity profiles of both bar lanes of our strongly barred galaxies are provided in Appendix~\ref{Appendix:RadialProfiles}.
The clumpiness of the molecular gas affects how continuously a bar lane can be identified. 
Therefore, we take the first (last) radius, where the bar lane intensity intersects with the $3\sigma$ noise level
as the minimum (maximum) length estimate and normalize these radii by deprojected bar radii $\mathrm{r}_\mathrm{bar}$ from \citet{diaz-garcia_characterization_2016}.
For each bar lane (ID~1 or ID~2), we consider the average of minimum and maximum normalized radii as bar lane length estimate $R_\mathrm{cov, ID} = \frac{1}{2} (\mathrm{r}_\mathrm{max} - \mathrm{r}_\mathrm{min}) / \mathrm{r}_\mathrm{bar}$. 
If no intersections are found, the maximum possible bar lane length is set by the bar radius ($R_\mathrm{cov, ID}=1$). 
The distribution of $R_\mathrm{cov, ID}$ for all galaxies can be found in Appendix~\ref{sec:Appendix:Barlanemasks}.

\paragraph{Trends with measured bar lane lengths}
\label{sec:BarlaneMethods:Trends}

\begin{figure}[t]
    \centering
    \includegraphics[width =0.45 \textwidth]{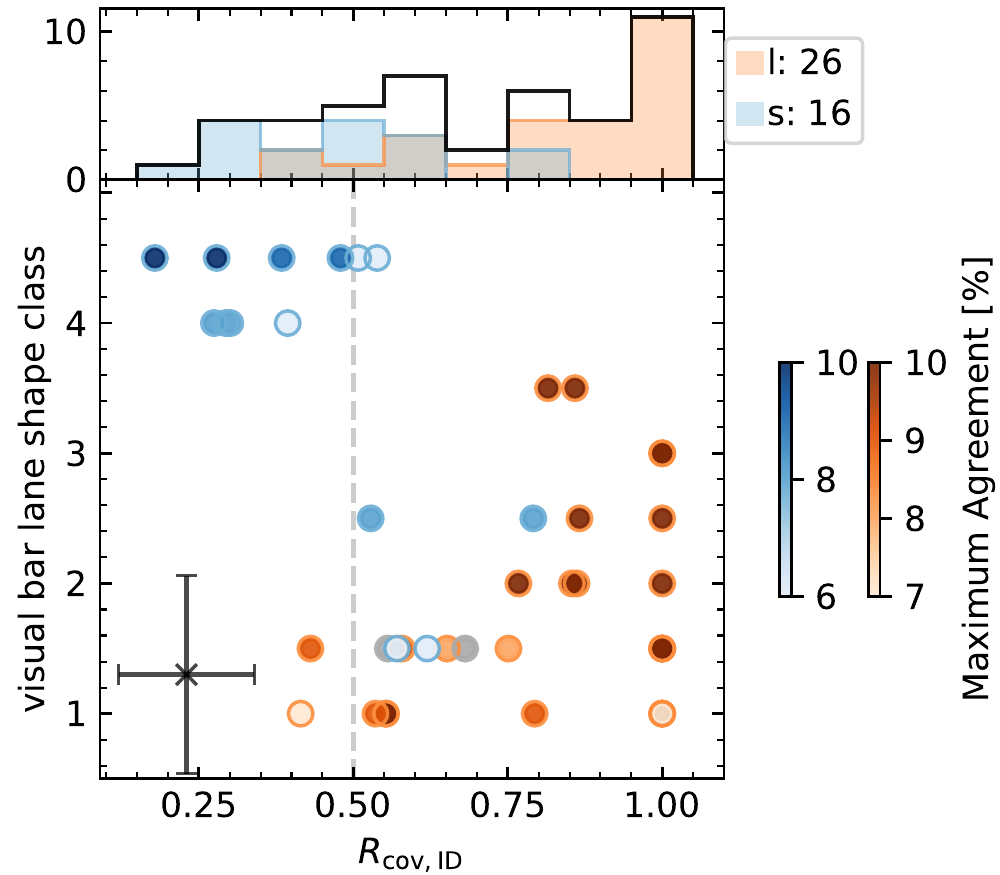}
    \caption{CO bar lane length ratios $R_\mathrm{cov, ID}$ and their dependency on  visual bar lane shape classes \textbf{c1}-\textbf{c5} from the morphological classification, colored by visual bar lane length classes \textbf{s} (blue) and \textbf{l} (orange). We indicate $R_\mathrm{cov, ID}=0.5$ (grey dashed line) and the agreement for the visual bar lane length classes is indicated by opacity.}
    \label{fig:BLnewlengths}
\end{figure}

We compare the measured bar lane lengths $R_{\mathrm{cov, ID}}$ to the corresponding visual length class in Figure~\ref{fig:BLnewlengths} (upper histogram).
Most measured $R_{\mathrm{cov, ID}}$ values agree with the visual ones, considering - per definition - a value of $R_\mathrm{cov,ID}\sim 0.5$ as separation between class \textbf{s} and \textbf{l}. 
KS-test reveals a significant difference between $R_{\mathrm{cov, ID}}$ of galaxies classified as \textbf{s} and those classified as \textbf{l}. 
Hereby, the average bar lane length ratios $R_{\mathrm{cov, ID}}$ have a standard deviation of $0.11$, derived from the distribution of $r_\mathrm{max}$ and $r_\mathrm{min}$.

KS-tests reveal a significant difference between bar lane shapes of galaxies with $R_{\mathrm{cov, ID}} < 0.5$ and those with $R_{\mathrm{cov, ID}}>0.5$ (KS-test values of $\kappa_{RcovID,c} = 0.68$, p-value of $p = 0.001\pm 0.05$), in agreement with our previous findings from Section~\ref{sec:MorphResults:Bltrends}, that bars with curved bar lanes (\textbf{c4}-\textbf{c5}) tend to be classified as short, and more straight lanes tend to be classified as long (\textbf{l}). However, we do not find a significant Spearman correlation coefficients between the visual bar lane shape classification and $R_{\mathrm{cov, ID}}$. 
Instead of a linear trend, our bar lane lengths seem to follow a bimodal distribution into either small (short bar lane) or large (long) values.

\begin{figure}[t]
    \centering
    \includegraphics[width = 0.44 \textwidth]{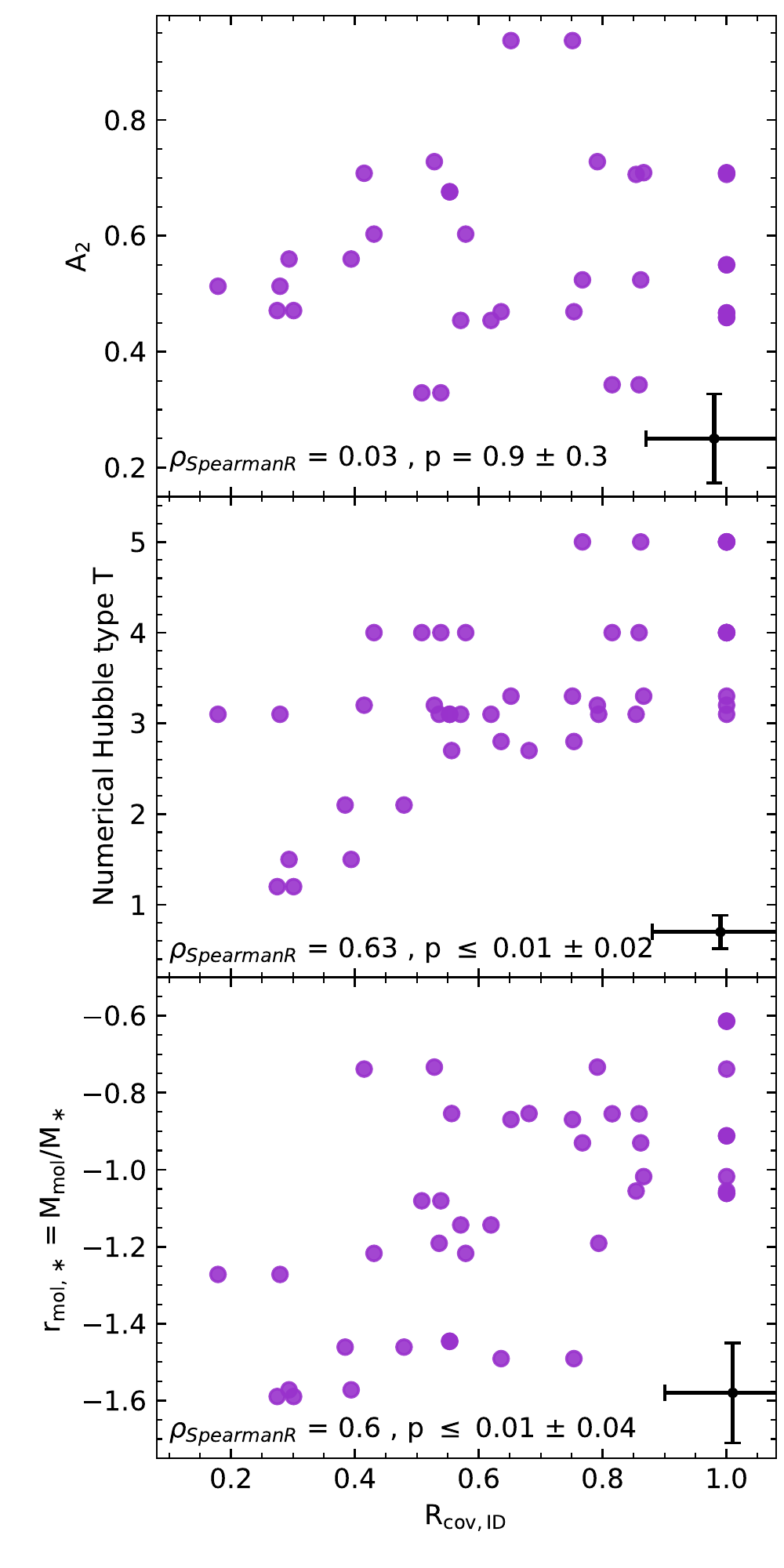}
    \caption{CO bar lane length ratios $R_\mathrm{cov, ID}$ and their dependency on bar strengths A$_2$ from \citet{diaz-garcia_characterization_2016} (top), Hubble type T (center) and molecular to stellar mass ratio $r_{\mathrm{mol}, \ast}$ (bottom). 
    Spearman correlation coefficients as well as errorbars are provided at the bottom.}
    \label{fig:BLresults:BLtrends}
\end{figure}

In Section~\ref{sec:MorphResults:Bltrends:shapelength} we found that the shape class of bar lanes correlates strongly with the bar lane length class and weakly with bar strength Q$_b$.  
We re-evaluate these findings with the bar lane length measurements $R_\mathrm{cov, ID}$. 
Figure~\ref{fig:BLresults:BLtrends} shows the bar strengths Q$_b$ from \citet{diaz-garcia_characterization_2016}, the numerical Hubble types T  and the ratio of molecular to stellar mass $r_{\mathrm{mol}, \ast}$ as a function of $R_{\mathrm{cov, ID}}$.
We use the 15\%-error for the bar strengths and the error of the molecular to stellar mass ratio values is $0.13$\,dex. 
Spearman correlation coefficients $\rho_{\mathrm{Spearman}}$ are provided in the bottom of each panel of Figure~\ref{fig:BLresults:BLtrends}.
We derive the coefficient uncertainties by perturbing the bar lane lengths with a uniform distribution between plus or minus their error, then performing 1000 times bootstrapping on these values. 
For each of these 1000 iterations we calculate the Spearman correlation coefficients on the bootstrapped values. The standard deviation of these coefficients yields the uncertainty.

While there is no significant trend between bar strength A$_2$ and $R_{\mathrm{cov, ID}}$, there might be a temptative lack of data points at smaller values of $R_{\mathrm{cov, ID}}$ and higher bar strengths.
The bar lane lengths $R_{\mathrm{cov, ID}}$ significantly correlate with the molecular gas fraction and numerical Hubble type. The latter is expected as Hubble type correlates with molecular gas content).
A higher content of molecular gas in the full galaxy disk might also reflect a higher amounts of gas  in the area of influence of the bar, that can be funnelled along the bar lanes and create potentially longer CO bar lanes. 
We note that higher resolution data might increase the measured length of bar lanes, as smaller molecular clouds -- if present in the bar lane -- could be detected at higher signal-to-noise level, leading to a potentially longer consecutive lane of molecular gas. However, since our sample populates a comparably small distance parameter space, all our measurements would be affected equally.

\paragraph{Emerging picture of bar evolution}
\label{sec:BarlaneMethods:BLconclusion}

Our results from Section~\ref{sec:MorphResults:Bltrends:shapelength} and \ref{sec:BarlaneMethods:Trends} indicate that bars with short bar lanes are weaker, tend to have more curved bar lanes and are preferentially in gas-poor galaxies when compared to bars with long and straighter bar lanes that also tend to be stronger. 
As a bar evolves, it grows in strength 
\citep{athanassoula_what_2003, diaz-garcia_characterization_2016}, and it accumulates mass in the centre due to funneling of gas along the bar lanes. Stronger bars tend to have less extended inner bar orbits perpendicular to the bar major axis (so-called x2-orbits) compared to the bar orbits parallel to the bar major axis (so-called x1-orbits), but more centrally concentrated bars tend to have more extended $x_2$ orbits \citep{athanassoula_morphology_1992}. 
Observations of nuclear discs indicate that the second effect prevails, so that the size of nuclear rings (which correlates with the extent of the $x_2$ orbits) grows over time \citep{gadotti_kinematic_2020,bittner_inside-out_2020}. Stronger bars also have more elongated $x_1$ orbits, and therefore they are expected to have straighter bar lanes.
 
Stronger bars might also be able to induce gas inflows for a longer period \citep{gadotti_kinematic_2020}, resulting in bar lanes fully covered in CO, thus being classified as longer.
The correlation found that bar lanes are longer in galaxies with higher molecular gas mass fractions might indicate that the general availability of gas in the disk plays an important role. 
We note however, that simulations find higher gas fractions to result in weaker bars \citep{athanassoula_bar_2013}, in contrast to our findings. 
We conclude that the molecular morphology of bars and their bar lanes might be able to add important insights on bar evolution, but larger sample sizes are required.

\subsection{Trends with rings}
\subsubsection{Central rings}

Our central rings (often in the literature referred to as \textit{nuclear rings}) are preferentially identified in barred galaxies ($59 \pm9\%$ 13, compared to $27\pm14\%$ in unbarred galaxies) in agreement with literature findings based on stellar light \citep[][]{knapen_structure_2005, comeron_ainur_2010}.
In barred galaxies the inner bar x2-orbits, if occupied by gas, are able to create such ring-features in the gas distribution. 
As stars form from the funnelled gas the ring is visible in stellar light as well \citep{gadotti_kinematic_2020, bittner_inside-out_2020}.
This increased presence of central rings in barred galaxies leads to secondary correlations: Central rings are more frequent at higher stellar masses and in systems with a grand-design spiral. 
Among our 29 barred galaxies, we find no significant differences in stellar mass, SFR or other properties between those with ($13$, $45\pm7$\%) and without ($12$, $41\pm10\%$) central rings.

In contrast, unbarred galaxies with a central ring ($6$, $17\pm8\%$) are at significantly lower Hubble types (at T$_r \sim -0.5$) and smaller offset from the main sequence (at $\Delta \mathrm{MS}_r \sim -1.0$) than those without central ring ($28$, $78\pm13\%$, T$_{nr} \sim 4.$, $\Delta \mathrm{MS}_{nr} \sim -0.1$).
These central rings in unbarred galaxies might be produced by tightly wound spiral arms that are more common at lower Hubble types. 
We do not find any correlation with molecular/atomic gas mass, SFR or distance. 

Out of 19 of our galaxies that are in the largest catalogue of nuclear rings, AINUR, \citep[surveying nuclear rings with HST at angular resolutions as high as $\sim 0.1$\arcsec;][]{comeron_ainur_2010}, three are not detected in our survey. Two are classified as \nAG, and one misclassified as \textbf{nr}, classified as having no central ring). 
This is likely due to our limiting resolution of 1\arcsec, that is not able to resolve rings with radii as small as 80\,pc in galaxies at these distances \citep{comeron_ainur_2010}. 
Resolution is therefore a limiting factor for the completeness of our identification of central rings.

\subsubsection{Non-central rings}

Our survey design is not optimized for the identification of non-central rings: 
the CO emission is weaker at larger radii, with an increasing phase transition from molecular to atomic gas at larger radii. This is combined with the limited FoV of the images that might exclude outer rings and the generally small fraction of rings in star-forming galaxies (compared to e.g., green valley galaxies, see \citealt{kelvin_galaxy_2018})

We find that nearly all galaxies with non-central rings (5/6) are clearly barred (\textbf{C}) and rings roughly coincide with the stellar bar radius (4/5, bar radii from \citealt{diaz-garcia_stellar_2016}). 
One ring is located well inside the bar radii, potentially indicating the misclassification of a central ring in our classification. 
Thus, we translate our rings into 4 literature so-called \textit{inner} rings located roughly at the bar radius, one ambiguous ring, and one without a bar to help distinguish between \textit{inner} or \textit{outer} ring which should be at $\sim$ twice the bar radius.  
This high bar fraction among non-central rings is in agreement with  \citet[][]{comeron_arrakis_2014}, but in disagreement with \citet{herrera-endoqui_catalogue_2015} who find inner rings preferentially in weakly and non-barred galaxies from S$^4$G.
We conclude that there are several reasons why our sample might not be well suited to perform a proper analysis of outer rings.

\section{Summary}
\label{sec:Summary}
Morphology provides crucial insights into the formation, evolution and properties of a galaxy.  
Surveys classifying the presence or absence of stellar bars, spiral arms and rings have been conducted for decades, but were limited to observations in the optical or near-infrared. 
For the first time, we visually classify the morphologies of 79 nearby main sequence galaxies based solely on their molecular gas distribution as traced by \mbox{CO(2--1)} line emission and observed with a resolution of about 1$^{\prime\prime}$ (${\sim}100\,$pc) by PHANGS-ALMA \citep{leroy_phangsalma_2021}.

We investigate the correspondence between our CO morphologies based on the gas distribution and near-IR/optical morphologies based on stellar light as a way to identify the main factors responsible for the formation and decoupling of different features in the disk. 
Our main findings include: 

\begin{enumerate}[noitemsep,topsep=0pt,leftmargin=2\parindent]
    \item The generally good agreement to near-IR literature classifications demonstrates the suitability of CO emission as morphological tracer.

    \item The multi-arm appearance of galaxies appears to be very sensitive to the tracer used, whereas grand-design spiral and flocculent morphologies are consistent in both gas and stellar tracers. 
    
    \item Bar classification at low stellar masses is less robust as shown by the large discrepancy between CO and optical/NIR classifications. 
    
    \item In our sample barred galaxies tend to populate the higher stellar mass regime of our sample, as do grand-design spirals.

    \item About one third of our galaxies analysed ($31\pm4\%$ or $22/72$) host central rings and are preferentially barred (59\%), as the inner bar orbits - when occupied by gas - can create such features. Central rings in unbarred galaxies are likely pseudorings created by high pitch angles of inner spiral arms. 
    
    \item In addition to the visual classifications we develop a method to measure bar lane properties. 
    Short CO bar lanes tend to have highly curved shapes, are weaker and preferentially reside in gas-poorer galaxies, compared to longer CO bar lanes that tend to be straighter. The radial coverage of CO emission along bar lanes likely depends on the overall cold gas reservoir as we see a trend with gas mass fraction, and potentially also by the bar strength.

\end{enumerate}

\noindent We conclude that CO is a good tracer for morphological features and also provides a number of advantages over traditionally used optical/infrared tracers.
This is especially relevant at higher redshift where massive galaxies typically contain large amounts of obscuring dust. In such environments, morphological features could be better recovered by molecular gas tracers that are not affected by dust.
As interferometers such as ALMA achieve unprecedented resolution for CO observations, this might help identify morphologies even in distant galaxies.


\begin{acknowledgements}
SKS acknowledges financial support from the German Research Foundation (DFG) via Sino-German research grant SCHI 536/11-1. 
ES acknowledges funding from the European Research Council (ERC) under the European Union’s Horizon 2020 research and innovation programme (grant agreement No. 694343).
TGW acknowledges funding from the European Research Council (ERC) under the European Union’s Horizon 2020 research and innovation programme (grant agreement No. 694343).
MQ acknowledges support from the Spanish grant PID2019-106027GA-C44, funded by MCIN/AEI/10.13039/501100011033.
AB acknowledges funding from the European Research Council (ERC) under the European Union’s Horizon 2020 research and innovation programme (grant agreement No.726384/Empire).
RSK, MCS and SCOG acknowledge financial support from the European Research Council via the ERC Synergy Grant ``ECOGAL'' (project ID 855130), from the Deutsche Forschungsgemeinschaft (DFG) via the Collaborative Research Center ``The Milky Way System''  (SFB 881 -- funding ID 138713538 -- subprojects A1, B1, B2 and B8), from the Heidelberg Cluster of Excellence (EXC 2181 - 390900948) ``STRUCTURES'', funded by the German Excellence Strategy, and from the German Ministry for Economic Affairs and Climate Action in project ``MAINN'' (funding ID 50OO2206). RSK also thanks for computing resources provided by the Ministry of Science, Research and the Arts (MWK) of the State of Baden-W\"{u}rttemberg through bwHPC and DFG through grant INST 35/1134-1 FUGG and for data storage at SDS@hd through grant INST 35/1314-1 FUGG.
The work of AKL is partially supported by the National Science Foundation under Grants No. 1615105, 1615109, and 1653300.
MC gratefully acknowledges funding from the Deutsche Forschungsgemeinschaft (DFG) through an Emmy Noether Research Group (grant number CH2137/1-1). COOL Research DAO is a Decentralized Autonomous Organization supporting research in astrophysics aimed at uncovering our cosmic origins.
K.G. is supported by the Australian Research Council through the Discovery Early Career Researcher Award (DECRA) Fellowship DE220100766 funded by the Australian Government.
JMDK acknowledges funding from the European Research Council (ERC) under the European Union's Horizon 2020 research and innovation programme via the ERC Starting Grant MUSTANG (grant agreement number 714907). 
HAP acknowledges support by the National Science and Technology Council of Taiwan under grant 110-2112-M-032-020-MY3.
JPe acknowledges support by the French Agence Nationale de la Recherche
through the DAOISM grant ANR-21-CE31-0010 and by the Programme National
``Physique et Chimie du Milieu Interstellaire'' (PCMI) of CNRS/INSU with
INC/INP, co-funded by CEA and CNES.
TS acknowledges funding from the European Research Council (ERC) under the European Union’s Horizon 2020 research and innovation programme (grant agreement No. 694343).
AU acknowledges support from the Spanish grants PGC2018-094671-B-I00, funded by MCIN/AEI/10.13039/501100011033 and by ``ERDF A way of making Europe'', and PID2019-108765GB-I00, funded by MCIN/AEI/10.13039/501100011033. 
EJW acknowledges funding from the Deutsche 
Forschungsgemeinschaft (DFG, German Research Foundation) in the form of 
an Emmy Noether Research Group (grant number KR4598/2-1, PI Kreckel).
FB acknowledges funding from the European Research Council (ERC) under the European Union’s Horizon 2020 research and innovation programme (grant agreement No.726384/Empire).

This paper makes use of the following ALMA data: \linebreak
ADS/JAO.ALMA\#2012.1.00650.S, \linebreak 
ADS/JAO.ALMA\#2013.1.00803.S, \linebreak 
ADS/JAO.ALMA\#2013.1.01161.S, \linebreak 
ADS/JAO.ALMA\#2015.1.00121.S, \linebreak 
ADS/JAO.ALMA\#2015.1.00782.S, \linebreak 
ADS/JAO.ALMA\#2015.1.00925.S, \linebreak 
ADS/JAO.ALMA\#2015.1.00956.S, \linebreak 
ADS/JAO.ALMA\#2016.1.00386.S, \linebreak 
ADS/JAO.ALMA\#2017.1.00392.S, \linebreak 
ADS/JAO.ALMA\#2017.1.00766.S, \linebreak 
ADS/JAO.ALMA\#2017.1.00886.L, \linebreak 
ADS/JAO.ALMA\#2018.1.01321.S, \linebreak 
ADS/JAO.ALMA\#2018.1.01651.S. \linebreak 
ADS/JAO.ALMA\#2018.A.00062.S. \linebreak 
ALMA is a partnership of ESO (representing its member states), NSF (USA) and NINS (Japan), together with NRC (Canada), MOST and ASIAA (Taiwan), and KASI (Republic of Korea), in cooperation with the Republic of Chile. The Joint ALMA Observatory is operated by ESO, AUI/NRAO and NAOJ.

\textit{Software:}
Astropy \citep{robitaille_astropy_2013, astropy_collaboration_astropy_2018, astropy_collaboration_astropy_2022}, NumPy \citep{harris_array_2020}, SciPy \citep{virtanen_scipy_2020}, Scikit-image \citep{van_der_walt_scikit-image_2014}.

\end{acknowledgements}

\bibliographystyle{aa} 
\bibliography{Bibliography.bib}

\begin{appendix}

\section{Additional morphological classification examples}
\label{Appendix:Morphexamples}
Here we provide additional examples that capture all different morphological features present among our sample in Figure~\ref{fig:Morphexamples}: \\
IC~1954, an unbarred (\textbf{A}), multi-arm (\textbf{M}) galaxy without central (\textbf{nr}) and without non-central ring (\textbf{nR}). \\
NGC~1512, a barred (\textbf{C}) galaxy without verified arm class, with short and curved bar lanes (\textbf{s}, \textbf{c4.0}) and both a central ring (\textbf{r}) and non-central ring (\textbf{R)}.\\
NGC~3059, a weakly barred (\textbf{B}), flocculent (\textbf{F}) galaxy with straight bar lanes (\textbf{c1.0}) and without central (\textbf{nr}) and without non-central ring (\textbf{nR}). 
No high quality bar shape class is available.\\
NGC~5128,  an example of a galaxy unsuited for classification. For this galaxy visibility is impaired and it was not used for further classification.
\begin{figure}[h]
    \centering
    \includegraphics[width=0.49\textwidth]{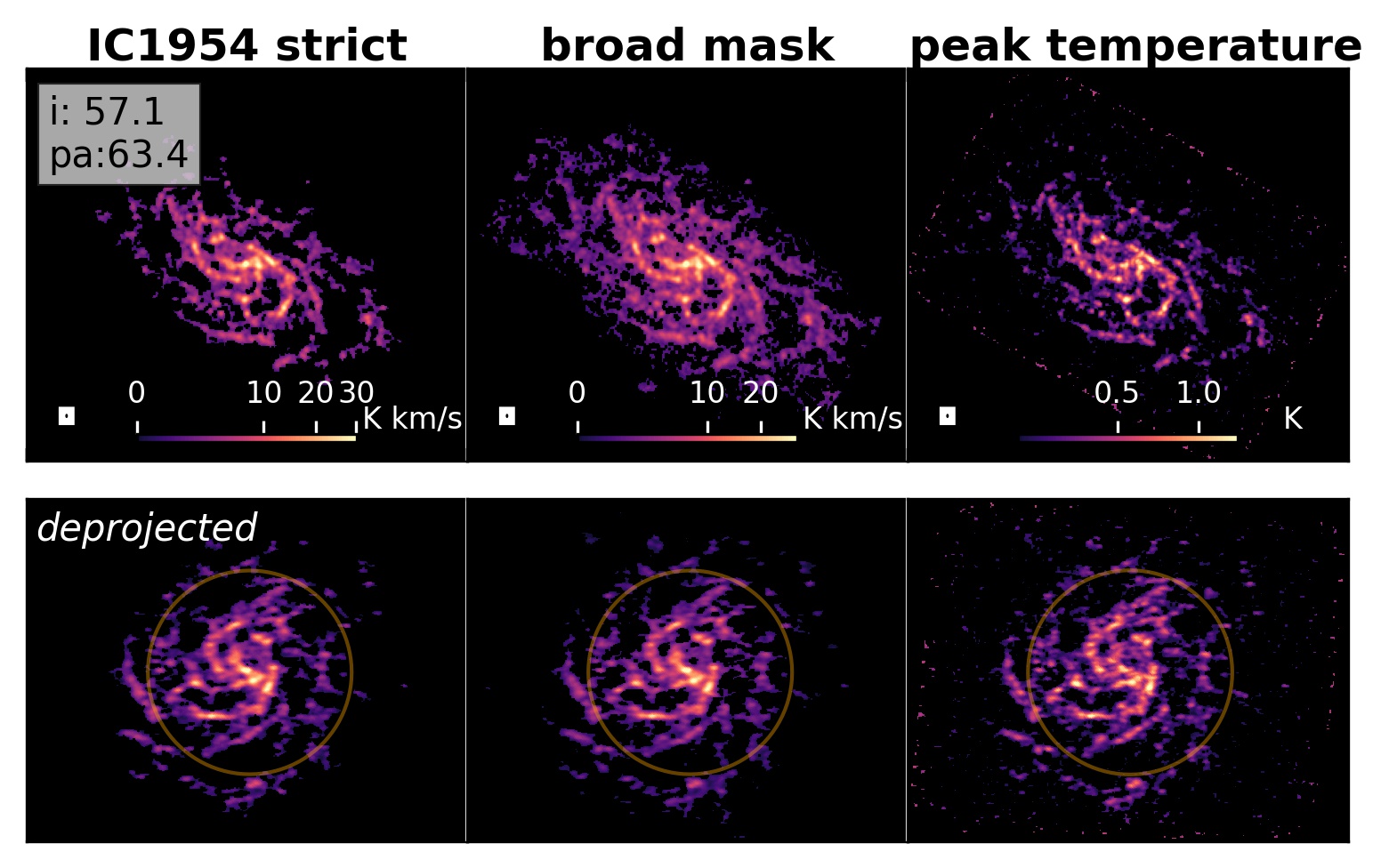}
    \includegraphics[width=0.49\textwidth]{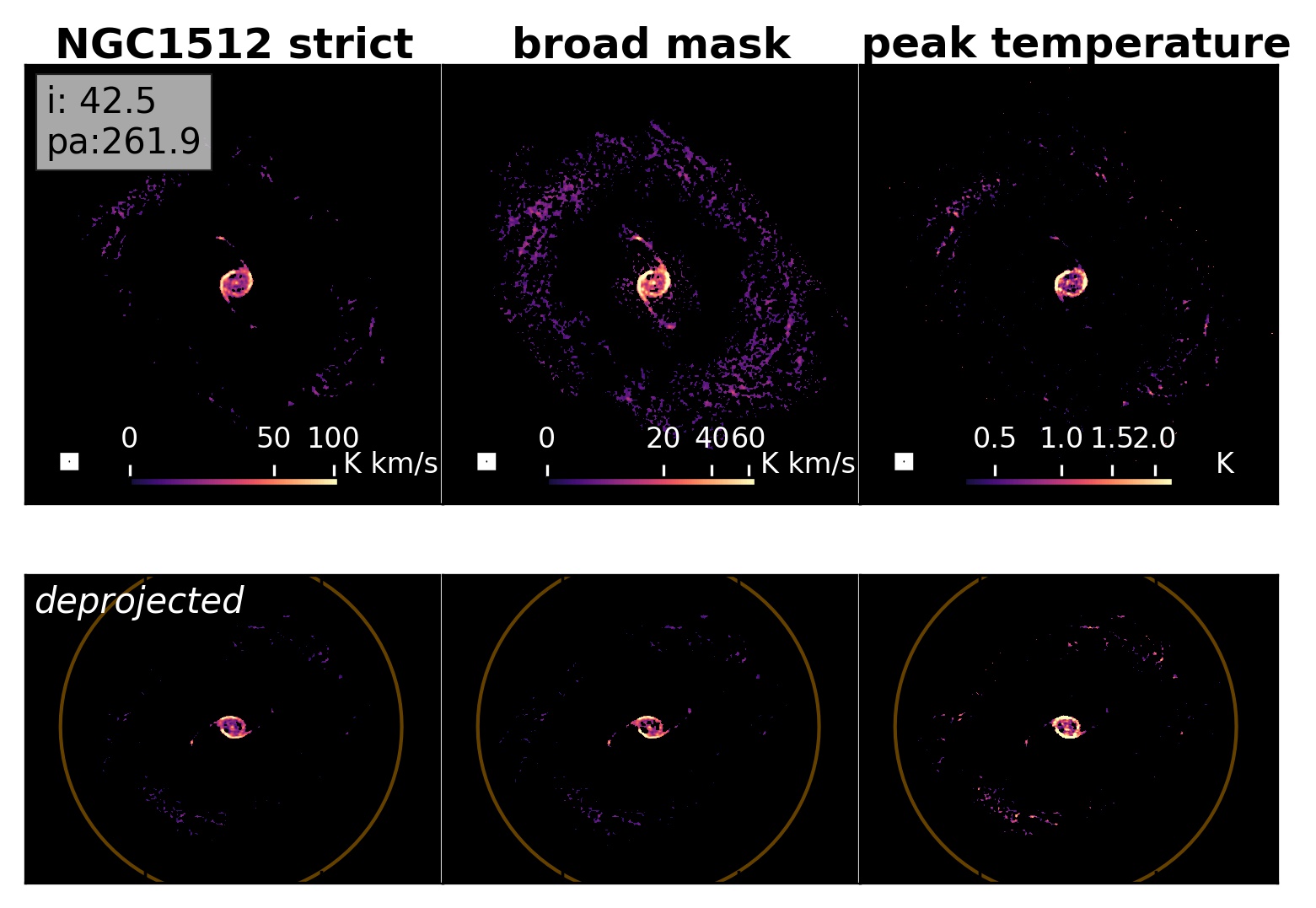}
    \caption{Same as Figure~\ref{fig:exampleCOmom0} but for selected PHANGS galaxies with various morphologies (see text).}
    \label{fig:Morphexamples}
\end{figure}

\begin{figure}
    \centering
    \ContinuedFloat
    \includegraphics[width=0.49\textwidth]{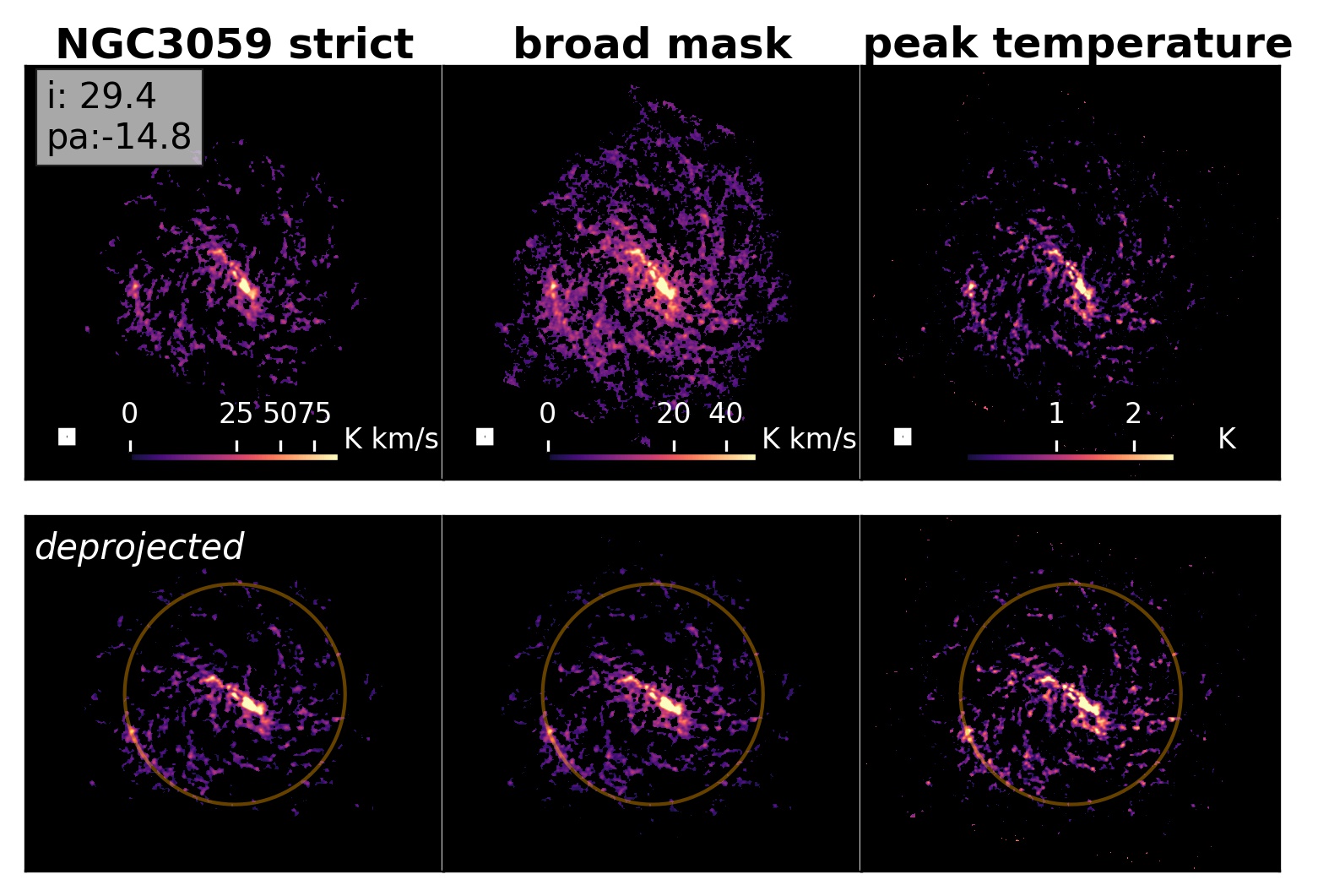}
    \includegraphics[width=0.49\textwidth]{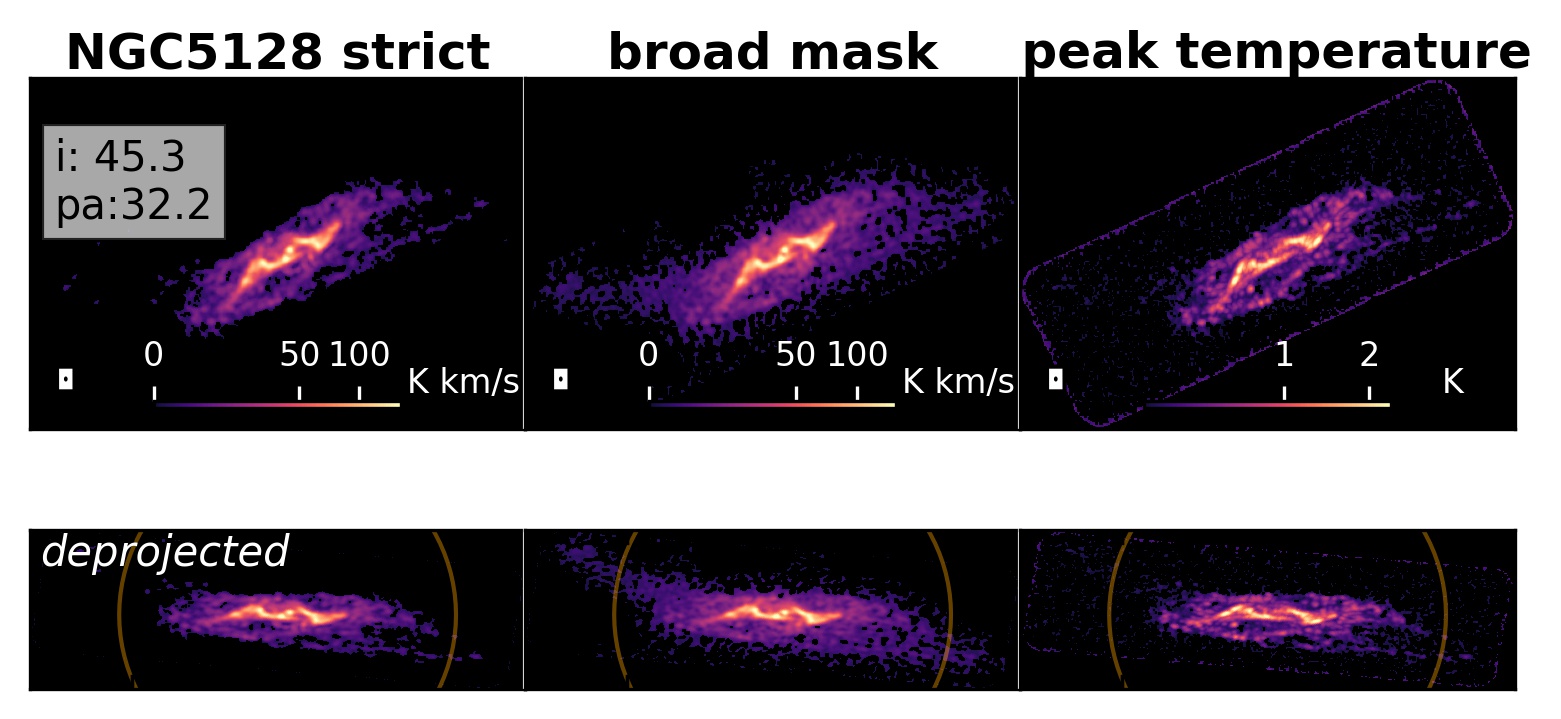}
    \caption{Continued}
\end{figure}

\FloatBarrier

\section{PHANGS galaxy properties}
\label{Appendix:PHANGSProperties}

We provide an overview of the properties of all 79 PHANGS-ALMA galaxies in Table~\ref{tab:AppendixPHANGSProperties}. 
A detailed description of galaxy orientation can be found in \citet[][]{leroy_z_2019,leroy_phangsalma_2021}.

\onecolumn
\begin{landscape}
\footnotesize 
\begin{longtable}{ l r r r r r r r r r r r r c c}
\caption{79 PHANGS galaxies and their properties}\label{tab:AppendixPHANGSProperties}\\
\hline\hline 
\noalign{\smallskip}
Galaxy & RA & DEC & PA & incl & d & log$_{10}\mathrm{M}_\ast$ & log$_{10} \mathrm{SFR}$  & $\Delta$MS & log$_{10}$M$_\mathrm{HI}$ & log$_{10}$M$_\mathrm{H_2}$ &  T & R$_{25}$ & C$_\mathrm{bar\,lit}$ & C$_\mathrm{arm\,lit}$\\
\noalign{\smallskip}
\hline
\noalign{\smallskip}
 & [$^\circ$] & [$^\circ$] & [$^\circ$]  & [$^\circ$] & [Mpc] & [log$_{10} M_\odot$] & [log$_{10} M_\odot\,\mathrm{yr}^{-1}$] & & [log$_{10} M_\odot$] & [log$_{10} M_\odot$] &  & [$^{\prime\prime}$] &  \\
\noalign{\smallskip}
(1) & (2) & (3) & (4) & (5) & (6) & (7) & (8) & (9) & (10) & (11) & (12) & (13) & (14) & (15) \\
\noalign{\smallskip}
\hline 
\noalign{\smallskip}
Circinus   & 213.29124 & -65.3391 & 36.7 & 64.3 & 4.20 & 10.53 & 0.61 & 0.42 & 9.81 & ... & 3.3 & 262.5 & ...    & ...   \\
IC~1954    & 52.87971 & -51.9049 & 63.4 & 57.1 & 12.80 & 9.67 & -0.44 & -0.04 & 8.85 & 8.73 & 3.3 & 89.8 & \SB    & \textit{M}    \\
IC~5273    & 344.86118 & -37.7028 & 234.1 & 52.0 & 14.18 & 9.72 & -0.27 & 0.09 & 8.95 & 8.65 & 5.6 & 91.9 & \SB    & \textit{M}    \\
NGC~0253   & 11.88797 & -25.2884 & 52.5 & 75.0 & 3.70 & 10.64 & 0.70 & 0.44 & 9.33 & 9.69 & 5.1 & 803.8 & ...    &\textit{G}    \\
NGC~0300   & 13.72302 & -37.6845 & 114.3 & 39.8 & 2.09 & 9.26 & -0.82 & -0.15 & 9.32 & 7.77 & 6.9 & 582.3 & ...    & \textit{M}    \\
NGC~0628   & 24.17385 & 15.7836 & 20.7 & 8.9 & 9.84 & 10.34 & 0.24 & 0.18 & 9.70 & 9.43 & 5.2 & 296.6 & \SA    & \textit{M}    \\
NGC~0685   & 26.92845 & -52.7620 & 100.9 & 23.0 & 19.94 & 10.06 & -0.38 & -0.25 & 9.57 & 8.77 & 5.4 & 90.4 & \SB    &\textit{F}    \\
NGC~1068   & 40.66973 & -0.0133 & 72.7 & 34.7 & 13.97 & 10.91 & 1.64 & 1.19 & 9.06 & ... & 3.0 & 183.3 & ...    &\textit{G}    \\
NGC~1087   & 41.60492 & -0.4987 & 359.1 & 42.9 & 15.85 & 9.93 & 0.12 & 0.33 & 9.10 & 9.20 & 5.2 & 89.1 & \SB    &\textit{F}    \\
NGC~1097   & 41.57896 & -30.2747 & 122.4 & 48.6 & 13.58 & 10.76 & 0.68 & 0.33 & 9.61 & 9.74 & 3.3 & 317.0 & \SB    &\textit{G}    \\
NGC~1300   & 49.92081 & -19.4111 & 278.0 & 31.8 & 18.99 & 10.62 & 0.07 & -0.18 & 9.38 & 9.40 & 4.0 & 178.3 & \SB    &\textit{G}    \\
NGC~1317   & 50.68454 & -37.1038 & 221.5 & 23.2 & 19.11 & 10.62 & -0.32 & -0.57 & ... & 8.91 & 0.8 & 92.1 & \SB    & ...   \\
NGC~1365   & 53.40152 & -36.1404 & 201.1 & 55.4 & 19.57 & 10.99 & 1.23 & 0.72 & 9.94 & 10.26 & 3.2 & 360.7 & \SB    &\textit{G}    \\
NGC~1385   & 54.36901 & -24.5012 & 181.3 & 44.0 & 17.22 & 9.98 & 0.32 & 0.50 & 9.19 & 9.23 & 5.9 & 102.1 & \SB    &\textit{F}    \\
NGC~1433   & 55.50620 & -47.2219 & 199.7 & 28.6 & 18.63 & 10.87 & 0.05 & -0.36 & 9.40 & 9.29 & 1.5 & 185.8 & \SB    &\textit{G}    \\
NGC~1511   & 59.90246 & -67.6339 & 297.0 & 72.7 & 15.28 & 9.91 & 0.36 & 0.59 & 9.57 & 9.17 & 2.0 & 110.9 & \SB    & ...   \\
NGC~1512   & 60.97557 & -43.3487 & 261.9 & 42.5 & 18.83 & 10.71 & 0.11 & -0.21 & 9.88 & 9.12 & 1.2 & 253.0 & \SB    &\textit{G}    \\
NGC~1546   & 63.65122 & -56.0609 & 147.8 & 70.3 & 17.69 & 10.35 & -0.08 & -0.15 & 8.68 & 9.29 & -0.4 & 111.2 & \SA    &\textit{G}    \\
NGC~1559   & 64.40238 & -62.7834 & 244.5 & 65.4 & 19.44 & 10.36 & 0.58 & 0.50 & 9.52 & 9.59 & 5.9 & 125.6 & \SB    &\textit{F}    \\
NGC~1566   & 65.00159 & -54.9380 & 214.7 & 29.5 & 17.69 & 10.78 & 0.66 & 0.29 & 9.80 & 9.70 & 4.0 & 216.8 & \SAB   &\textit{G}    \\
NGC~1637   & 70.36741 & -2.8580 & 20.6 & 31.1 & 11.70 & 9.95 & -0.19 & 0.01 & 9.20 & 8.83 & 5.0 & 95.3 & \SAB   & \textit{M}    \\
NGC~1672   & 71.42704 & -59.2473 & 134.3 & 42.6 & 19.40 & 10.73 & 0.88 & 0.56 & 10.21 & 9.86 & 3.3 & 184.6 & \SABB  &\textit{G}    \\
NGC~1792   & 76.30969 & -37.9806 & 318.9 & 65.1 & 16.20 & 10.61 & 0.57 & 0.32 & 9.25 & 9.82 & 4.0 & 166.8 & \SA    & \textit{M}    \\
NGC~1809   & 75.52066 & -69.5679 & 138.2 & 57.6 & 19.95 & 9.77 & 0.76 & 1.08 & 9.60 & 8.97 & 5.0 & 112.2 & \SA    & ...   \\
NGC~2090   & 86.75787 & -34.2506 & 192.5 & 64.5 & 11.75 & 10.04 & -0.39 & -0.25 & 9.37 & 8.66 & 4.5 & 134.6 & \SA    & ...   \\
NGC~2283   & 101.46997 & -18.2108 & -4.1 & 43.7 & 13.68 & 9.89 & -0.28 & -0.04 & 9.70 & 8.61 & 5.9 & 82.8 & \SB    & ...   \\
NGC~2566   & 124.69003 & -25.4995 & 312.0 & 48.5 & 23.44 & 10.71 & 0.94 & 0.63 & 9.37 & 9.86 & 2.7 & 127.7 & \SB    & ...   \\
NGC~2903   & 143.04211 & 21.5008 & 203.7 & 66.8 & 10.00 & 10.63 & 0.49 & 0.23 & 9.54 & 9.57 & 4.0 & 358.2 & \SB    & \textit{M}    \\
NGC~2997   & 146.41164 & -31.1911 & 108.1 & 33.0 & 14.06 & 10.73 & 0.64 & 0.31 & 9.86 & 9.83 & 5.1 & 307.7 & \SAB   & ...   \\
NGC~3059   & 147.53400 & -73.9222 & -14.8 & 29.4 & 20.23 & 10.38 & 0.38 & 0.29 & 9.75 & 9.39 & 4.0 & 113.8 & \SB    & ...   \\
NGC~3137   & 152.28116 & -29.0643 & -0.3 & 70.3 & 16.37 & 9.88 & -0.31 & -0.06 & 9.68 & 8.56 & 5.9 & 166.4 & \SA    & ...   \\
NGC~3351   & 160.99065 & 11.7037 & 193.2 & 45.1 & 9.96 & 10.36 & 0.12 & 0.05 & 8.93 & 9.09 & 3.1 & 216.8 & \SB    &\textit{G}    \\
NGC~3489   & 165.07736 & 13.9012 & 70.0 & 63.7 & 11.86 & 10.28 & -1.63 & -1.65 & 7.40 & 7.68 & -1.2 & 103.3 & ...    & ...   \\
NGC~3507   & 165.85573 & 18.1355 & 55.8 & 21.7 & 23.55 & 10.40 & -0.00 & -0.10 & 9.32 & 9.25 & 3.1 & 87.5 & \SABB  &\textit{G}    \\
NGC~3511   & 165.84921 & -23.0867 & 256.8 & 75.1 & 13.94 & 10.03 & -0.09 & 0.06 & 9.37 & 9.04 & 5.1 & 181.2 & \SAB   & \textit{M}    \\
NGC~3521   & 166.45239 & -0.0359 & 343.0 & 68.8 & 13.24 & 11.02 & 0.57 & 0.05 & 9.83 & 9.77 & 4.0 & 249.5 & \SA    & \textit{M}    \\
NGC~3596   & 168.77580 & 14.7871 & 78.4 & 25.1 & 11.30 & 9.66 & -0.52 & -0.12 & 8.85 & 8.70 & 5.2 & 109.2 & \SA    & \textit{M}    \\
NGC~3599   & 168.86229 & 18.1104 & 41.9 & 23.0 & 19.86 & 10.04 & -1.33 & -1.18 & ... & 7.20 & -2.0 & 72.0 & ...    & ...   \\
NGC~3621   & 169.56792 & -32.8126 & 343.8 & 65.8 & 7.06 & 10.06 & -0.00 & 0.13 & 9.66 & 9.06 & 6.9 & 286.5 & \SA    & ...   \\
NGC~3626   & 170.01588 & 18.3568 & 165.2 & 46.6 & 20.05 & 10.46 & -0.67 & -0.82 & 8.89 & 8.55 & -0.8 & 88.3 & \SAB   &\textit{G}    \\
NGC~3627   & 170.06252 & 12.9915 & 173.1 & 57.3 & 11.32 & 10.83 & 0.58 & 0.19 & 9.09 & 9.78 & 3.1 & 308.4 & \SB    &\textit{G}    \\
NGC~4207   & 183.87682 & 9.5849 & 121.9 & 64.5 & 15.78 & 9.71 & -0.72 & -0.35 & 8.58 & 8.71 & 7.7 & 45.1 & \SB    & ...   \\
NGC~4254   & 184.70680 & 14.4164 & 68.1 & 34.4 & 13.10 & 10.42 & 0.49 & 0.37 & 9.48 & 9.85 & 5.2 & 151.1 & \SA    & \textit{M}    \\
NGC~4293   & 185.30347 & 18.3826 & 48.3 & 65.0 & 15.76 & 10.51 & -0.29 & -0.46 & 7.67 & 8.99 & 0.3 & 187.1 & \SB    & ...   \\
NGC~4298   & 185.38651 & 14.6061 & 313.9 & 59.2 & 14.92 & 10.02 & -0.34 & -0.18 & 8.87 & 9.19 & 5.1 & 76.1 & \SA    &\textit{F}    \\
NGC~4303   & 185.47888 & 4.4737 & 312.4 & 23.5 & 16.99 & 10.52 & 0.73 & 0.54 & 9.67 & 9.91 & 4.0 & 206.6 & \SAB   & \textit{M}    \\
NGC~4321   & 185.72887 & 15.8223 & 156.2 & 38.5 & 15.21 & 10.75 & 0.55 & 0.21 & 9.43 & 9.89 & 4.0 & 182.9 & \SAB   &\textit{G}    \\
NGC~4424   & 186.79820 & 9.4206 & 88.3 & 58.2 & 16.20 & 9.91 & -0.52 & -0.29 & 8.30 & 8.43 & 1.3 & 91.2 & \SB    & ...   \\
NGC~4457   & 187.24593 & 3.5706 & 78.7 & 17.4 & 15.10 & 10.42 & -0.51 & -0.63 & 8.36 & 8.99 & 0.3 & 83.8 & \SAB   & ...   \\
NGC~4459   & 187.25018 & 13.9785 & 108.8 & 47.0 & 15.85 & 10.68 & -0.65 & -0.94 & ... & 8.42 & -1.6 & 125.1 & ...    & ...   \\
NGC~4476   & 187.49622 & 12.3486 & 27.4 & 60.1 & 17.54 & 9.81 & -1.39 & -1.10 & ... & 7.85 & -2.9 & 50.9 & ...    & ...   \\
NGC~4477   & 187.50917 & 13.6364 & 25.7 & 33.5 & 15.76 & 10.59 & -1.10 & -1.33 & ... & 7.58 & -1.7 & 110.9 & ...    & ...   \\
NGC~4535   & 188.58459 & 8.1980 & 179.7 & 44.7 & 15.77 & 10.53 & 0.33 & 0.14 & 9.56 & 9.60 & 5.0 & 244.4 & \SAB   & \textit{M}    \\
NGC~4536   & 188.61278 & 2.1882 & 305.6 & 66.0 & 16.25 & 10.40 & 0.54 & 0.44 & 9.54 & 9.42 & 4.3 & 212.4 & \SAB   & \textit{M}    \\
NGC~4540   & 188.71193 & 15.5517 & 12.8 & 28.7 & 15.76 & 9.79 & -0.78 & -0.46 & 8.44 & 8.62 & 6.2 & 65.8 & \SAB   &\textit{F}    \\
NGC~4548   & 188.86024 & 14.4963 & 138.0 & 38.3 & 16.22 & 10.69 & -0.28 & -0.58 & 8.84 & 9.25 & 3.1 & 166.4 & \SB    &\textit{G}    \\
NGC~4569   & 189.20760 & 13.1629 & 18.0 & 70.0 & 15.76 & 10.81 & 0.12 & -0.26 & 8.84 & 9.66 & 2.4 & 273.6 & \SABB  &\textit{G}    \\
NGC~4579   & 189.43138 & 11.8182 & 91.3 & 40.2 & 21.00 & 11.15 & 0.34 & -0.27 & 9.02 & 9.65 & 2.8 & 150.4 & \SB    &\textit{G}    \\
NGC~4596   & 189.98308 & 10.1762 & 120.0 & 36.6 & 15.76 & 10.59 & -0.96 & -1.19 & ... & 7.60 & -0.8 & 117.8 & ...    & ...   \\
NGC~4654   & 190.98575 & 13.1267 & 123.2 & 55.6 & 21.98 & 10.57 & 0.58 & 0.36 & 9.75 & 9.68 & 5.9 & 141.6 & \SB    & \textit{M}    \\
NGC~4689   & 191.93990 & 13.7627 & 164.1 & 38.7 & 15.00 & 10.22 & -0.39 & -0.37 & 8.54 & 9.05 & 4.7 & 114.6 & \SA    & \textit{M}    \\
NGC~4731   & 192.75504 & -6.3928 & 255.4 & 64.0 & 13.28 & 9.48 & -0.22 & 0.30 & 9.44 & 8.63 & 5.9 & 189.7 & \SB    &\textit{G}    \\
NGC~4781   & 193.59917 & -10.5371 & 290.0 & 59.0 & 11.31 & 9.64 & -0.32 & 0.09 & 8.94 & 8.81 & 7.0 & 111.2 & \SB    &\textit{F}    \\
NGC~4826   & 194.18184 & 21.6831 & 293.6 & 59.1 & 4.41 & 10.24 & -0.69 & -0.68 & 8.26 & 8.61 & 2.2 & 315.6 & \SA    & \textit{M}    \\
NGC~4941   & 196.05461 & -5.5515 & 202.2 & 53.4 & 15.00 & 10.17 & -0.35 & -0.30 & 8.49 & 8.71 & 2.1 & 100.7 & \SA    & ...   \\
NGC~4945   & 196.36377 & -49.4679 & 43.8 & 90.0 & 3.47 & 10.36 & 0.19 & 0.11 & 8.92 & ... & 6.1 & 700.0 & ...    & ...   \\
NGC~4951   & 196.28214 & -6.4938 & 91.2 & 70.2 & 15.00 & 9.79 & -0.45 & -0.14 & 9.21 & 8.57 & 6.0 & 94.2 & \SA    & \textit{M}    \\
NGC~5068   & 199.72807 & -21.0387 & 342.4 & 35.7 & 5.20 & 9.40 & -0.56 & 0.02 & 8.82 & 8.42 & 6.0 & 224.5 & \SB    &\textit{F}    \\
NGC~5128   & 201.36507 & -43.0191 & 32.2 & 45.3 & 3.69 & 10.97 & 0.09 & -0.40 & 8.43 & ... & -2.1 & 767.6 & ...    & ...   \\
NGC~5134   & 201.32726 & -21.1342 & 311.6 & 22.7 & 19.92 & 10.41 & -0.34 & -0.45 & 8.92 & 8.83 & 2.9 & 81.3 & \SAB   & ...   \\
NGC~5236   & 204.25391 & -29.8656 & 225.0 & 24.0 & 4.89 & 10.53 & 0.63 & 0.44 & 9.98 & ... & 5.0 & 408.4 & ...    & \textit{M}    \\
NGC~5248   & 204.38336 & 8.8852 & 109.2 & 47.4 & 14.87 & 10.41 & 0.36 & 0.25 & 9.50 & 9.66 & 4.0 & 122.2 & \SAB   &\textit{G}    \\
NGC~5530   & 214.61380 & -43.3883 & 305.4 & 61.9 & 12.27 & 10.08 & -0.48 & -0.37 & 9.11 & 8.87 & 4.2 & 144.9 & \SA    & ...   \\
NGC~5643   & 218.16991 & -44.1746 & 318.7 & 29.9 & 12.68 & 10.34 & 0.41 & 0.36 & 9.12 & 9.42 & 5.0 & 157.4 & \SAB   & ...   \\
NGC~6300   & 259.24780 & -62.8205 & 105.4 & 49.6 & 11.58 & 10.47 & 0.28 & 0.13 & 9.13 & 9.28 & 3.1 & 160.0 & \SB    & ...   \\
NGC~7456   & 345.54306 & -39.5694 & 16.0 & 67.3 & 15.70 & 9.64 & -0.43 & -0.02 & 9.28 & ... & 6.0 & 123.3 & \SAB   &\textit{F}    \\
NGC~7496   & 347.44702 & -43.4278 & 193.7 & 35.9 & 18.72 & 10.00 & 0.35 & 0.53 & 9.07 & 9.26 & 3.2 & 100.5 & \SB    &\textit{G}    \\
NGC~7743   & 356.08804 & 9.9340 & 86.2 & 37.1 & 20.32 & 10.36 & -0.67 & -0.74 & 8.50 & 8.56 & -0.9 & 78.4 & ...    &\textit{G}    \\
NGC~7793   & 359.45758 & -32.5911 & 290.0 & 50.0 & 3.62 & 9.36 & -0.58 & 0.03 & 8.70 & ... & 7.4 & 310.5 & ...    &\textit{F}    \\
\end{longtable}
\tablefoot{
List of 79 PHANGS-ALMA galaxies (1) used in this work. 
We provide galaxy right ascension (2), declination (3) and position angle (4) as described in \citet{lang_phangs_2020}.
Distances (4) have 0.02\,dex uncertainties for values derived by TRGB method, 0.06\,dex for other quality distances and 0.125\,dex, other distances are described in \citet[][]{anand_extragalactic_2021}.
We list stellar masses (7), global SFRs (8) and offset from the main sequence according to \citet[][]{leroy_z_2019}(9).
The stellar masses have $\sim 0.1\,$dex uncertainties and SFRs $\sim 0.15\,$dex.
Further, we provide atomic gas mass M$_\mathrm{HI}$ (10) and molecular gas mass M$_{\mathrm{H}_2}$ (11) \citep{leroy_z_2019}. 
Numerical Hubble type T (12), R$_{25}$ for galaxy size (13) are given \citep{leroy_phangsalma_2021-1}, as well as literature classifications for stellar bar features (C$_\mathrm{bar\,lit}$, 14) and spiral arm type (C$_\mathrm{arm\,lit}$, 15). Literature classes are from \citet{buta_classical_2015} (infrared) if available,  from \citet[][]{de_vaucouleurs_third_1991} (RC3, optical). For some galaxies no secure literature classes are available (...). 
}
\end{landscape} 

\twocolumn

\section{Notes on individual galaxy classes}
\label{sec:Appendix:MorphologySpecialcases}

Although generally co-aligned, the gas response to the underlying gravitational potential might be able to create features different from what is seen in stellar light. 
Galaxies with the largest deviations between literature and our gas-based classifications, might help better understand what physical properties favour such differences in the features observed.

\subsection{\textit{G}-\textbf{F} galaxies}
\label{sec:Appendix:MorphologySpecialcasesG-F}

We investigate galaxies classified as flocculent (\textbf{F}) in our data, but as grand-design (\textit{G}) according to the literature data (see Section~\ref{sec:MorphologyResults:ArmsLiterature}) in more detail to understand the origin of these class deviations. 
These \textit{G}-\textbf{F} galaxies have classification agreements of 60\%, 70\%, 80\% and 100\% for 
NGC~3351, NGC~7496, NGC~3626 and NGC~4569 respectively. 
We show their CO and 3.6$\mu$m distribution in Figure~\ref{fig:SelectionFGGalaxies} and list our findings for each galaxy below: 

\begin{figure*}
    \centering
    \includegraphics[width = 0.9\textwidth]{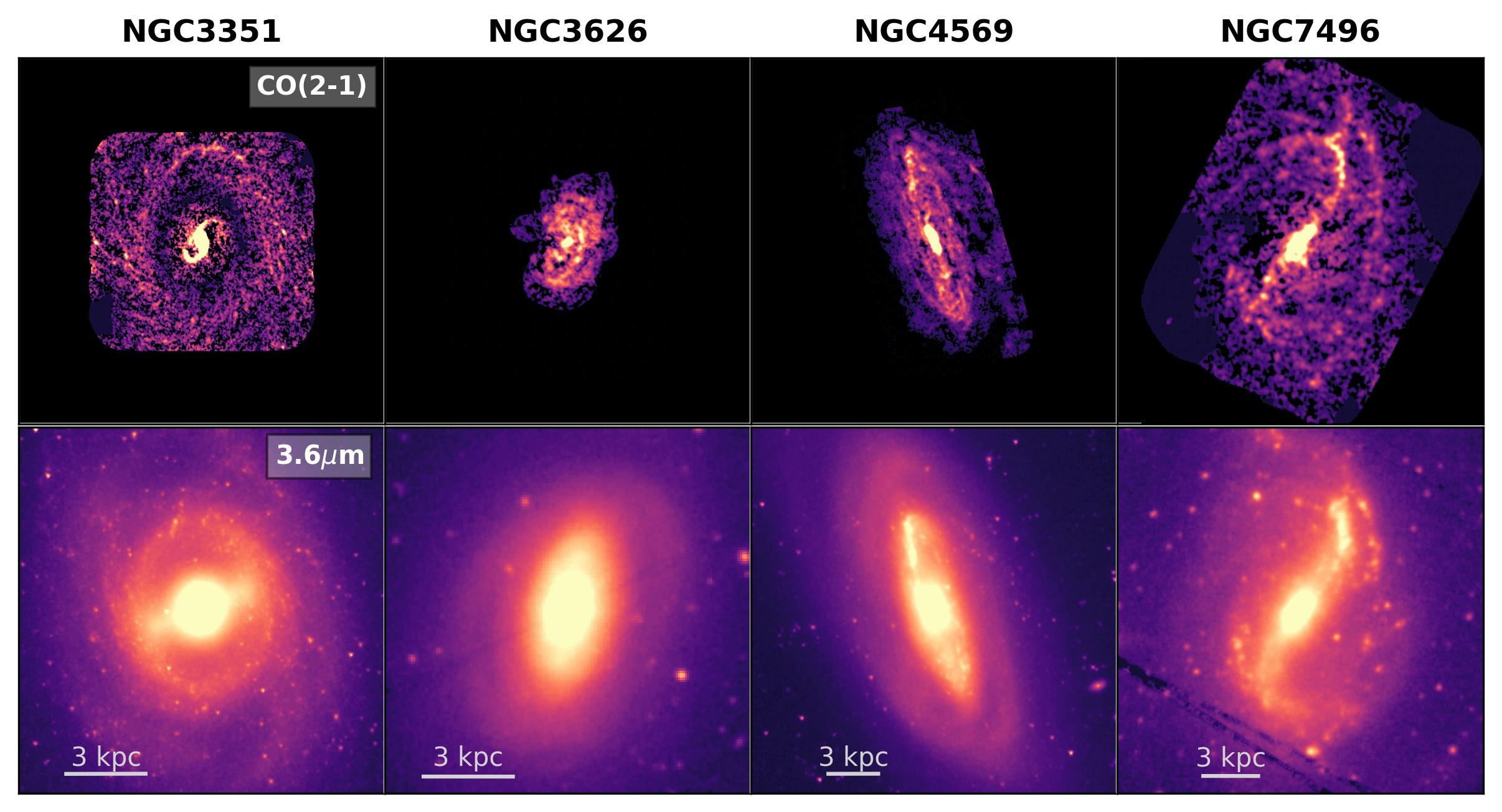}
    \caption{\textit{G}-\textbf{F} 
    galaxies classified as flocculent based on the molecular gas mom-0 maps (\textbf{F}, top panels), but as having grand-design in the \textit{Spitzer}3.6$\mu$m images (\textit{G}, bottom panels). The galaxy centers are saturated to emphasize fainter structures in the outer disk.}
    \label{fig:SelectionFGGalaxies}
\end{figure*}

\begin{itemize}[noitemsep,topsep=0pt,leftmargin=2\parindent]
\item \textbf{NGC~3351} hosts one of the few outer rings (\textbf{R}) that we were able to identify in our data, and upon visual confirmation with optical images, the spiral arms seem to be located at larger radii further out in the disk. 
Our CO data, however, does not probe this very outer regions of the disk and does in fact covers the bar and ends at the outer ring structure. 

\item \textbf{NGC~3626} is classified as S0/a galaxy in \citet{buta_classical_2015}, which corresponds to a class between a classical S0 galaxy without spiral arms and a Sa galaxy with arms. Upon visual confirmation, the spiral arms appear subtle even in optical images, and relatively tightly wound.

\item \textbf{NGC~4569} has been classified as a prototype for an anemic spiral \citep[][]{van_den_bergh_new_1976} or as passive spiral \citep[][]{moran_wide-field_2007}, thus, the arm-interarm contrast is very low, the spiral arms are smooth and mostly without any star-formation and the overall gas content is low as well and confined to the center of the disk.  
The deviation between our classification and the one from the literature likely captures an important physical difference in the overall structure. 

\item \textbf{NGC~7496} has similarly to NGC~3351 spiral arms in the outer regions of the disk that is not probed by our CO data, but results are ambiguous. 
    
\end{itemize}

\subsection{\SB+\textbf{A}}
\label{sec:Appendix:MorphologySpecialcasesSB-A}

\begin{figure*}
    \centering
    \includegraphics[width = 0.99 \textwidth]{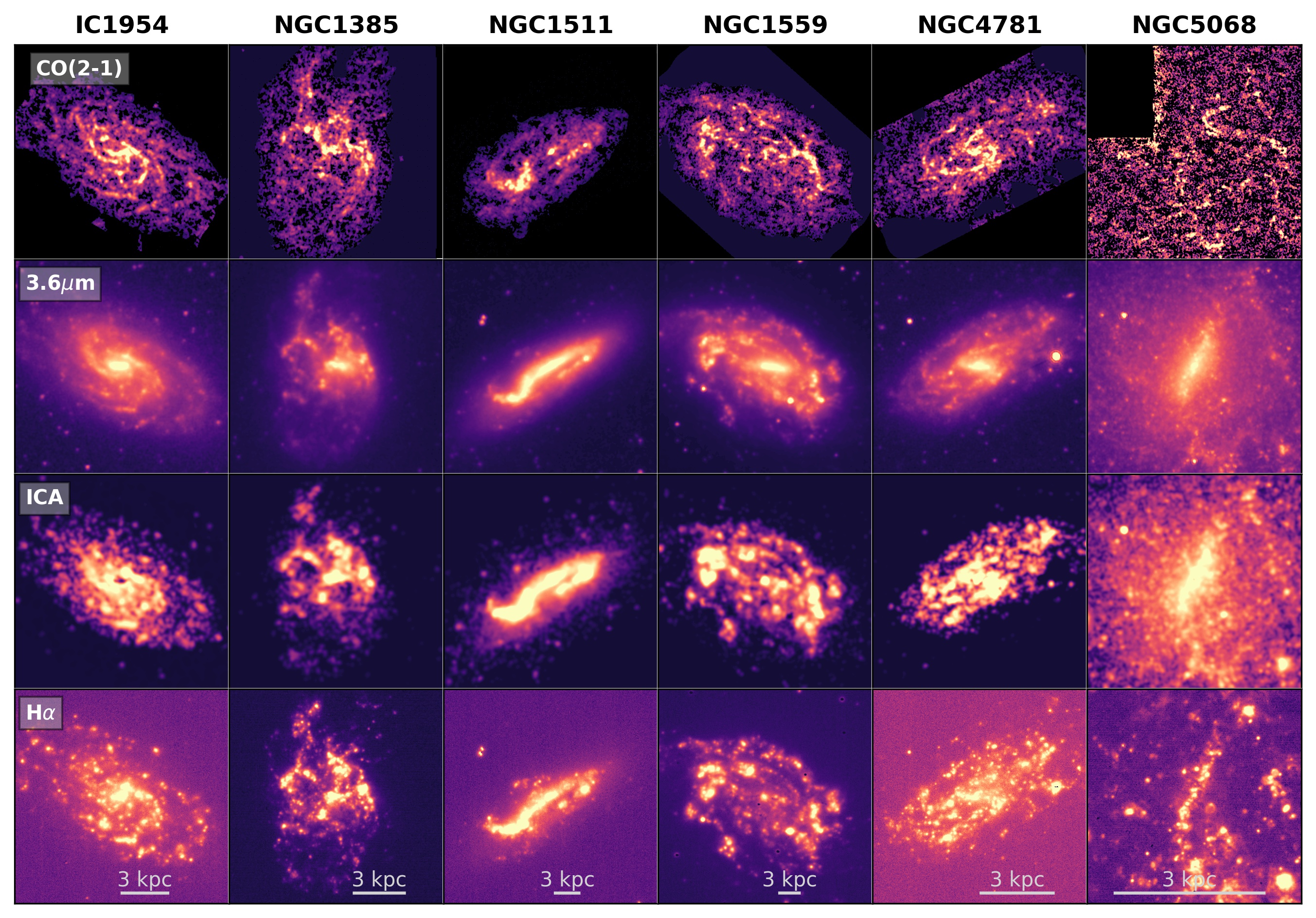}
    \caption{Image of \SB+\textbf{A} galaxies at different wavelengths. From top to bottom: broad CO mom-0 map, \textit{Spitzer} $3.6\,\mu$m image, ICA s2 image, H$\alpha$ image.}
    \label{fig:SBAmultiICA}
\end{figure*}

We find six galaxies that are classified as strongly barred in the literature (\SB) but show no signs of a bar in CO (\textbf{A}). 
We provide their properties in Table~\ref{tab:SB+AgalaxiesApp}, as well as show multi-wavelength images in Figure~\ref{fig:SBAmultiICA} and list literature studies on the individual galaxies below. 

\begin{table*}
\caption{Properties of individual \SB+\textbf{A} galaxies}\label{tab:SB+AgalaxiesApp}
\centering
\begin{tabular}{lllllll}
\hline\hline
\noalign{\smallskip}
Galaxy & log$_{10}$ M$_\ast$ & log$_{10}$ SFR & AG$_\mathrm{Bar}$ & Arm & AG$_\mathrm{Arm}$ & $T_\mathrm{Hubble}$ \\
 \noalign{\smallskip}
 & [M$_\odot$] & [M$_\odot$ yr$^{-1}$] &  [\%] & & [\%] \\
\noalign{\smallskip}
\hline
\noalign{\smallskip}
IC~1954      & 9.67 & -0.438 & 70  & M & 80 & $3.3\pm 0.9$\\ 
NGC~1385     & 9.98 & 0.320  & 100 & F & 100 & $5.9\pm 0.5$\\
NGC~1511     & 9.91 & 0.356  & 90  & F & 80 & $2.0\pm 2.0$\\
NGC~1559     & 10.36& 0.576  & 100 & F & 100 & $5.9 \pm 0.3$\\
NGC~4781     & 9.64 & -0.322 & 80  & F & 80 & $7.0 \pm 0.3$\\
NGC~5068     & 9.40 & -0.560 & 70  & F & 70 & $6.0 \pm 0.4$\\
\noalign{\smallskip}
\hline
\noalign{\smallskip}
SB+A & 9.94 & 0.188  & 85 & - & 85 & $5.2 \pm 0.4$\\
PHANGS & 10.49& 0.318  & 86 & - & 76 & $3.5 \pm 0.1$\\
\hline 
\end{tabular}
\tablefoot{We list properties of individual \SB+\textbf{A} galaxies compared to averages of all \SB+\textbf{A} galaxies and all PHANGS galaxies with verified bar classes. Stellar masses and SFRs have a statistical uncertainty of ${\sim}0.1\,$dex, and for the bar and arm agreement of ${\sim}5-10$\%. }
\end{table*}

\begin{itemize}[noitemsep,topsep=0pt,leftmargin=2\parindent]
\item \textbf{NGC~4781:}  
Interestingly, for this galaxy our high-resolution CO data reveals a spiral arm that connects to the very center of the galaxy. 
In the other wavelength images this central spiral appears as an elongated feature in the center. 
We can not rule out the possibility that the distribution of CO clumps by chance appears like a spiral, e.g., some overshooting from gas entering the bar. 
We consider this an interesting candidate for further research. 

\item \textbf{IC~1954:} 
No clear bar is evident in the $3.6\,\mu$m image (compare to Figure~\ref{fig:SBAmultiICA}). 
Likely, young star-forming regions evident in H$\alpha$ affect the appearance and imitate a bar-like feature.  
This galaxy further shows a multi-arm pattern as both classified in the literature and our study. 
Several recent studies on this galaxy make use of literature classes by \citet[][]{buta_classical_2015} \citep[e.g.,][]{herrera-endoqui_catalogue_2015}, thus use the \SB classification for this galaxy. 
Older results such as \citet{phillips_nuclei_1996} find a short bar consisting of several knots in HST images. 
The work by \citet{querejeta_stellar_2021} agree with the \citet[][]{buta_classical_2015} literature classification, but find that the bar is clearly offset from the galaxy center. 

\item \textbf{NGC~1385:}
We do not find any bar-focused study on this galaxy. 
Our images do not reveal any evidence for a bar in the disk. 
The CO map as well as H$\alpha$ image do show a random distribution of clumps that connect in a non-structured way. 
Also, \citet{querejeta_stellar_2021} mark this galaxy as unbarred. 

\item \textbf{NGC~1511:} 
Its inclination of 72 degree makes it difficult to assess the actual geometry of the disk. 
The CO map shows no elongated bar-like object, while a bright elongated central feature can be identified in the \textit{Spitzer} and H$\alpha$ images. 
\citet{querejeta_stellar_2021} do not consider this galaxy as barred. 

\item \textbf{NGC~1559:} 
Similar to NGC~1511, the 3.6\,$\mu$m image shows an elongated feature that can not be seen in other tracers. 
\citet{querejeta_stellar_2021} agree with the literature classification. 
This could physically hint to a quiescent and weak bar, that has lost most of its gas content and does not actively form stars. 
More likely, some bright star forming regions blend together in the $3.6\,\mu$m image, as otherwise the bar would be significantly brighter on one side. 

\item \textbf{NGC~5068}: 
Its CO classification likely suffers from low sensitivity, given that most emission has low S/N.
All other wavelengths show an elongated bar-like object and the presence of a bar is agreed on by \citet{querejeta_stellar_2021}. 
We conclude a lack of visibility of the gas disk that lead to an incorrect CO-based classification. 
\end{itemize}
\FloatBarrier

\section{Details on the generation of bar lane masks}
\label{sec:Appendix:Barlanemasks}

For our quantitative assessment of bar lane properties, we require masks containing the bar lanes. 
This allows us to measure the distribution of CO emission along the masks as pendant to the visual bar lane length classes. 

First, we apply environmental masks \citep{querejeta_stellar_2021} to our deprojected and de-rotated broad mom-0 maps to mask out the emission of spiral arms, the central region and, if no spiral arms are present, the outer disk region outside the bar radius. 
As bar lanes can often be found towards the outer bar ends, we only mask features (e.g., spiral arms and centers) not related to the bar. 
We blank pixels at radii larger than $r_\mathrm{max}$, which is the bar radius $R_\mathrm{bar, proj}$ from \citet{diaz-garcia_characterization_2016} deprojected according to Equation~\ref{eq:Barsizedeprojection} from \citet{williams_applying_2021}.

\begin{equation}\label{eq:Barsizedeprojection}
    r_\mathrm{max} = R_\mathrm{bar, proj} \cdot \sqrt{ \cos{\Delta \mathrm{PA}}^2 + \left(\sin{\Delta \mathrm{PA}} / \cos{i} \right)^2}
\end{equation}

$\Delta$PA corresponds to the difference between galaxy position angle and bar position angle in the sky, and $i$ is the inclination. 
The deprojected values are listed in Table~\ref{tab:Barlanelist}.
For one galaxy, no literature bar size is available, so we use a visual estimate based on our CO data.

Next, the masked maps are converted to a polar projection, with a radial pixel size of $50\,$pc (roughly a third of the beam size of the unprojected data mom-0 maps) in which bar lanes appear as two equidistant curved features (see Figure~\ref{fig:PolarimageFullmask}) that are best described by an exponential function.
We visually isolate the bar lanes azimuthaly (values see Table~\ref{tab:Barlanelist}) and mask out pixels with S/N$< 5\sigma$. 
The bar lane located at lower azimuth is named ID~1, the bar lane at higher azimuth or at the polar image edge ID~2. 

Next, we fit the pixels with maximum intensity per row, per column and the overlap of both with an exponential profile shifted in radial and azimuthal direction. 
This results in initial guesses of the location of the bar lane emission, which is then visually confirmed and adapted if necessary. 
Our mask then consists of two exponentials of this shape that are shifted along the normal of the best-fit exponential function, as shown in Figure~\ref{fig:PolarimageFullmask}.

\begin{table}
\centering\footnotesize  
\caption{Strongly barred galaxies and their parameters used to study bar lanes}
\begin{tabular}{lclllll}
\hline\hline
\noalign{\smallskip}
 Galaxy  & BL$_\mathrm{length}$ & BL$_\mathrm{shape}$ & r$_{\mathrm{max}}$ & $\theta_{bl1}$ & $\theta_{bl2}$ &   \\
  & (s,l) & (c1-5) & [kpc] & [deg] & [deg] &  \\
\noalign{\smallskip}
(1) & (2) & (3) & (4) & (5) & (6) \\
\noalign{\smallskip}
\hline
\noalign{\smallskip}
NGC~1097  & l & 2.5 & 6.6 & 45-135 & 225-324\\
NGC~1300  & l & 1.5  & 7.4 & 36-126 & 225-288 & \\
NGC~1365  & s & 2.5  & 14.2 & 108-189 & 270-360\\
NGC~1433  & s & 4.0  & 6.7 & 72-162 & 252-360\\
NGC~1512  & s & 4.0  & 7.5 & 153-225 & 315-54 & \\
NGC~1566  & s & 4.5  & 3.2 & 135-198 & 306-18 & \\
NGC~1672 & l & 1.5  & 8.1 & 0-126 & 180-315\\
NGC~2566  & \nAG & 1.5  & 9.8 & 144-198 & 306-54\\
NGC~2903 & l & 1.5  & 3.4 & 36-144 & 234-342 & \\
NGC~3351  & s & 4.5  & 3.5 & 9-99 & 180-261 & b\\
NGC~3507 & s & 1.5 & 3.1 & 90-162 & 270-342 \\
NGC~3627  & l & 2.0 & 3.9 & 45-126 & 234-279 & \\
NGC~4303  & l & 3.0  & 3.1 & 63-162 & 252-342  & \\
NGC~4321  & l & 3.5  & 4.6 & 18-126 & 198-288 & \\
NGC~4535 & l & 2.0  & 3.5 & 126-198 & 306-36 & \\
NGC~4548  & l & 1.0  & 5.9 & 126-216 & 333-18\\
NGC~4579  & s & \nAG  & 4.7 & 0-108 & 198-324  & \\
NGC~4941 & s & 4.5  & 6.8 & 72-126 & 225-360 &  \\
NGC~5236  & l & 1.5  & 2.8 & 36-126 & 198-288 & c\\
NGC~5643 & l & 1.0  & 3.7 & 18-63 & 180-243\\
NGC~6300  & l & 1.0  & 2.4 & 0-72 & 162-234\\
NGC~7496 & l & 1.0  & 3.9 & 0-54 & 180-252 & \\
\noalign{\smallskip}
\hline\hline
\noalign{\smallskip}
\end{tabular}
\tablefoot{The 22 strongly barred galaxies (\textbf{C}) (1) with visual bar lane shape classification used for the detailed bar lane analysis presented in Section~\ref{sec:BarlaneAnalysis}. We list the visual bar lane length class (length of the CO bar lane compared to the full bar length) BL$_\mathrm{length}$ (2) and bar lane shape class BL$_\mathrm{shape}$ (3) (compare Table~\ref{tab:Allclasses}). 
We list parameters used for generating the bar lane masks: maximum radius r$_\mathrm{max}$ (4) which corresponds to the deprojected bar radius from \citet{herrera-endoqui_catalogue_2015} when possible, and \citet{menendezdelmestre_nearinfrared_2007} otherwise, or a visual estimate based on the CO images (see text). Columns (5,6) contain the azimuthal range we use to isolate both bar lanes ($\theta_{bl1}$, $\theta_{bl2}$ respectively).
b: no environmental mask available. c: no literature bar lane radius available. }\label{tab:Barlanelist}
\end{table}
\FloatBarrier

\section{Additional radial intensity profiles}
\label{Appendix:RadialProfiles}

We provide mean radial intensity profiles along the created bar lane masks for both bar lanes of 21 galaxies, as described in Section~\ref{sec:BarlaneMethods:Radialprofiles}. 

\begin{figure*}
    \centering
    \includegraphics[width = 0.99\textwidth]{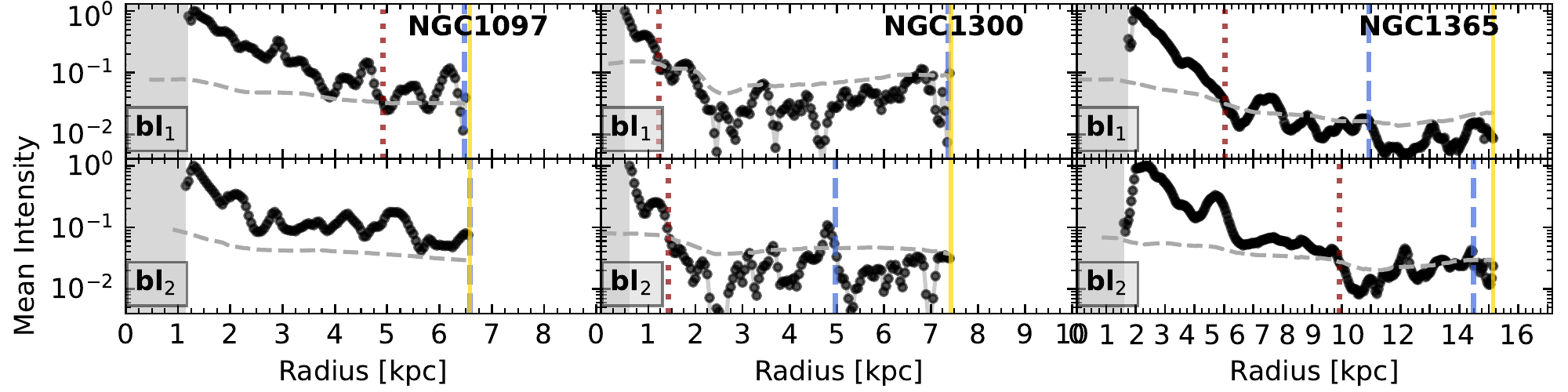}
    \includegraphics[width = 0.99\textwidth]{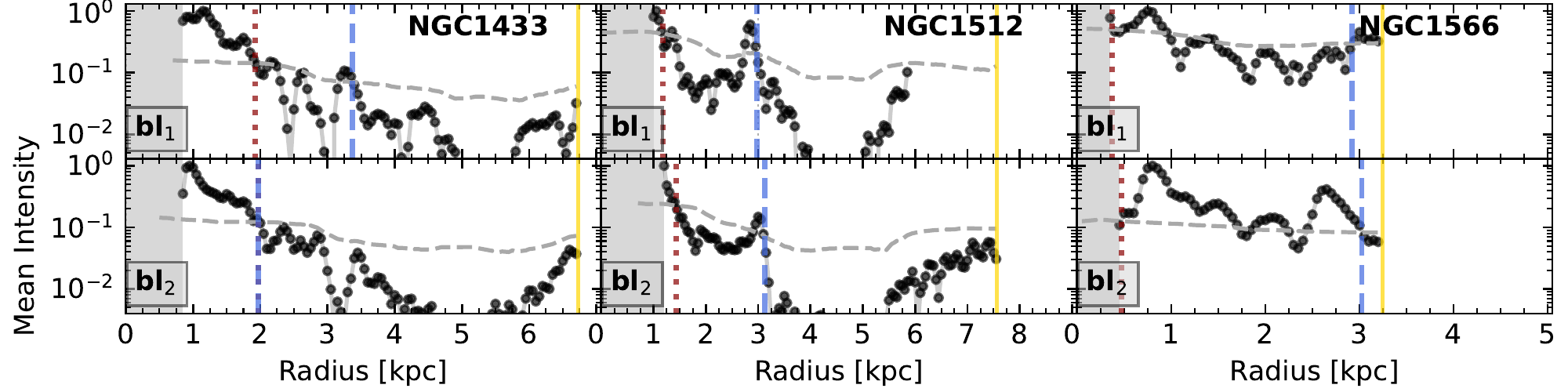}
    \includegraphics[width = 0.99\textwidth]{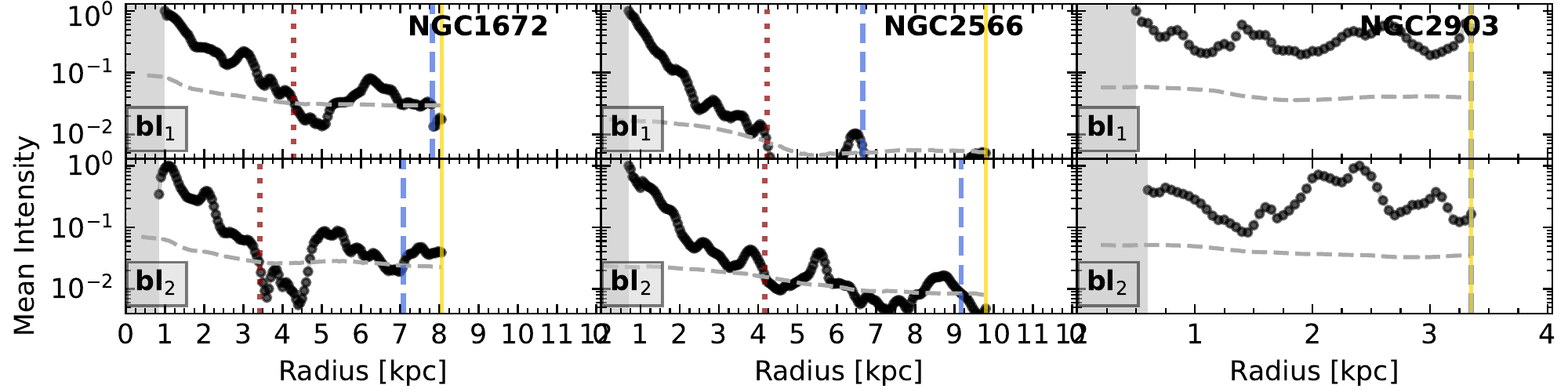}
    \includegraphics[width = 0.99\textwidth]{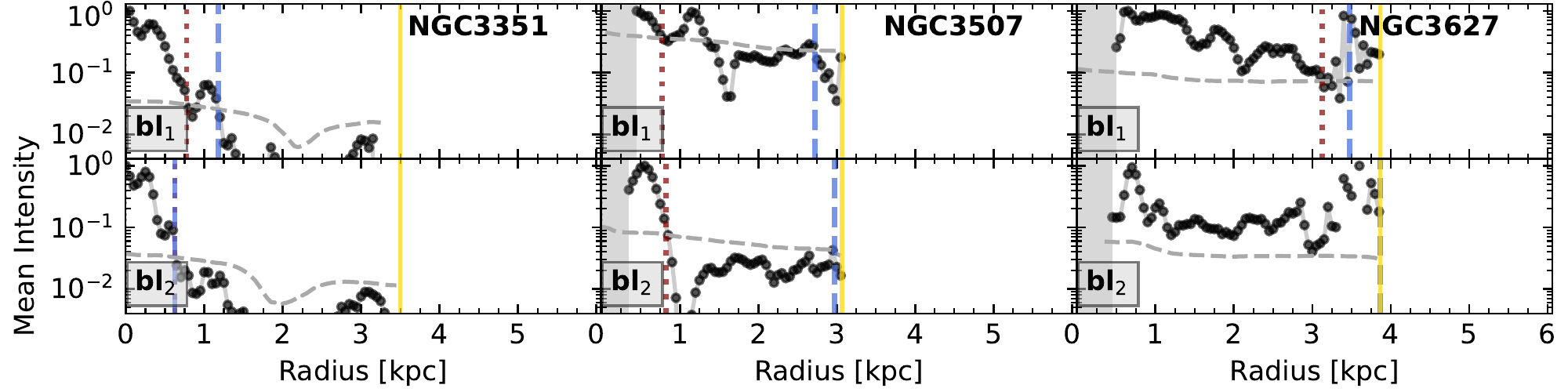}
    \includegraphics[width = 0.99\textwidth]{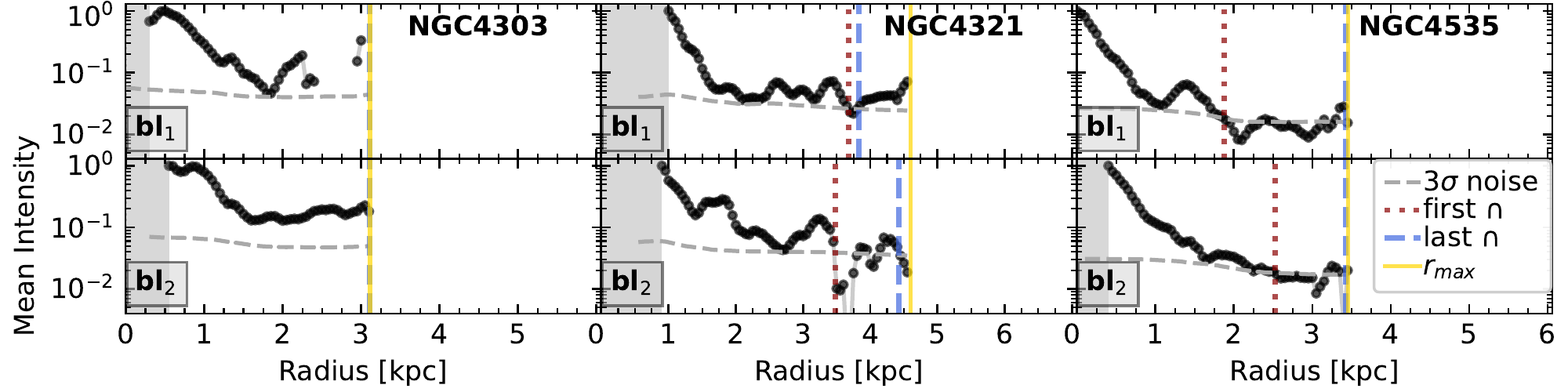}
    \caption{Normalized radial mean intensity profile for the two bar lanes of strongly barred PHANGS galaxies. 
    The corresponding average radial noise at $3\sigma$ is shown (grey dashed line). 
    The first (red dotted vertical line) and last (blue dashed vertical line) intersection of the bar lane emission with $3\sigma$, referred to as $r_\mathrm{bl, min}$ and $r_\mathrm{bl, max}$ respectively, provides the minimum and maximum bar lane length.
    The yellow solid line indicates the bar radius r$_{max}$. 
    For most galaxies, the central region is masked out (grey shaded area).}
    \label{fig:RadialprofilesApp}
\end{figure*}

\begin{figure*}
    \centering
    \ContinuedFloat
    \includegraphics[width = 0.99\textwidth]{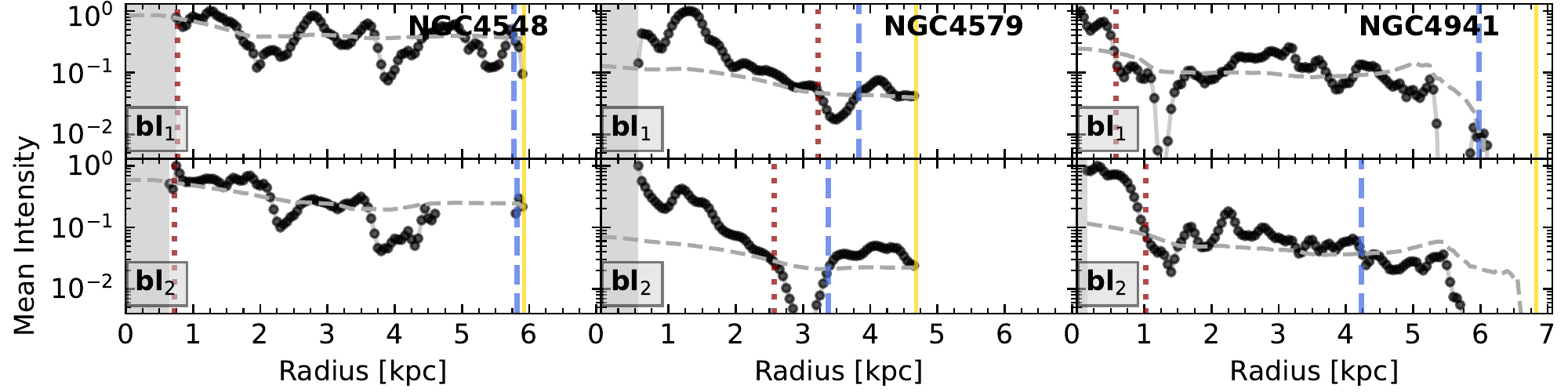}
    \includegraphics[width = 0.99\textwidth]{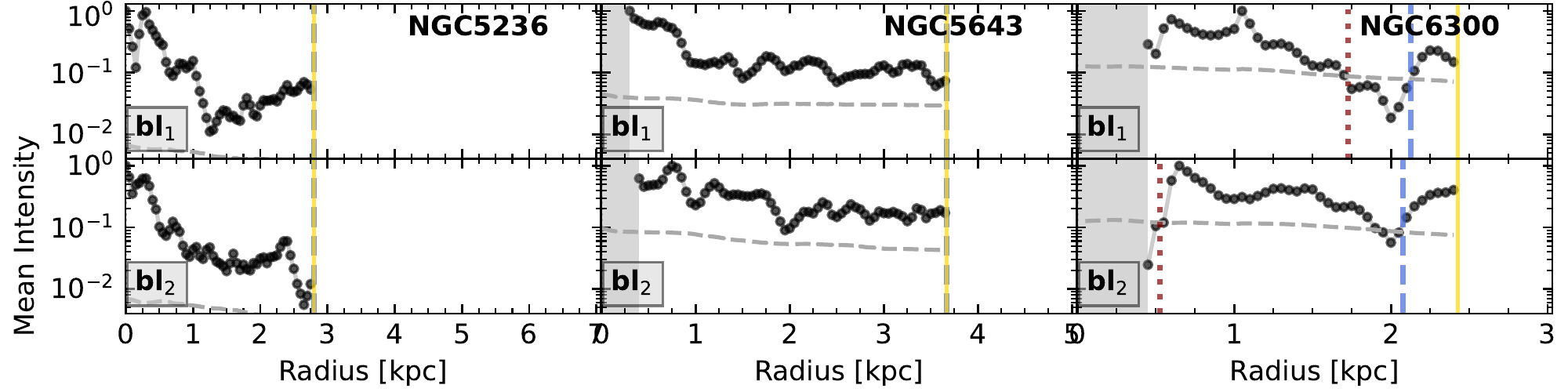}
    \includegraphics[width = 0.99\textwidth]{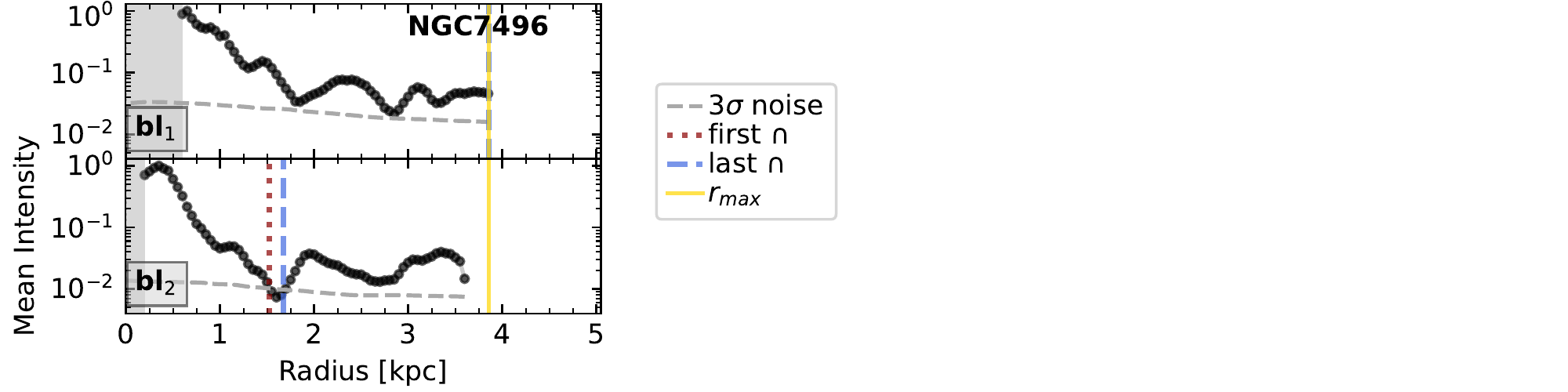}
    \caption{Continued}
\end{figure*}

\end{appendix}

\end{document}